\documentclass[10.5pt,compsoc]{JSC}

\usepackage[utf8]{inputenc}
\usepackage[T1]{fontenc}
\usepackage{hyphenat}
\usepackage{xspace}
\usepackage{amsmath}
\usepackage{amsfonts}
\usepackage{hyperref}
\usepackage{url}
\usepackage{booktabs}
\usepackage{multirow}
\usepackage{makecell}
\usepackage{caption}
\usepackage{minibox}
\usepackage{bbm}
\usepackage{graphicx}
\usepackage{balance}
\usepackage{mathtools}
\usepackage{color}
\usepackage{marvosym}
\usepackage{ifthen}
\usepackage{textcomp}
\usepackage{enumitem}
\usepackage{verbatim}
\usepackage{algorithm}
\usepackage{algorithmic}
\usepackage{numprint}
\usepackage{balance}

\usepackage{amsthm}
\theoremstyle{plain}

\newcommand{\chatoDisplayMode}[1]{#1}

\definecolor{MyRed}{rgb}{0.6,0.0,0.0} 
\definecolor{MyBlack}{rgb}{0.1,0.1,0.1} 
\newcommand{\inred}[1]{{\color{MyRed}\sf\textbf{\textsc{#1}}}}
\newcommand{\frameit}[2]{
  \begin{center}
  {\color{MyRed}
  \framebox[.9\columnwidth][l]{
    \begin{minipage}{.85\columnwidth}
    \inred{#1}: {\sf\color{MyBlack}#2}
    \end{minipage}
  }\\
  }
  \end{center}
}

\newcommand{\note}[2][]{\chatoDisplayMode{\def\@tmpsig{#1}\frameit{{\Pointinghand} Note}{#2\ifx \@tmpsig \@empty \else \mbox{ --\em #1}\fi}}}
\newcommand{\todo}[2][]{\chatoDisplayMode{\def\@tmpsig{#1}\frameit{{\Writinghand} To-do}{#2\ifx \@tmpsig \@empty \else \mbox{ --\em #1}\fi}}}

\newcommand{\abbrevStyle}[1]{#1}

\newcommand{\ie}{\abbrevStyle{i.e.}\xspace}
\newcommand{\eg}{\abbrevStyle{e.g.}\xspace}
\newcommand{\cf}{\abbrevStyle{cf.}\xspace}
\newcommand{\etal}{\abbrevStyle{et al.}\xspace}
\newcommand{\vs}{\abbrevStyle{vs.}\xspace}

\newcommand{\Secref}[1]{Sec.~\ref{#1}}
\newcommand{\Eqnref}[1]{Eq.~(\ref{#1})}

\newcommand{\Tabref}[1]{Table~\ref{#1}}
\newcommand{\Figref}[1]{Fig.~\ref{#1}}

\newcommand{\xhdr}[1]{\vspace{1.7mm}\noindent{{\bf #1.}}}

\newcommand{\textcite}[1]{\citeauthor{#1} \shortcite{#1}}

\newcommand{\hide}[1]{}

\makeatletter
\newcommand{\iffont}[2]{\ifthenelse{\equal{\f@family}{#1}}{#2}{}}
\makeatother

  \usepackage{mathptmx}

  \DeclareSymbolFont{greek}{OML}{cmm}{m}{n}
  \DeclareMathSymbol{\alpha}{\mathalpha}{greek}{"0B}
  \DeclareMathSymbol{\beta}{\mathalpha}{greek}{"0C}
  \DeclareMathSymbol{\gamma}{\mathalpha}{greek}{"0D}
  \DeclareMathSymbol{\delta}{\mathalpha}{greek}{"0E}
  \DeclareMathSymbol{\epsilon}{\mathalpha}{greek}{"0F}
  \DeclareMathSymbol{\zeta}{\mathalpha}{greek}{"10}
  \DeclareMathSymbol{\eta}{\mathalpha}{greek}{"11}
  \DeclareMathSymbol{\theta}{\mathalpha}{greek}{"12}
  \DeclareMathSymbol{\iota}{\mathalpha}{greek}{"13}
  \DeclareMathSymbol{\kappa}{\mathalpha}{greek}{"14}
  \DeclareMathSymbol{\lambda}{\mathalpha}{greek}{"15}
  \DeclareMathSymbol{\mu}{\mathalpha}{greek}{"16}
  \DeclareMathSymbol{\nu}{\mathalpha}{greek}{"17}
  \DeclareMathSymbol{\xi}{\mathalpha}{greek}{"18}
  \DeclareMathSymbol{\pi}{\mathalpha}{greek}{"19}
  \DeclareMathSymbol{\rho}{\mathalpha}{greek}{"1A}
  \DeclareMathSymbol{\sigma}{\mathalpha}{greek}{"1B}
  \DeclareMathSymbol{\tau}{\mathalpha}{greek}{"1C}
  \DeclareMathSymbol{\upsilon}{\mathalpha}{greek}{"1D}
  \DeclareMathSymbol{\phi}{\mathalpha}{greek}{"1E}
  \DeclareMathSymbol{\chi}{\mathalpha}{greek}{"1F}
  \DeclareMathSymbol{\psi}{\mathalpha}{greek}{"20}
  \DeclareMathSymbol{\omega}{\mathalpha}{greek}{"21}
  \DeclareMathSymbol{\varepsilon}{\mathalpha}{greek}{"22}
  \DeclareMathSymbol{\vartheta}{\mathalpha}{greek}{"23}
  \DeclareMathSymbol{\varpi}{\mathalpha}{greek}{"24}
  \DeclareMathSymbol{\varrho}{\mathalpha}{greek}{"25}
  \DeclareMathSymbol{\varsigma}{\mathalpha}{greek}{"26}
  \DeclareMathSymbol{\varphi}{\mathalpha}{greek}{"27}
  \DeclareSymbolFont{otone}{OT1}{cmr}{m}{n}
  \DeclareMathSymbol{\Gamma}{\mathalpha}{otone}{0}
  \DeclareMathSymbol{\Delta}{\mathalpha}{otone}{1}
  \DeclareMathSymbol{\Theta}{\mathalpha}{otone}{2}
  \DeclareMathSymbol{\Lambda}{\mathalpha}{otone}{3}
  \DeclareMathSymbol{\Xi}{\mathalpha}{otone}{4}
  \DeclareMathSymbol{\Pi}{\mathalpha}{otone}{5}
  \DeclareMathSymbol{\Sigma}{\mathalpha}{otone}{6}
  \DeclareMathSymbol{\Upsilon}{\mathalpha}{otone}{7}
  \DeclareMathSymbol{\Phi}{\mathalpha}{otone}{8}
  \DeclareMathSymbol{\Psi}{\mathalpha}{otone}{9}
  \DeclareMathSymbol{\Omega}{\mathalpha}{otone}{10}
  \DeclareSymbolFont{syms}{OML}{cmm}{m}{it}
  \DeclareMathSymbol{\partial}{\mathord}{syms}{"40}
  \DeclareMathAlphabet{\mathbold}{OML}{cmm}{b}{it}
  \DeclareSymbolFont{largesymbols}{OMX}{cmex}{m}{n}

\usepackage{float}
\usepackage{subcaption}
\usepackage{cuted}  

\graphicspath{{Plots/}}
\usepackage[noadjust]{cite}

\newcommand{\est}{\mathrm{est}}

\newcommand{\studyOne}{study~1\xspace}
\newcommand{\studyTwo}{study~2\xspace}

\headevenname{\zihao{-5}{\textbf{\emph{Journal of Social Computing, \quad \quad }}} 20??, ?(?): ???-???}%
\headoddname{{\sf Martynov et al.:}\quad {\textbf{\emph{Darks and Stripes: Effects of Clothing on Weight Perception}}}}%

\makeatletter
\renewcommand{\@biblabel}[1]{[#1]\hfill}
\makeatother

\begin{document}

\tolerance=1000
\hyphenpenalty=1000

\makeatletter
\newcommand\mysmall{\@setfontsize\mysmall{7}{9.5}}

\newenvironment{tablehere}
  {\def\@captype{table}}
  {}
\newenvironment{figurehere}
  {\def\@captype{figure}}
  {}

\thispagestyle{plain}%
\thispagestyle{empty}%

\let\temp\footnote
\renewcommand \footnote[1]{\temp{\zihao{-5}#1}}
{}
\vspace*{-40pt}
\noindent{\zihao{5-}\textbf{\scalebox{0.95}[1.0]{\makebox[5.93cm][s]
{JOURNAL\hfil OF\hfil SOCIAL\hfil COMPUTING}}}}

\vskip .2mm
{\zihao{5-}
\textbf{
\hspace{-5mm}
\scalebox{1}[1.0]{\makebox[5.6cm][s]{%
I\hspace{0.70pt}S\hspace{0.70pt}S\hspace{0.70pt}N\hspace{0.70pt}{\color{black}%
\hspace{2pt}\hspace{5.70pt}2\hspace{0.70pt}6\hspace{0.70pt}8\hspace{0.70pt}8\hspace{0.70pt}-\hspace{0.70pt%
}5\hspace{0.70pt}2\hspace{0.70pt}5\hspace{0.00pt}5\hspace{0.70pt}\hspace{0.70pt%
}\hspace{0.70pt}{\color{white}l\hspace{0.70pt}l\hspace{0.70pt}}0\hspace{0.70pt}%
?\hspace{0.70pt}/\hspace{0.70pt}?\hspace{0.70pt}?\hspace{0.70pt}{\color{white}%
l\hspace{0.70pt}l\hspace{0.70pt}}p\hspace{0.70pt}p\hspace{0.70pt}?\hspace{0.70pt}?\hspace{0.70pt}?%
--\hspace{0.70pt}?\hspace{0.70pt}?\hspace{0.70pt}?}}}}
}
\vskip .2mm\noindent
{\zihao{5-}\textbf{\scalebox{1}[1.0]{\makebox[5.6cm][s]{%
V\hspace{0.4pt}o\hspace{0.4pt}l\hspace{0.4pt}u\hspace{0.4pt}m\hspace{0.4pt}%
e\hspace{0.4em}?\hspace{0.4pt},\hspace{0.8em}N\hspace{0.4pt}u\hspace{0.4pt}%
m\hspace{0.4pt}b\hspace{0.4pt}e\hspace{0.4pt}r\hspace{0.4em}?,\hspace{0.8em}%
\color{white}{J\hspace{0.4pt}a\hspace{0.4pt}n\hspace{0.4pt}u\hspace{0.4pt}a\hspace{0.4pt}%
\hspace{0.4pt}r\hspace{0.4pt}y\hspace{0.4em}2\hspace{0.4pt}0\hspace{0.4pt}1\hspace{0.4pt}8}}}}}

\begin{strip}
{\center
{\zihao{3}\textbf{
Darks and Stripes: Effects of Clothing on Weight Perception
}}
\vskip 9mm}

{\center {\sf \zihao{5}
Kirill Martynov, Kiran Garimella, Robert West$^*$
}
\vskip 5mm}
%

\centering{
\begin{tabular}{p{160mm}}

{\zihao{-5}
\linespread{1.6667} %
\noindent
\bf{Abstract:}
{\sf
In many societies, appearing slim (corresponding to a small body-size index) is considered attractive.
The fashion industry has been attempting to cater to this trend by designing outfits that can enhance the appearance of slimness.
Two anecdotal rules, widespread in the world of fashion, are
(1)~to choose dark clothes and
(2)~to avoid horizontal stripes,
in order to appear slim.
Thus far, empirical evidence has been unable to conclusively determine the validity of these rules, and there is consequently much controversy regarding the impact of both color and patterns on the visual perception of weight.
In this paper, we aim to close this gap by presenting the results from a series of large-scale crowdsourcing studies that investigate the above two claims.
We gathered a dataset of around 1,000 images of people from the Web together with their ground-truth weight and height, as well as clothing attributes about colors and patterns.
To elicit the effects of colors and patterns, we asked crowd workers to estimate the weight in each image.
For the analysis, we controlled potential confounds by matching images in pairs where the two images differ with respect to color or pattern, but are similar with respect to other relevant aspects.
We created image pairs in two ways:
first, observationally, i.e., from two real images;
and second, experimentally, by manipulating the color or pattern of clothing in a real image via photo editing.
Based on our analysis, we conclude that
(1)~dark clothes indeed decrease perceived weight slightly but statistically significantly, and
(2)~horizontal stripes have no discernible effect compared to solid light\hyp colored clothes.
These results contribute to advancing the debate around the effect of specific clothing colors and patterns and thus provide empirical grounds for everyday fashion decisions.
Moreover, our work gives an outlook on the vast opportunities of using crowd sourcing in the modern fashion industry.
}
}

\end{tabular}
}
\vskip 6mm

\vskip -3mm
\zihao{6}
\end{strip}

\thispagestyle{plain}%
\thispagestyle{empty}%
\makeatother
\pagestyle{tstheadings}

\begin{figure}[b]
\vskip -6mm
\begin{tabular}{p{44mm}}
\toprule\\
\end{tabular}
\vskip -4.5mm
\noindent
\setlength{\tabcolsep}{1pt}
\begin{tabular}{p{1.5mm}p{79.5mm}}
&\\
$\bullet$& Kirill Martynov, Google, D-80636 Munich, Germany (work done at EPFL and TUM). Email: kirmartynov@gmail.com \\
$\bullet$& Kiran Garimella, MIT, Cambridge, MA 02139, USA. Email: garimell@mit.edu \\
$\bullet$& Robert West, EPFL, CH-1004 Lausanne, Switzerland. Email: robert.west@epfl.ch \\
$\sf{^*}$&
To whom correspondence should be addressed. \\
          &          Published at the IEEE Journal for Social Computing. Please cite the journal version. \url{https://ieeexplore.ieee.org/stamp/stamp.jsp?tp=&arnumber=9241513}

\end{tabular}
\end{figure}\zihao{5}

\section{Introduction}

Western female beauty standards are dominated by an ideal that favors slimness, or, more technically, a small body-mass index \cite{adaval2019seeing,tovee1998}.
Historically, the ideal body-mass index has become ever smaller over time \cite{voracek2003}.
While slimness plays a lesser role in male beauty standards, studies suggest that obesity can have a negative impact on male attractiveness \cite{maisey1999}.
Based on the widespread belief that particular choices of clothing can enhance or reduce perceived body size~\cite{frith2004clothing}, a major question for the fashion industry, as well as for individuals worldwide, revolves around the impact of clothing on body-size perception.
The present paper addresses two particularly well-known pieces of anecdotal fashion advice:
the claim that horizontally striped clothes increase body-size perception~\cite{swami2012}, and
the claim that dark clothes decrease body-size perception~\cite{winakor1987effect}.

On the one hand, the purported advantage of dark clothes has been widely assumed in the fashion industry \cite{klepp2017} as well as the research community \cite{kremkow2014}.
There is a general agreement that a dark object is perceived as smaller compared to a light-colored object of the same size.
The assumed effect in the context of clothing is, however, mostly based on anecdotal evidence and, compared to the ubiquity of folk wisdom around dark clothing, there is a scarcity of large\hyp scale empirical studies that could confirm or quantify the effect of dark clothes on weight perception.

On the other hand, there is much controversy about the effect of horizontal stripes.
Folk wisdom states that horizontal stripes increase perceived body size \cite{feldon2000}, in stark contrast to predictions of physiological optics, such as the Helmholtz \cite{helmholtz1867} and Oppel--Kundt \cite{oppel,kundt} illusions, which state that horizontal stripes make rectangular shapes appear both taller and thinner.
Even recent empirical studies disagree about the effect of horizontal stripes: while some have claimed that horizontal stripes increased perceived weight \cite{swami2012}, others have suggested that the difference between striped and solid\hyp colored clothes was negligible \cite{thompson2011}, whereas Helmholtz---echoing the predictions of the aforementioned optical illusion named after him---even claimed that horizontal stripes made a figure appear taller~\cite{helmholtz1867}.
Prior studies are, however, limited by their small scale---usually involving fewer than 100 participants, all rating a single image~\cite{swami2012}---, which makes it difficult to draw generalizable conclusions.

This work takes a novel approach to the problem, relying on crowdsourcing \cite{chittilappilly2016survey, kittur2008crowdsourcing, luz2015survey, yuen2011survey} to study the effect of clothing on weight perception.
We conducted a series of studies in which crowd workers were shown images of people and estimated their weight and height, or indicated which of two shown people they considered to weigh less.

The image dataset consisted of around 1,000 photographs depicting people wearing various styles of clothing, taken under natural circumstances and posted on an online weight-loss forum.
All images were annotated with ground-truth weight and height labels by the users who uploaded them to the forum, and were annotated with color and stripe labels by the authors of this paper, using custom algorithms developed for this work.

Based on these images, we designed a matched observational study for estimating the effects of colors and stripes (\Secref{sec:Research design:Observational study}).
To overcome the limitations imposed by potential unobserved confounds, we augmented the original dataset of real images with carefully manipulated versions.
In particular, starting from original images showing people wearing horizontal stripes, a graphic\hyp design expert used Adobe Photoshop to produce versions where the originally striped clothes were replaced by solid light and solid dark clothes, respectively, while keeping everything else in the image fixed.
Based on the resulting images, we designed two studies for estimating the effects of colors and stripes (\Secref{sec:Research design:Experimental study 1}--\ref{sec:Research design:Experimental study 2}) that, due to the careful manipulation of images, are experimental in nature and thus not hampered by the same potential unobserved confounds as the observational study.

By analyzing more than 75k estimates from around 6.5k crowd workers, we arrived at two main conclusions:
\begin{enumerate}
\item Solid dark clothes indeed
make a person appear to weigh slightly but statistically significantly less,
such that a person switching from solid light to solid dark clothes can increase their chance of appearing to weigh less than a similar\hyp looking reference person by 2.7 percentage points ($p=0.0069$).
\item Horizontal stripes and solid light colors are indistinguishable in terms of weight perception ($p=0.58$).
\end{enumerate}

In a casual but catchy formula, if $D$ stands for solid dark, $L$ for solid light, and $S$ for horizontal stripes, our results about weight perception may be summarized as
$$\text{``}D < L \approx S\text{''}$$
(``the weight of solid dark is perceived as lower than that of solid light, which is indistinguishable from that of horizontal stripes'').

Taken together, this research contributes to advancing the longstanding debate around the effect of specific clothing types on body-size perception and thus helps to lay solid empirical grounds for everyday fashion advice.
Given the importance of slimness for the Western ideal of beauty \cite{maisey1999,tovee1998,voracek2003}, the reach of our findings goes considerably beyond the world of academia.
Furthermore, it showcases the vast opportunities of crowdsourcing for the modern fashion industry.

\section{Annotated image data}
\label{sec:Annotated image data}

We begin by describing the collection of images used in this research (\Secref{sec:Weight- and height-labeled images}), followed by a description of the algorithms we designed for annotating images with
stripe (\Secref{sec:stripes detection})
and color (\Secref{sec:color classification}) labels.


\subsection{Weight- and height-labeled images}
\label{sec:Weight- and height-labeled images}

The image data used in this paper was collected from a Reddit forum (``subreddit'') called \textit{r/progresspics},\footnote{\url{https://www.reddit.com/r/progresspics}}
where users who intend to lose (or sometimes gain) weight post pictures of themselves before and after their weight transformation.
Each sample contains the ID of the Reddit post, the height and gender of the user, their weight before and after the transformation, as well as one or several images. 
In order to be able to attach unique weight labels to all images, we automatically removed posts with more than two images (``before'' and ``after'') and ensured that each image shows exactly one fully dressed person.

\begin{table}
    \caption{Classification of the images used in our studies per BMI-based body types.
    ``Real images'' refers to the images used in the observational study (\Secref{sec:Research design:Observational study}); ``Manipulated images'', to the images used in the experimental studies (\Secref{sec:Research design:Experimental study 1}--\ref{sec:Research design:Experimental study 2}).
    Body-type boundaries in terms of BMI points as defined by the Centers for Disease Control and Prevention: 
    18.5, 
    25, 
    30 kg$/$m$^2$. 
    }
    \centering
    \begin{tabular}{lrr}
                & \textbf{Real images}  & \textbf{Manipulated images} \\
                & \textbf{(\Secref{sec:Research design:Observational study})}  & \textbf{(\Secref{sec:Research design:Experimental study 1}--\ref{sec:Research design:Experimental study 2})} \\
    \hline
    Underweight & 1\%   & 1\% \\
    Normal      & 18\%  & 15\% \\
    Overweight  & 25\%  & 24\% \\
    Obese       & 56\%  & 60\% \\
    \hline
	\end{tabular}
    \label{tab:BMI}
\end{table}

The dataset used in our analyses was assembled from two parts, A and B:
part~A consists of 10k samples that were generously provided by Kocabey \etal\ \cite{kocabey2017face};
part~B consists of 20k further samples that were crawled by the authors themselves.
From part~A, we extracted 600 images (348 female, 252 male) of people wearing solid dark or solid light colors (details in \Secref{sec:color classification}).
From the union of parts A and B, we extracted 100 images (54 female, 46 male) of people wearing horizontally striped clothes (details in \Secref{sec:stripes detection}).
Taken together, we worked with 700 annotated images.

\begin{figure}[t]
    \centering  
    \begin{subfigure}{.49\linewidth}
        \centering
        \includegraphics[width=\linewidth, clip=true, trim=0 260 20 150]{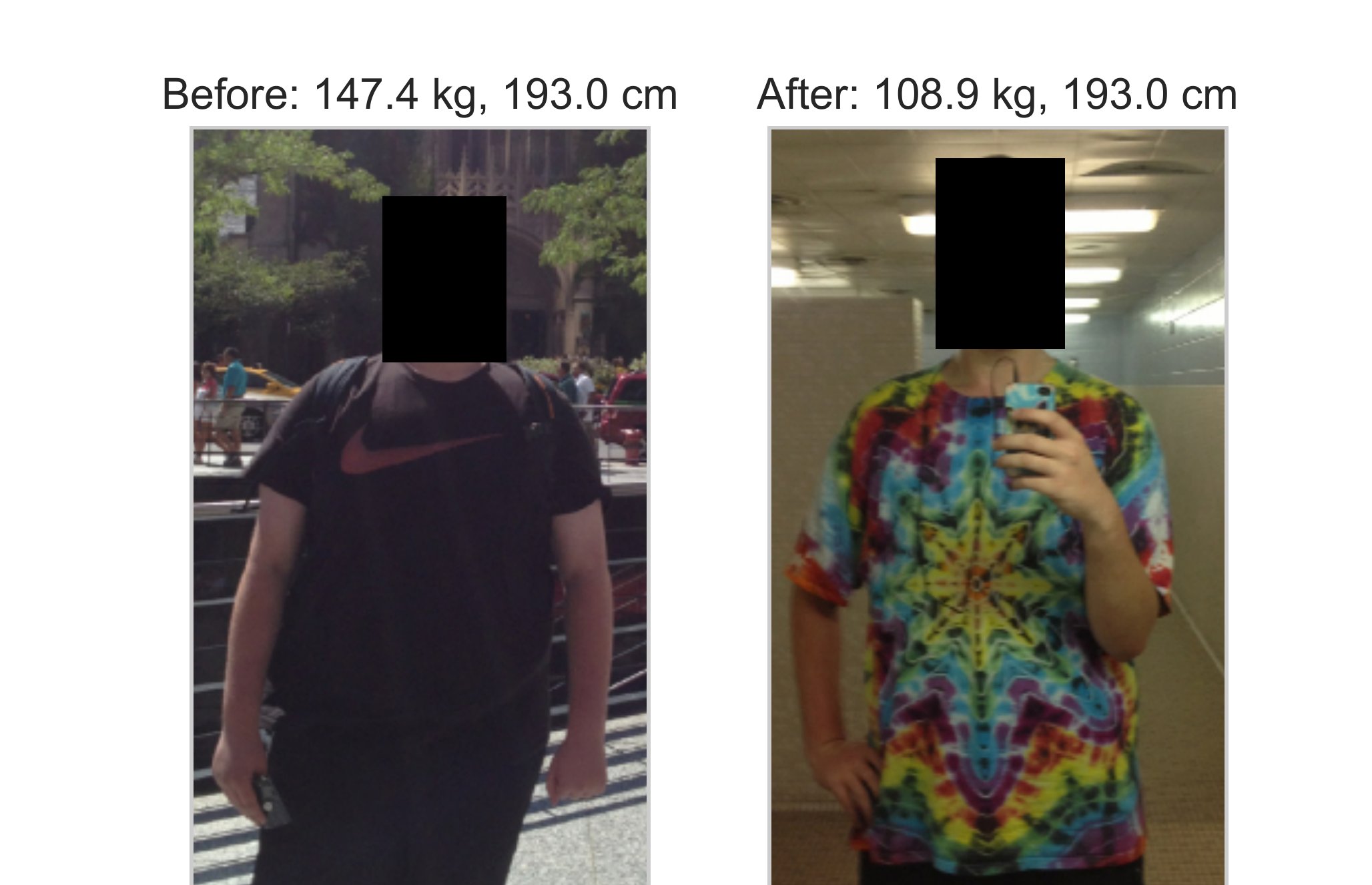}
        \label{fig:sample1}
    \end{subfigure}%
    \begin{subfigure}{.49\linewidth}
        \centering
        \includegraphics[width=\linewidth, clip=true, trim=0 260 20 150]{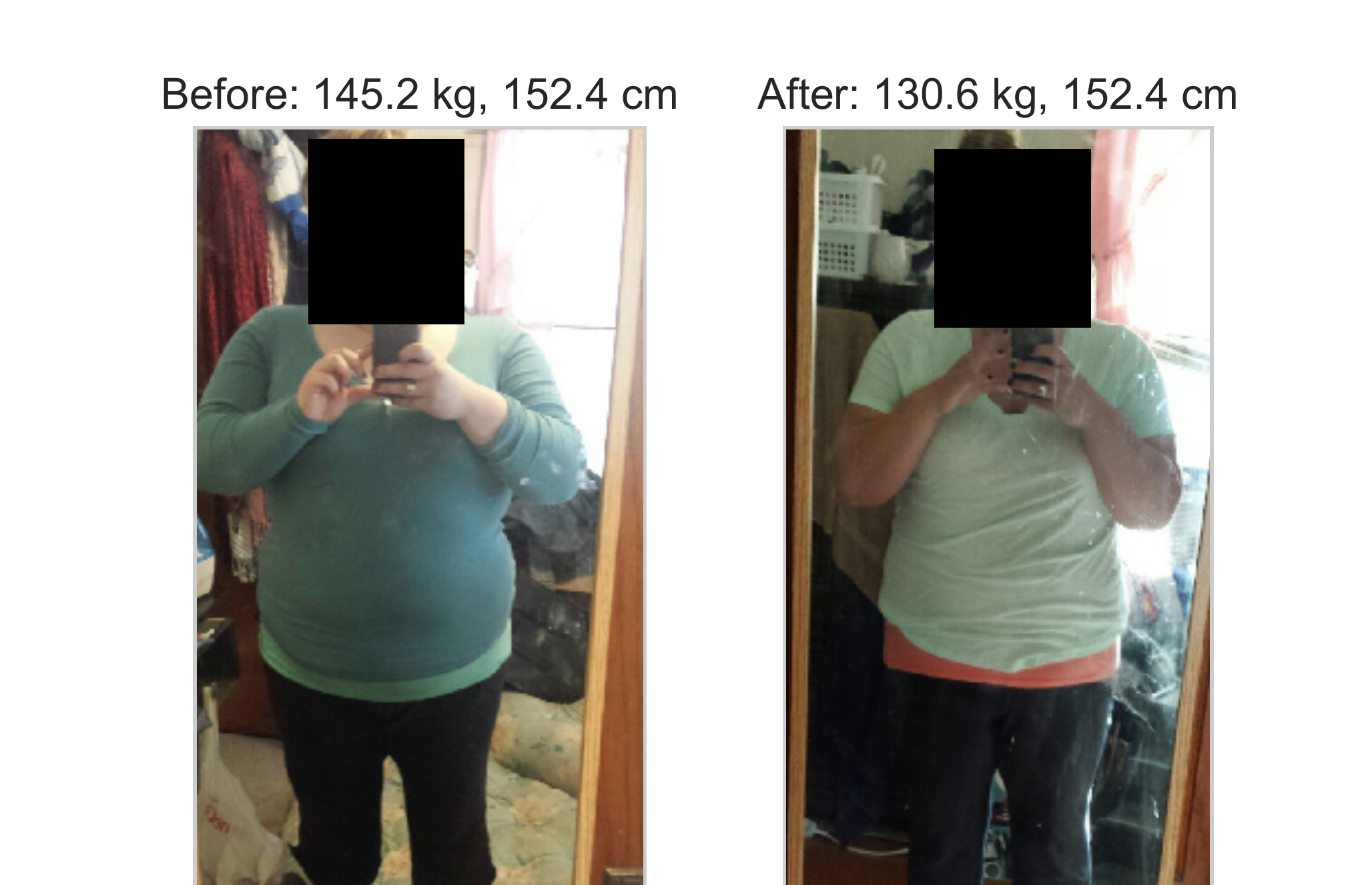}
        \label{fig:sample2}
    \end{subfigure}
\caption{Two samples from the dataset of weight- and height\hyp labeled images collected from Reddit. Each sample contains a ``before'' and an ``after'' image.
Faces censored only for paper.
}
\label{fig:samples}
\end{figure}

A summary of the the body\hyp type classification by body-mass index (BMI)
is presented in the left column (labeled ``Real images'') of \Tabref{tab:BMI}.
We see that a majority of the users in our dataset are obese.
Visual inspection of the dataset further revealed that 
around 35\% of the images are full-body pictures, whereas the rest only contain upper bodies (starting from the hips),
and that around 40\% of the images are ``selfies'' taken in a mirror.
The head is visible in over 90\% of the images.
Two samples from the dataset are shown in \Figref{fig:samples}.


We automatically annotated the clothing in each image with a color attribute and with a flag indicating the presence or absence of horizontal stripes.
Color and stripe annotations were exclusively based on upper-body clothing, for two reasons:
first, because upper-body measurements, in particular waist circumference, are a strong
indicator of obesity \cite{lean1995waist} and thus crucial for weight perception;
and second, because only 35\% of images in our dataset show full bodies (see above).
The methods used to obtain stripe and color annotations are described next.

\begin{figure*}
\centering
\begin{subfigure}{.18\linewidth}
  \centering
  \includegraphics[width=0.45\linewidth, clip=true, trim=50 10 100 10]{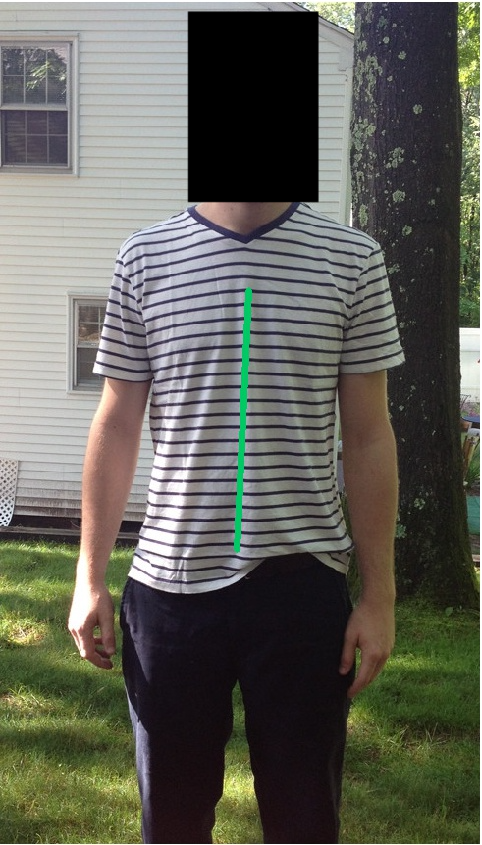}
  \caption{Image}
\end{subfigure}%
\begin{subfigure}{.27\linewidth}
  \centering
  \includegraphics[width=\linewidth]{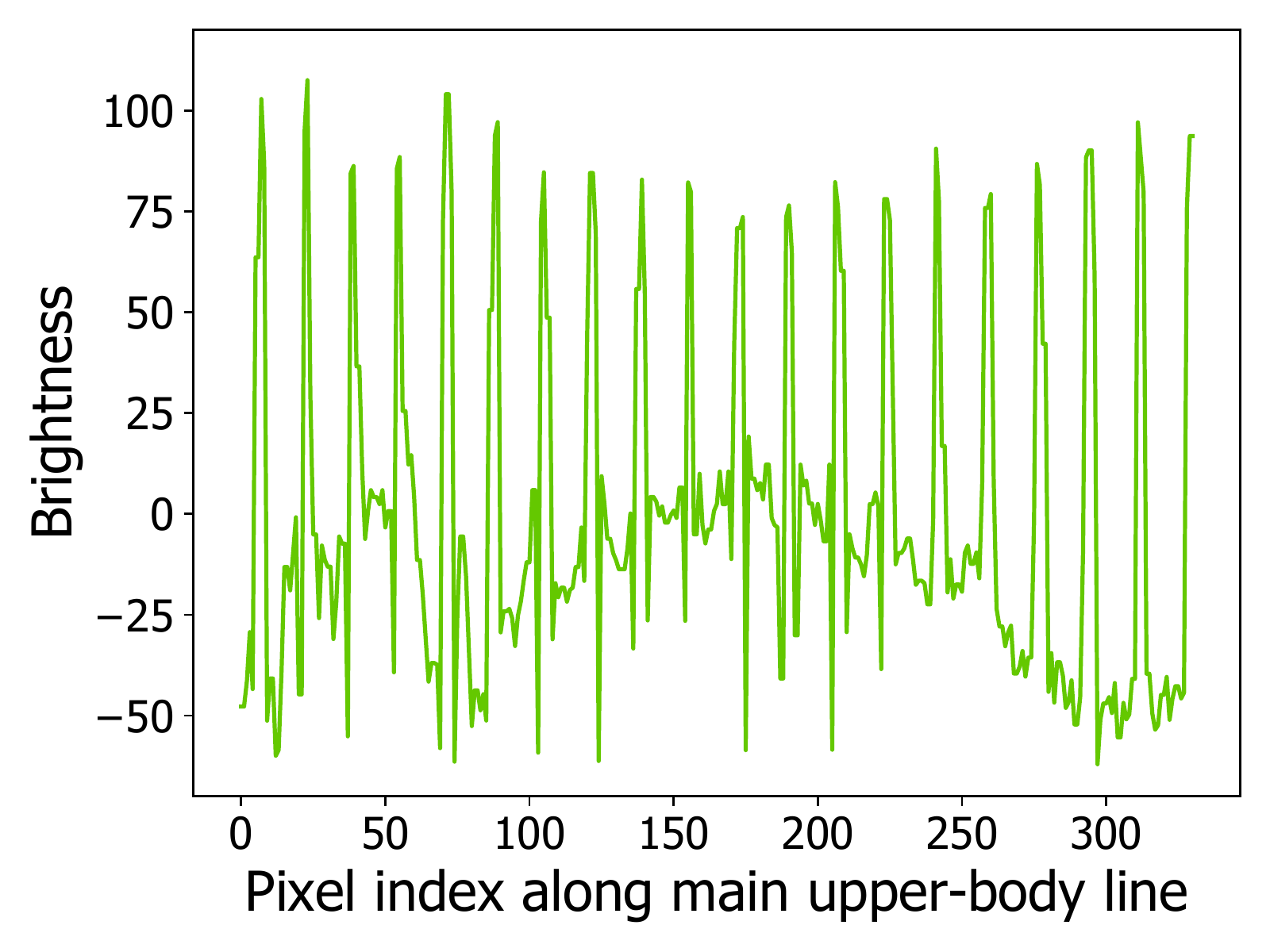}
  \caption{Brightness vector}
\end{subfigure}
\begin{subfigure}{.27\linewidth}
  \centering
  \includegraphics[width=\linewidth]{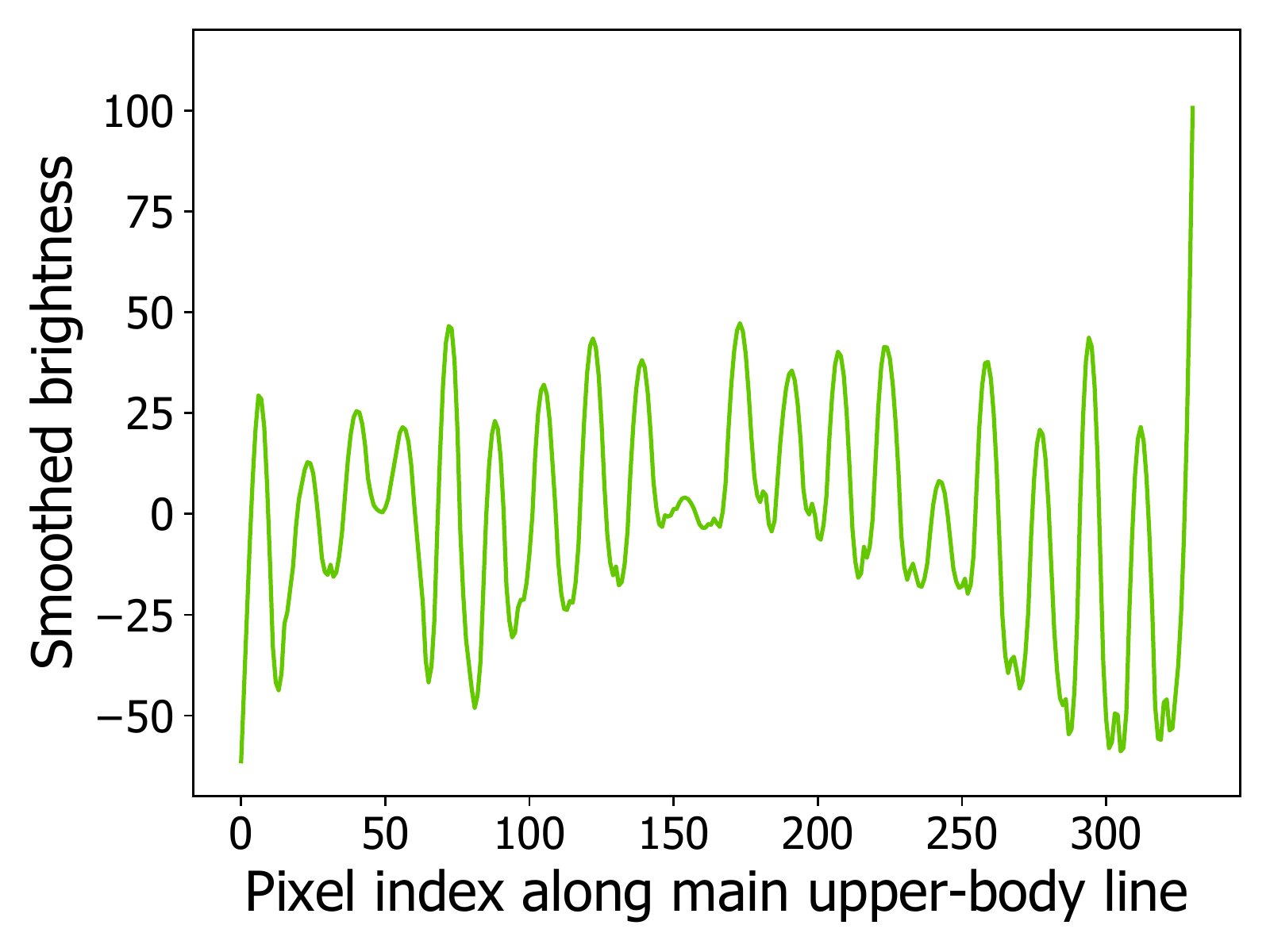}
  \caption{Smoothed brightness vector}
\end{subfigure}
\begin{subfigure}{.27\linewidth}
  \centering
  \includegraphics[width=\linewidth]{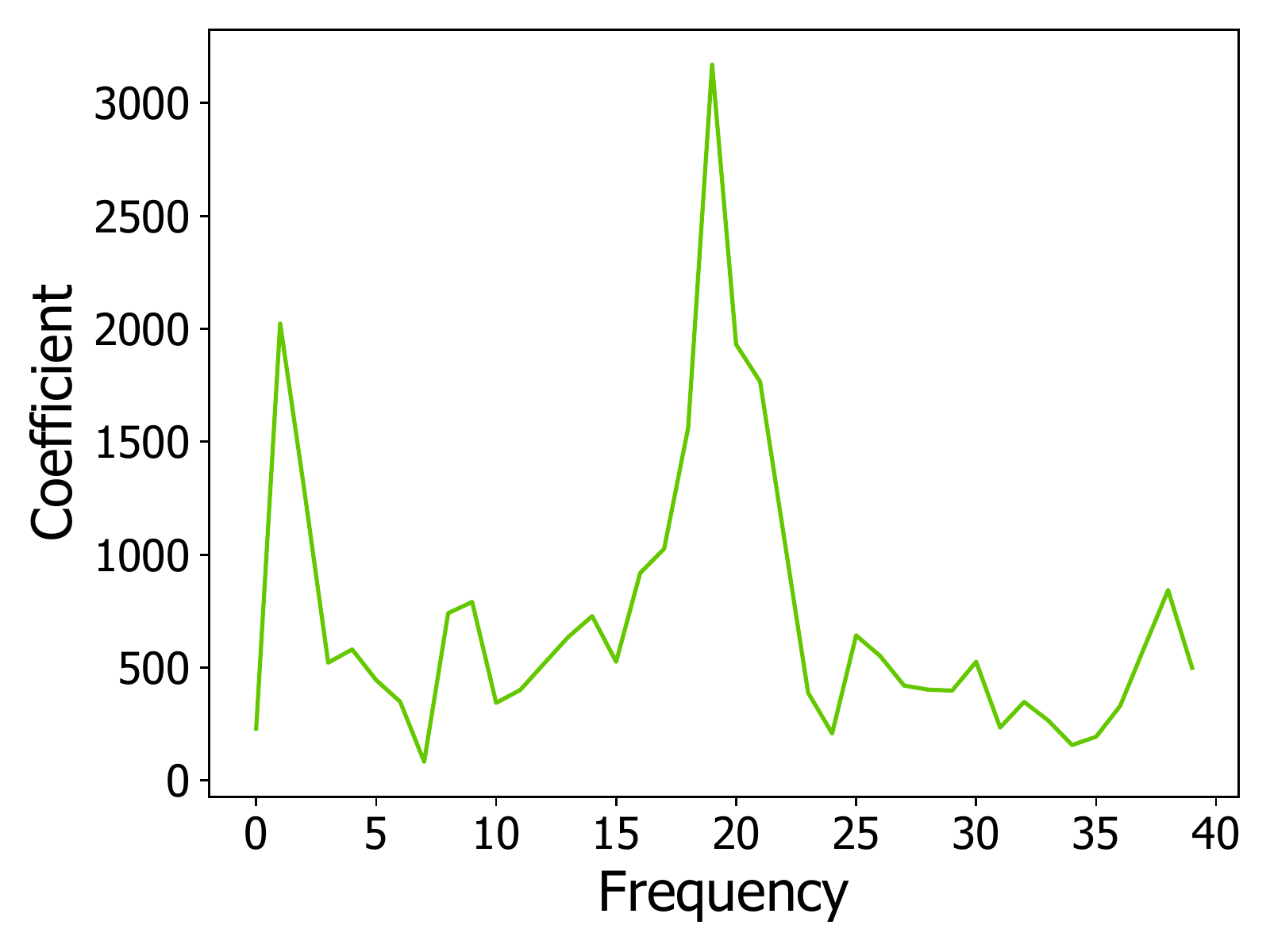}
  \caption{Fourier spectrum}
\end{subfigure}
\vspace{2mm}
\caption{
Steps of the stripe detection algorithm:
\textbf{(a)} identify main upper-body line (MUBL; drawn in green);
\textbf{(b)} construct vector of brightness values along MUBL, where brightness of a pixel equals $L_2$-norm of RGB triple (mean\hyp centered along MUBL);
\textbf{(c)} smooth the brightness vector;
\textbf{(d)} apply Fourier transform to obtain coefficient associated with each brightness frequency.
The sample is classified as positive (\ie, horizontally striped) because the frequency with the largest coefficient, 19, lies between 6 and 25.
Note that exactly 19 stripes are crossed by the MUBL.
}
\label{fig:stripes classifier}
\end{figure*}

\subsection{Stripe detection}
\label{sec:stripes detection}

To automatically detect horizontal stripes, we first determined the pose of the depicted body by leveraging PoseNet~\cite{cao2016}, a state-of-the-art pose detection algorithm that, given a body image, identifies all joints and body parts.
Using PoseNet, we extracted the main upper-body line, running from the neck to the underbelly, and constructed a vector of brightness values along the detected line, where the brightness of a pixel is defined as the $L_2$-norm of the pixel's RGB triple after mean\hyp centering by the average RGB triple along the line.
Next, we remove noise by processing the brightness vector with a median filter of width 5, followed by a third-order Savitzky--Golay filter~\cite{savitzky1964smoothing}. 
Finally, we apply a Fourier transform.
The magnitude of Fourier coefficients is crucial: our experiments suggest that horizontal stripes lead to a large coefficient for the frequency that corresponds to the actual number of visible stripes, which gives rise to the following heuristic rule:
classify as horizontally striped any outfit whose maximal Fourier coefficient is associated with a frequency between 6 and 25, corresponding to a typical number of horizontal stripes, as manually determined by the authors.
\Figref{fig:stripes classifier} illustrates the method on an image that was classified as positive.

The above\hyp described heuristic stripe detection algorithm has high recall but a relatively low precision of around 20\%.
It misclassifies as positive numerous other types of periodic variations of color, such as checkerboard patterns, letters, graphics, or shading patterns caused by certain lighting conditions.
Despite the high false\hyp positive rate, the algorithm worked well for our purpose, as it provided a cheap and fast way to sift through more than 30k images and returned a set of candidates that could then be rapidly filtered by the authors by visual inspection.

\begin{figure}[tb]
  \centering
  \includegraphics[width=0.45\textwidth]{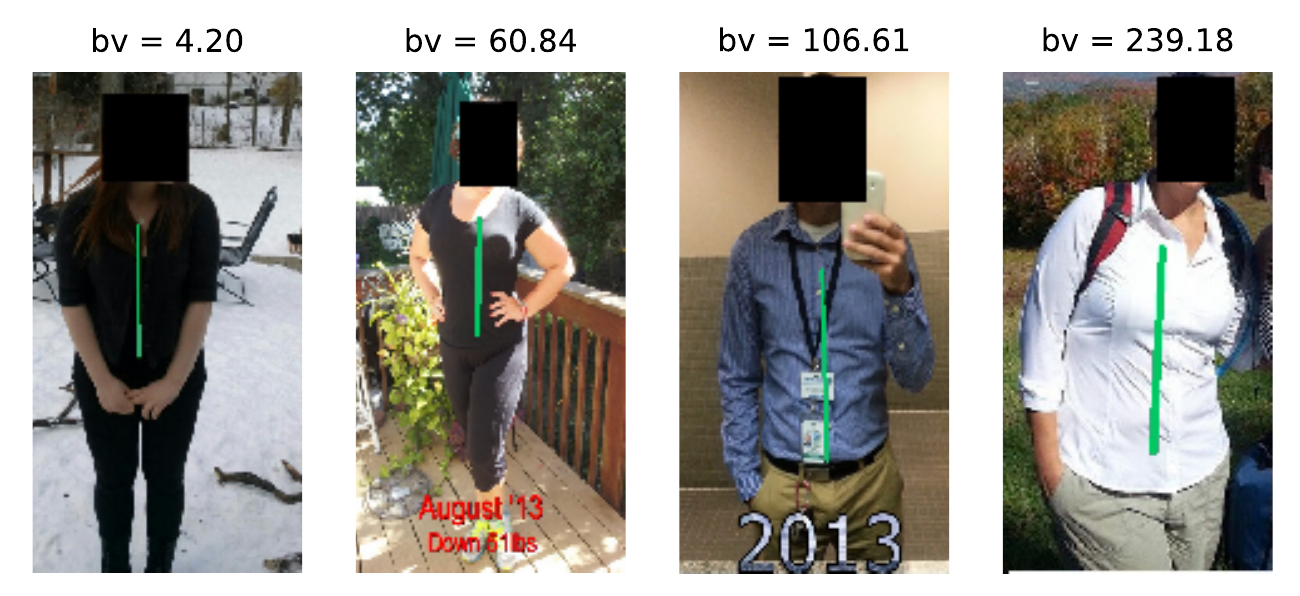}
  \caption{Sample images with main upper-body line and clothing brightness value (bv). In the full dataset, these images are (from left to right) at brightness\hyp value percentiles 0, 33, 66, and 100.}
  \label{fig:color quantiles}
\end{figure}

\subsection{Color classification}
\label{sec:color classification}

As in stripe detection, our approach to color detection starts by automatically identifying the main upper-body line using PoseNet~\cite{cao2016}.
We calculated the brightness value of an outfit as the average brightness of the pixels along the line, where the brightness of a pixel is defined as the average of the three RGB channels. 
That is, a brightness value of 0 corresponds to perfect black, and 255 to perfect white.

To detect solid light and solid dark clothing, we first picked 1,000 images of people with the most clearly visible upper-body lines (\ie, images with the highest PoseNet scores for the relevant body parts), removed images with horizontal stripes (\Secref{sec:stripes detection}), split all remaining images into three approximately equally\hyp sized groups according to their brightness value, and finally used the lower third as solid dark, and the upper third as solid light. This way, we obtained 300 images of dark-colored, and 300 images of light-colored outfits.
In a manual validation of 100 images, the images in the inspected sample had been perfectly classified as dark \vs\ light.
\Figref{fig:color quantiles} shows 4 sample images with the corresponding brightness values.

\section{Research design}
\label{sec:Research design}

We are interested in comparing three clothing types---solid dark, solid light, and horizontally striped---with respect to their effect on perceived weight.
In this section, we introduce three studies, observational as well as experimental, for estimating effects on weight perception.

\subsection{Observational study: weight estimation of real images}
\label{sec:Research design:Observational study}

Our first study aimed at measuring the effect of clothing on weight perception observationally, by analyzing the crowd's weight estimates for the naturally occurring images described in \Secref{sec:Weight- and height-labeled images}.
We first describe how we collected weight and height estimates via crowdsourcing and then how we matched images in pairs in order to control for potential confounds as much as possible.

\xhdr{Collecting weight and height estimates via crowdsourcing}
We used Amazon Mechanical Turk,
a popular crowdsourcing platform, to gather weight and height estimates from a diverse pool of crowd workers.
We divided the set of images in the dataset (\Secref{sec:Weight- and height-labeled images}) into tasks with 10 images each. For each image, crowd workers guessed the weight and height of the shown person and entered their estimates into corresponding input fields located under the image.
The field for weight was placed above the field for height.
To provide familiar units for crowd workers from diverse backgrounds, workers could choose between kilograms and pounds for weight, and between centimeters and feet\slash inches for height.
Automatic conversion between units was performed on the fly, such that, as workers were typing their guesses in their preferred unit, the value in the other unit was updated in real time.
We collected $n=45$ independent estimates for each of the 700 images (300 light-colored, 300 dark-colored, and 100 horizontally striped outfits), for a total of more than 30k estimates from 3,751 unique workers.

To obtain a single crowd estimate $w^i_\est$ for the weight of an image $i$, we averaged the $n$ individual estimates via the arithmetic mean:%
\footnote{One could also use the median instead of the mean. We found the difference in results to be negligible, so we use the mean in the rest of the paper.}
\begin{equation}
	w_{\est}^i =  \frac{1}{n} \sum_{j=1}^{n} w_{j}^i,
	\label{eq:w_est}
\end{equation}
where $w_j^i$ is worker $j$'s weight guess for image $i$.
Similarly, we denote the average height estimate for image $i$ as $h^i_{\est}$, and the average body-mass index (BMI) estimate as $b^i_{\est}$.
The BMI estimate $b^i_\est$ was computed as $w^i_\est / (h^i_\est)^2$ (unit: kg$/$m$^2$); it was not estimated directly by crowd workers.

\begin{figure}[tb]
    \centering
	\begin{subfigure}{.5\linewidth}
 		\centering
  		\includegraphics[width=\linewidth]{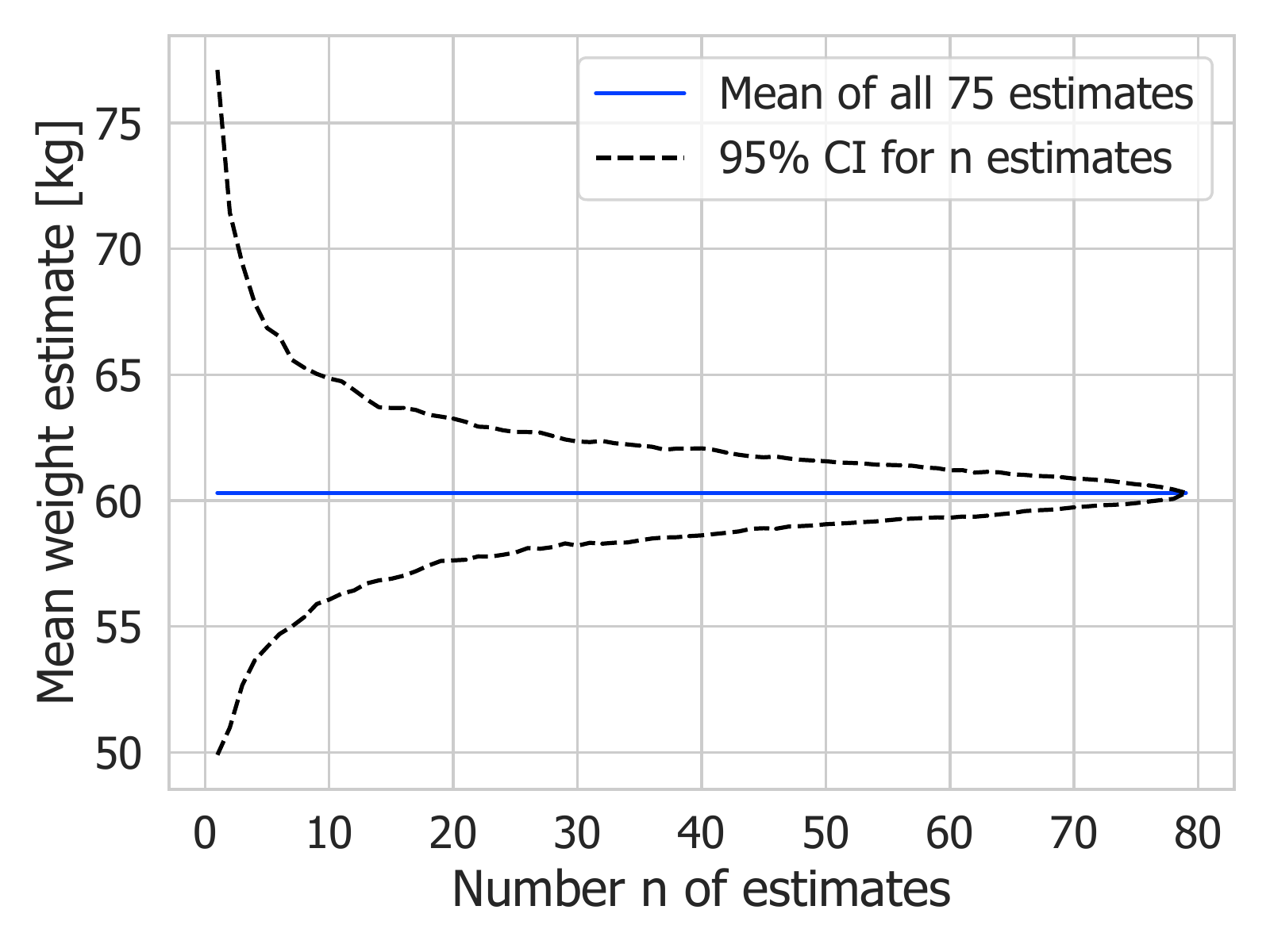}
	\end{subfigure}%
	\begin{subfigure}{.5\linewidth}
  		\centering
  		\includegraphics[width=\linewidth]{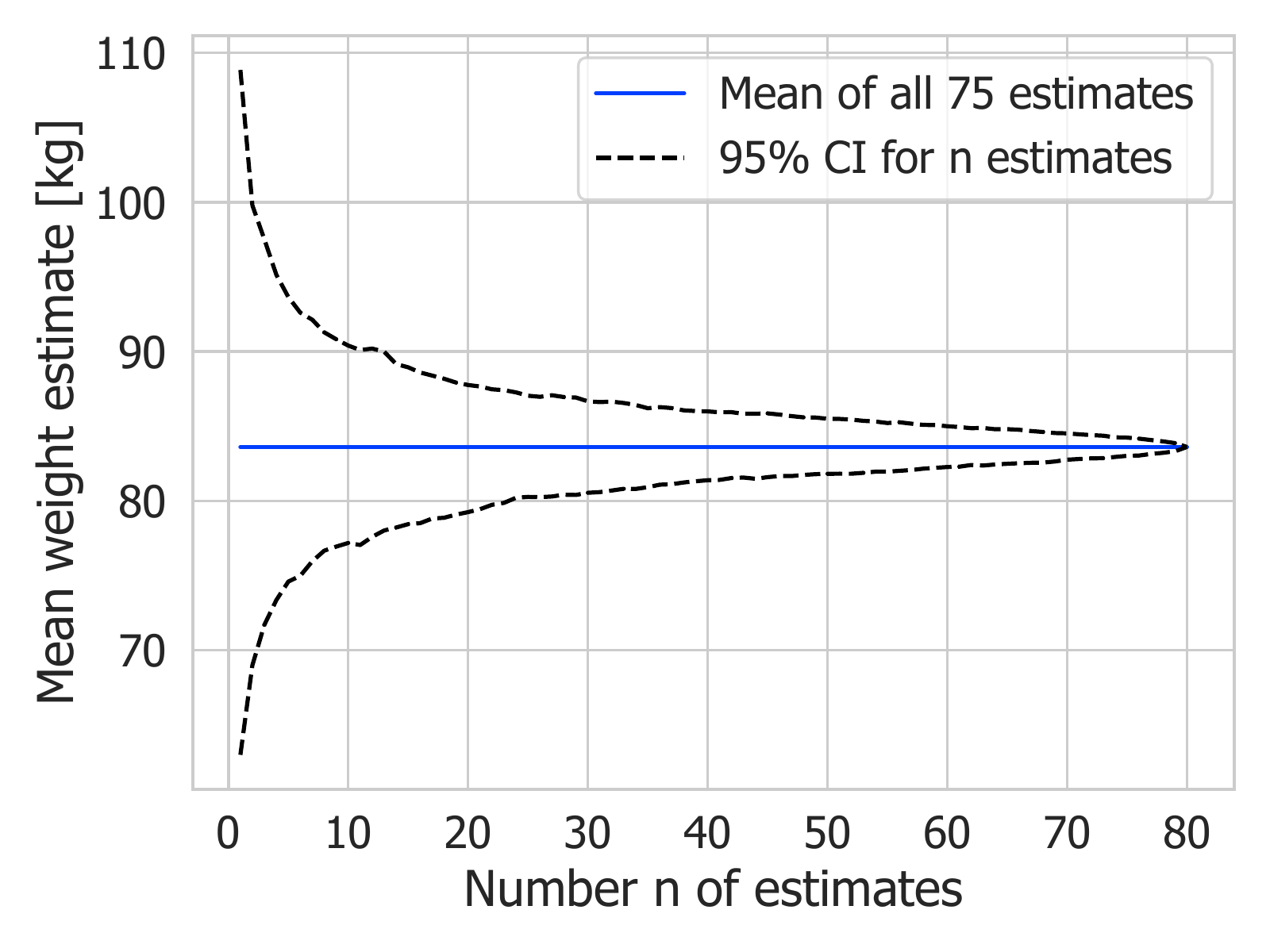}
  	\end{subfigure}
	\caption{
	Convergence of mean weight estimates for two sample images (one image left, one right) as function of number of estimates.
	95\% confidence intervals computed from 1,000 random permutations of all 75 estimates.
	}
	\label{fig:convergence}
\end{figure}

The number of $n=45$ guesses per image was determined in a pilot study where we collected the much larger number of 75 guesses for each of a small number of images and observed the convergence behavior of the mean estimates.
\Figref{fig:convergence} illustrates for two sample images, showing that the mean stabilizes quickly, well before reaching the eventually chosen sample size of $n=45$ guesses.

In addition to weight and height estimates or votes, we asked crowd workers to complete a survey to provide information about their own
weight,
height,
country of residence,
age,
education,
and gender.
These statistics are summarized in \Tabref{tab:worker statistics} for the more than 6.5k crowd workers who contributed to our three studies.

\begin{table}
    \caption{Summary statistics of crowd workers who contributed to the three studies.}
    \centering
    \begin{tabular}{lrr}
                & \textbf{Mean}  & \textbf{Median} \\
    \hline
    Weight & 78.9~kg   & 74.8~kg \\
    Height      & 169.7~cm  & 169.0~cm \\
    BMI  & 27.3~kg$/$m$^2$  & 25.7~kg$/$m$^2$ \\
    Country       &     \\
    \hspace{5mm}United Stated       & 81\%  &  \\
    \hspace{5mm}India       & 12\%  &  \\
    \hspace{5mm}Other       & 7\%  &  \\
    Age       &     & \\
    \hspace{5mm}Under 21       & 2\%  &  \\
    \hspace{5mm}21--30       & 32\%  &  \\
    \hspace{5mm}31--40       & 34\%  &  \\
    \hspace{5mm}41--50       & 17\%  &  \\
    \hspace{5mm}51--60       & 10\%  &  \\
    \hspace{5mm}Over 60       & 4\%  &  \\
    Education       &    & \\
    \hspace{5mm}High-school degree       & 30\%  &  \\
    \hspace{5mm}Bachelor's degree       & 52\%  &  \\
    \hspace{5mm}Master's degree       & 16\%  &  \\
    \hspace{5mm}Doctor's degree       & 2\%  &  \\
    Gender       &    & \\
    \hspace{5mm}Male       & 56\%  &  \\
    \hspace{5mm}Female       & 44\%  &  \\
    \hline
	\end{tabular}
    \label{tab:worker statistics}
\end{table}


Workers were awarded \$0.01 per image.
To encourage high-quality estimates, the 25\% most accurate workers per image were awarded a bonus that doubled their pay for the respective image (where accuracy was computed as the absolute difference between the guess and the ground-truth weight for the image, which is available for all images as described in \Secref{sec:Weight- and height-labeled images}).
To further increase data quality, we filtered and preprocessed the raw weight and height estimates by removing guesses from workers who
did not complete the demographic survey or who
appeared to have used scripts to generate random values,%
\footnote{Concretely, we computed, for each worker, the fraction of images for which their guess ranked among the top 50\%. If it happened for less than 10\% of the worker's guesses, we dropped all their guesses.}
as well as obviously erroneous guesses, such as typos or usage of wrong measurement units.%
\footnote{Concretely, we dropped guesses that were more than $z=3$ standard deviations away from the mean guess for the respective image. The results were robust with respect to the choice of $z$, with $z=2$ resulting in identical conclusions.}
These data cleaning steps removed around 20\% of the crowdsourced estimates.

\xhdr{Matching}
Our goal is to determine whether the color and pattern of clothing changes the perception of weight.
A na\"ive approach to addressing this question would be to compute the average weight estimate for each clothing type and determine whether the averages differ significantly across clothing types. 
The problem with this simple analysis is confounding:
factors that correlate with weight estimates, such as true weight, true height, and gender \cite{martynov2020human}, might also correlate with clothing choice, implying that differences between clothing types may be due to these confounding factors rather than clothing.
For instance, weight estimates for women are lower than for men (because women weigh less than men on average), and if women preferred dark clothes more than men do, then
if individuals wearing dark clothes appeared to weigh less, this
may simply be due to the fact that women are over\hyp represented in the dark group.

To mitigate these problems, we controlled for potential confounds by comparing people who are nearly identical with respect to a number of observed covariates while differing only with respect to clothing.
Specifically, we matched images in pairs $p=(p_1, p_2)$ such that $p_1$ wore clothes of type $c_1$ (\eg, dark) and $p_2$ wore clothes of a different type $c_2 \neq c_1$ (\eg, light), but $p_1$ and $p_2$ were as similar as possible with respect to everything else.
We then define the \textit{within-pair difference} $\Delta w^p_\est$ in weight estimates as
\begin{equation}
    \Delta w^p_\est = w^{p_1}_\est - w^{p_2}_\est,
    \label{eqn:pairwise difference}
\end{equation}
and the average within-pair difference across all pairs $p \in P$ as
\begin{equation}
    \Delta w_\est = \frac{1}{|P|} \sum_{p \in P} \Delta w^p_\est.
    \label{eqn:avg pairwise difference}
\end{equation}
Analogously, we define $\Delta h_\est$ (height) and~$\Delta b_\est$ (BMI).

If all confounds were balanced by the matching---a big \textit{if}, which led us to complement this observational study with the experimental studies introduced later---then the average within-pair difference across all pairs, $\Delta w_\est$, would yield the size of the effect on perceived weight that clothing type $c_1$ affords over clothing type~$c_2$.


Four measured covariates are available in our image dataset: gender, true weight, true height, and true BMI.
We used all of them for matching, as follows:
we considered a pair $(p_1,p_2)$ of images to be a valid candidate for matching if $p_1$ and $p_2$
were of the same gender,
differed in weight by at most $\epsilon_w$, 
differed in height by at most $\epsilon_h$,
and differed in BMI by at most $\epsilon_b$,
but wore different types of clothes (solid light, solid dark, or horizontal stripes).

For a given pair $(c_1,c_2)$ of clothing types, this setting can be modeled as a bipartite graph where every valid candidate pair of images is connected by an edge. To find a matching with the largest number of matched pairs, we ran an off-the-shelf maximum matching algorithm on the resulting bipartite graph.

When choosing the thresholds $\epsilon_w$, $\epsilon_h$, and $\epsilon_b$, we faced a trade-off between the number of matched pairs and the quality of the resulting matching.
Our choice of $\epsilon_b$ for BMI was guided by the literature, which has established 1 BMI point as the so-called \textit{just noticeable difference}~\cite{cornelissen2016visual}, \ie, the largest BMI difference that humans consistently cannot detect.
We hence chose $\epsilon_b = 1\text{ kg}/\text{m}^2$.
To find reasonable values for $\epsilon_w$ and $\epsilon_h$, we ran our analysis for various values, observing that the results were robust with respect to the specific choice.
We settled for $\epsilon_w = 2.5\text{ kg}$ and $\epsilon_h = 2.5\text{ cm}$.
\Tabref{tab:matching} shows that the matching process resulted in pairs of nearly identical images with respect to all observed covariates.
The table also contains the number of pairs created for each pairwise clothing type comparison.

\begin{table}[]
    \caption{
    Validation of pairwise image matching performed for observational study (\Secref{sec:Research design:Observational study}), in terms of mean within-pair differences, with bootstrapped 95\% confidence intervals and $p$-values from Wilcoxon's signed rank test for null hypothesis of no difference in means (\ie, large $p$-values are good, as they imply balanced pairs).
    Height is equal within each pair because granularity of Reddit images is 1~inch (2.54~cm), which exceeds the chosen matching threshold of $\epsilon_h = 2.5\text{ cm}$.
    }
    \label{tab:matching}
    \centering
    \begin{tabular}{lrrr}
        & \textbf{Dark/light} & \textbf{Light/striped} & \textbf{Dark/striped} \\
        \hline
        Num.\ pairs & 153 & 65 & 54 \\
        Male & 41\% & 40\% & 39\% \\
        \hline
        Actual & $-0.04\text{ kg}$ & $-0.13\text{ kg}$ & $-0.30\text{ kg}$ \\
        weight diff. & $[-0.30, 0.21]$ & $[-0.41, 0.36]$ & $[-0.60, 0.28]$ \\
        & ($p=0.71$) & ($p=0.60$) & ($p=0.16$) \\
        \hline
        Actual  & 0 cm & 0 cm & 0 cm \\
        height diff. & ($p=1$) & ($p=1$) & ($p=1$) \\
        \hline
        Actual & $-0.02$ & $-0.04$ & $-0.10$ \\
        BMI diff. & $[-0.11, 0.06]$ & $[-0.14, 0.13]$ & $[-0.20, 0.09]$ \\
        & ($p=0.67$) & ($p=0.63$) & ($p=0.19$) \\
        \hline
    \end{tabular}
\end{table}

\subsection{Experimental \studyOne: weight estimation of manipulated images}
\label{sec:Research design:Experimental study 1}

The matched observational study introduced in the previous section lets us estimate the causal effect of clothing on weight perception under the condition that all confounds were balanced by the matching.
As \Tabref{tab:matching} shows, the matching did indeed balance the 4 explicitly observed confounds (true weight, true height, true BMI; gender was balanced exactly by construction).
\Tabref{tab:matching} does not, however, speak to any of the potentially large number of additional confounds that still remain and that may be hard to measure because they are available only implicitly as visual information (\eg, background, face shape, body pose, camera angle, size and fit of clothes) or that are altogether unobservable (\eg, weight awareness, fashion awareness, mood).
For instance, people in outdoor settings might both dress in certain ways and be perceived to weigh less (as outdoor settings might be associated with healthiness in raters' minds, whether consciously or not);
or sharp dressers might both be more likely to wear dark (because they are more aware of the anecdotal advantages of dark clothes) and be perceived to weigh less, even when controlling for true weight.

The perfect dataset that would let us avoid such factors altogether would contain each person photographed multiple times under exactly identical conditions---including the size and fit of the clothes they wear---with only one difference: the color or pattern of the clothes they wear.
This way, all confounding factors would be eliminated, and the measured effect sizes could be attributed solely to clothing color or pattern, respectively.
Unfortunately, creating such an ideal dataset would require a considerable investment, which probably explains why previous research that has adopted a similar approach worked with one single judged person \cite{swami2012}.
Moreover, concerns regarding external validity would arise, as photographs staged for research purposes would be likely to lack the variety and naturalness of real photographs.

To circumvent these issues, we adopted a different approach: instead of modifying the color and pattern of clothing physically at the time photographs were taken, we did so \textit{post hoc} by manipulating photographs that had already been taken.
This process is cheaper and maintains the variety and naturalness of photographs taken without experimentation in mind.
Specifically, we chose, from the dataset of \Secref{sec:Weight- and height-labeled images}, 100 images of people wearing horizontal stripes and used Fiverr.com,
an online marketplace for freelance services, to hire a graphic\hyp design expert who manipulated each image in Adobe Photoshop by removing the horizontal stripes and producing two additional versions of the same image: one solid light, the other solid dark.
The expert was paid \$5 per original image.
To reduce the risk of bias, they were not informed about the purpose of the manipulated images.%
\footnote{Original and manipulated images available from the authors upon request.}

\begin{figure}[tb]
\centering
\begin{subfigure}{.15\textwidth}
  \centering
  \includegraphics[width=\linewidth]{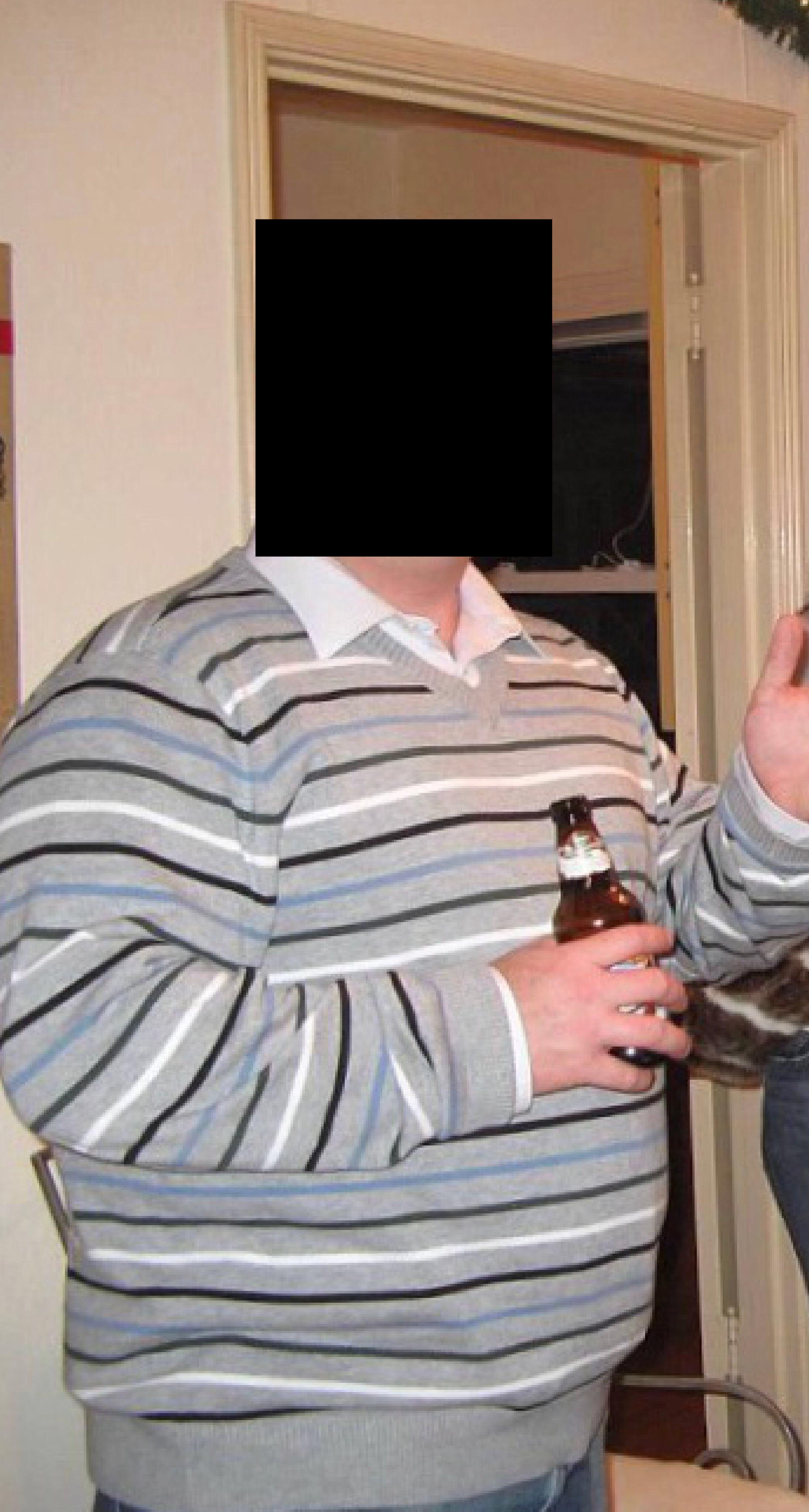}
\end{subfigure}
\begin{subfigure}{.15\textwidth}
  \centering
  \includegraphics[width=\linewidth]{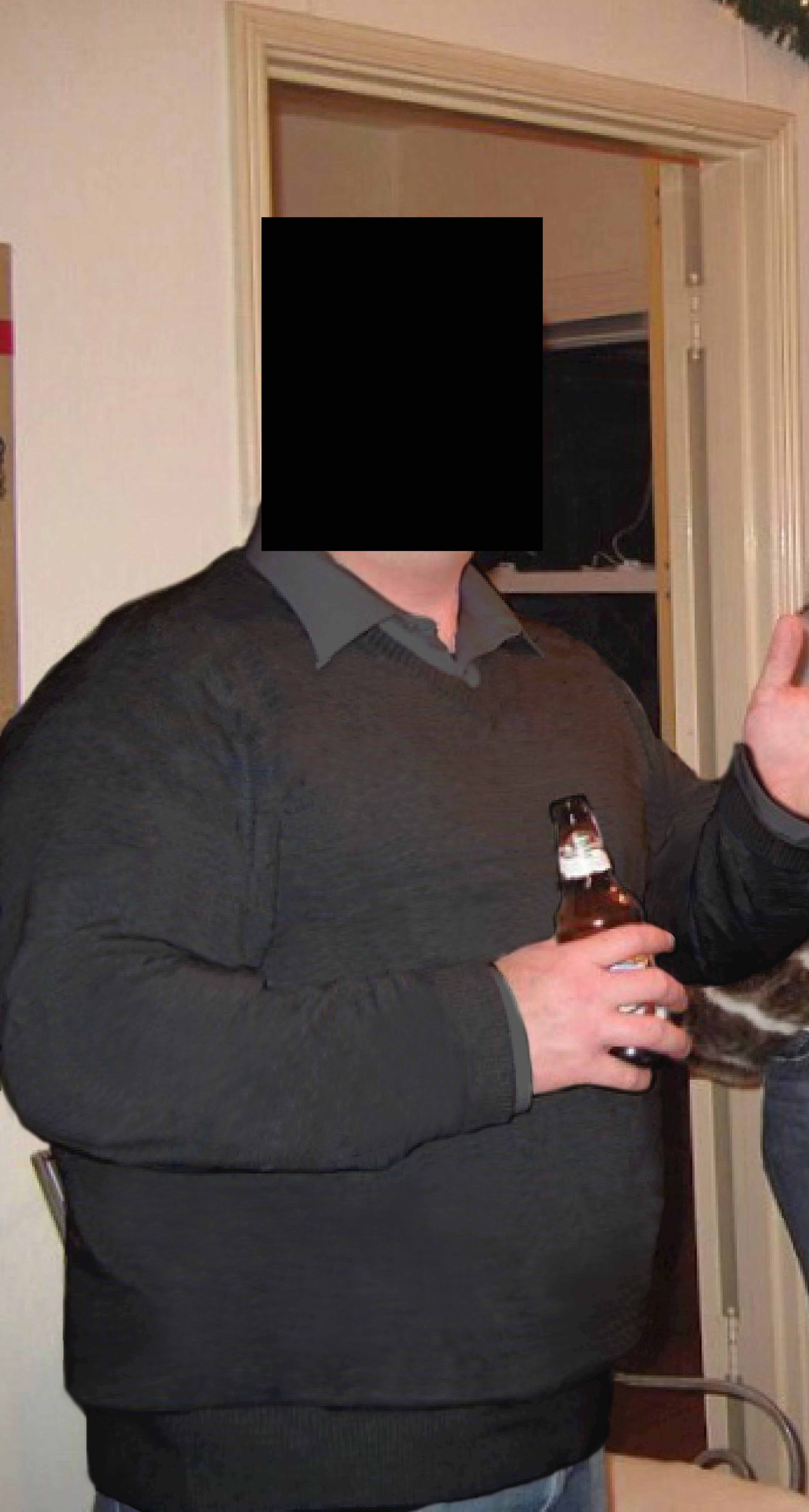}
\end{subfigure}
\begin{subfigure}{.15\textwidth}
  \centering
  \includegraphics[width=\linewidth]{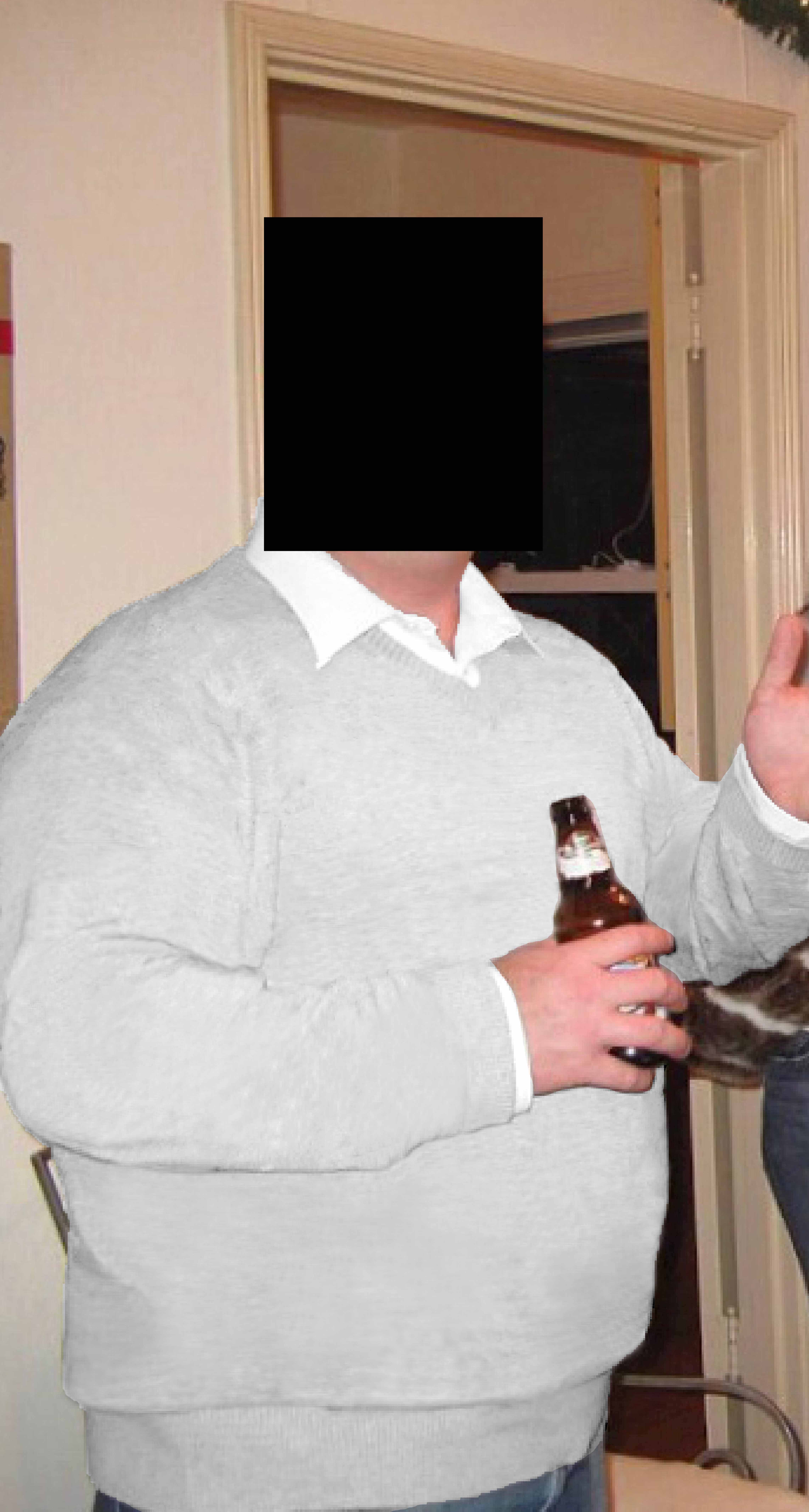}
\end{subfigure}

\vspace{.5mm}

\begin{subfigure}{.15\textwidth}
  \centering
  \includegraphics[width=\linewidth]{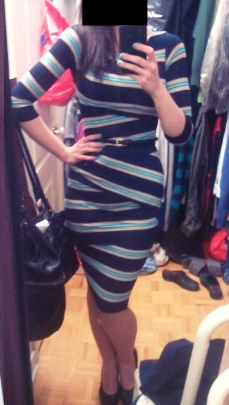}
  \caption{}
\end{subfigure}
\begin{subfigure}{.15\textwidth}
  \centering
  \includegraphics[width=\linewidth]{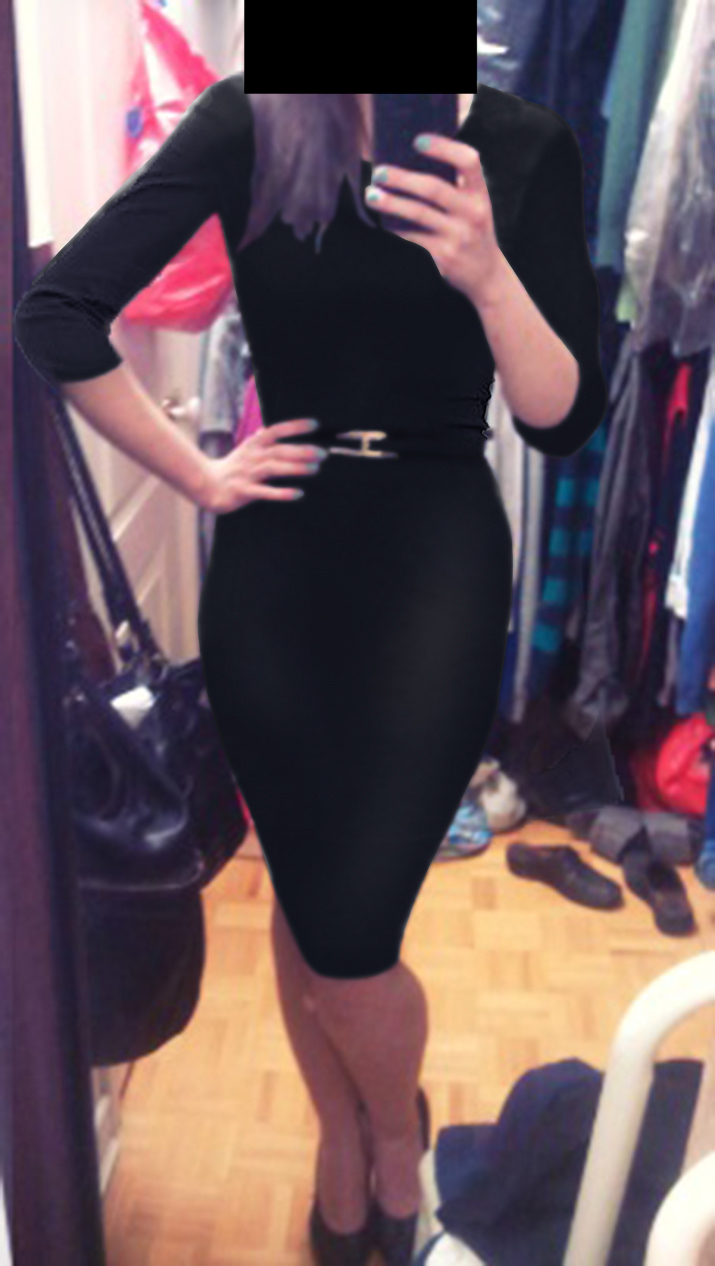}
  \caption{}
\end{subfigure}
\begin{subfigure}{.15\textwidth}
  \centering
  \includegraphics[width=\linewidth]{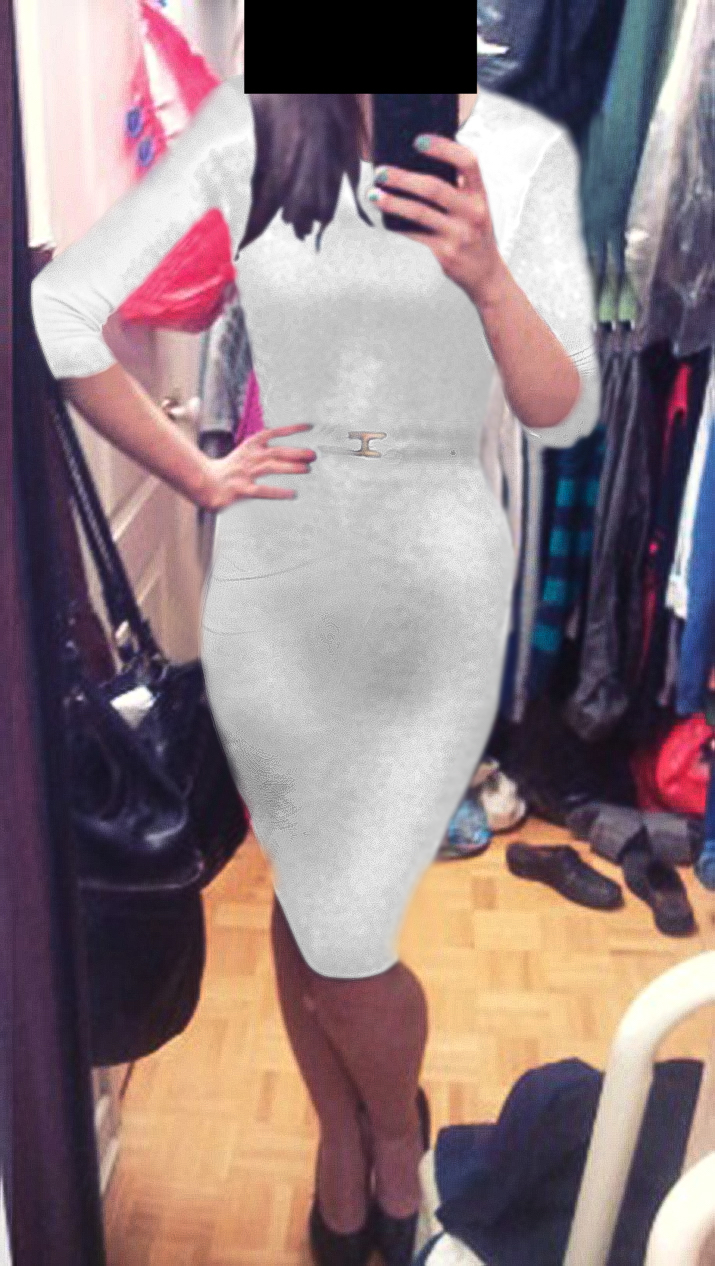}
  \caption{}
\end{subfigure}
\caption{
Two samples of manipulated images as used in experimental studies (\Secref{sec:Research design:Experimental study 1}--\ref{sec:Research design:Experimental study 2}).
Starting from
\textbf{(a)} real photographs with horizontally striped clothes, a graphic\hyp design expert manipulated them to obtain
\textbf{(b)} solid dark and
\textbf{(c)} solid light versions.
The manipulated versions in the top row appear realistic, whereas those in the bottom row appear artificial.
Images of the latter kind were manually removed from the study.
}
\label{fig:photoshop}
\end{figure}

Two examples of real images alongside their manipulated versions are shown in \Figref{fig:photoshop}. 
The example in the top row is well suited for our purposes: even when looking closely, it is hard to tell that photo editing took place.
In the example in the bottom row, on the contrary, the manipulated versions clearly look artificial.
We removed such bad samples \textit{post hoc,} after collecting the crowd estimates (as described below).

After image manipulation, experimental \studyOne proceeded in exactly the same way as the observational study of \Secref{sec:Research design:Observational study}:
for each image, $n=45$ crowd estimates were collected on Amazon Mechanical Turk,%
\footnote{Based on our experiences from the observational study, which was conducted before the experimental studies, we restricted the worker pool to residents of the United States, with the goal of avoiding country\hyp specific biases.}
and a matched analysis was performed.
In contrast to the observational setup, however, the matching did not need to be performed \textit{post hoc} in the experimental setup, since here all covariates were balanced---and thus all potential confounds controlled---from the very start, by construction.

After collecting all crowd estimates, we \textit{post hoc} addressed the issue of the low quality of some manipulated image versions by dropping all pairs of original and manipulated versions for which the estimated difference exceeded 10~kg in weight or 10~cm in height, as visual inspection revealed that in such pairs the manipulated images did not preserve the original body silhouette and were thus not well suited for our study. 
Although a Photoshop expert might still be able to identify the manipulated images in the remaining sample, we believe the difference is hardly noticeable for most people, especially in the context of the crowdsourcing task of height and weight estimation, on which most workers spent only a few seconds per image.

After filtering, the dataset comprised
98, 99, and 97
pairs for the
dark\slash light,
light\slash striped,
dark\slash striped
comparisons, respectively.
A summary of the BMI-based body type classification for the manipulated images is presented in the right column (labeled ``Manipulated  images'') of Table~\ref{tab:BMI}.
The distribution of body types in the manipulated images closely matches that of the full sample of images used in the observational study (left column of Table~\ref{tab:BMI}).

\subsection{Experimental \studyTwo: pairwise weight comparison of manipulated images}
\label{sec:Research design:Experimental study 2}

In both the observational study (\Secref{sec:Research design:Observational study}) and experimental \studyOne (\Secref{sec:Research design:Experimental study 1}) we collected absolute estimates of weight and height.
Absolute weight and height estimation is, however, a hard task for humans, and it is furthermore affected strongly by personal biases on the behalf of raters \cite{martynov2020human}.
Since relative judging tasks tend to be easier for humans than absolute judging tasks, we took a complementary approach in experimental \studyTwo, asking raters to provide pairwise comparisons between images.

Recall from the previous section that, for 100 original images, we obtained manipulated versions where clothing color and pattern---and nothing else---were changed.
That is we have 100 triples where the same person is shown in light, dark, and striped clothes.
Maybe the most direct way of comparing clothing types in a pairwise rating setting would be to show workers two versions of the same image and ask them in which image the person seems to weigh less.
This would, however, be a highly unnatural task: workers would realize that the weight in the two images must be identical, which would shift the focus to the meta level---``Do I think that light or dark clothing makes a person appear to weigh less?''---and lead us to measure the prevalence of anecdotal clothing advice, rather than immediate weight perception.

In our design, we therefore matched images of two \textit{different} people into pairs, while ensuring that both images appeared similar by requiring that the estimated (not necessarily the true) weight and height were nearly identical for both people (without loss of generality, in the solid light versions of the images).
Using the same maximum matching algorithm as in \Secref{sec:Research design:Observational study} (with $\epsilon_w = 2\text{ kg}, \epsilon_h = 2\text{ cm}, \epsilon_b = 1\text{ kg}/\text{m}^2$), we obtained 56 matched pairs of two different people.
Note that the matching ensured that every person participated in at most one pair.

In the following, we let $L$, $D$, and $S$ stand for light, dark, and striped, respectively.
Given the results of the observational study (\Secref{sec:Results:Observational study}) and experimental \studyOne (\Secref{sec:Results:Experimental study 1}), we were particularly interested in comparing light to dark clothing, $\{L,D\}$, and light to striped clothing, $\{L,S\}$.
We therefore conducted experimental \studyTwo twice, once for $\{L,D\}$, and once for $\{L,S\}$.
For ease of exposition, we shall describe the study for $\{L,D\}$, but the case $\{L,S\}$ is fully analogous.

As we have 2 images ($L$ and $D$) per person, there are 4 possible \textit{configurations} per pair of persons:
$LL$, $LD$, $DL$, $DD$.
For each configuration of each person pair, $n=40$ crowd workers guessed whether the first or the second person weighed less.
(The order of the two people in a pair was randomized once and subsequently kept fixed for all rating tasks.)
That is, we collected $2n=80$ ratings for each person $i$ in each of two clothing conditions $c \in \{L,D\}$: $i$ wearing $c=L$, and $i$ wearing $c=D$.
Let the \textit{vote share} $s_{ic} \in [0,1]$ capture in what fraction of the $2n=80$ pairwise comparisons $i$ wearing $c$ was judged to weigh less than the other image.
The \textit{vote-share difference}
$$\delta_i = s_{iD}-s_{iL}$$
quantifies the causal effect of clothing type on the perceived weight of person $i$ in terms of the fraction of votes gained by wearing dark rather than light.
Averaging $\delta_i$ over all people $i$, we obtain the overall causal effect $\delta$ of wearing dark rather than light.

In practice, each crowdsourcing task consisted of 10 comparisons, shown sequentially on separate screens.
Pairs from all 4 configurations appeared in random order.
Each crowd worker saw each pair in at most one of the 4 configurations, so they would not be biased by previous information.
Across both runs of the experiment (for $\{L,D\}$ and $\{L,S\}$, respectively), we collected a total of 19,200 votes from 750 crowd workers.



As a quality assessment measure, most workers rated up to 4 pairs twice, which allowed us to determine test--retest reliability: 85\% of workers consistently judged the same image in the pair to weigh less, a satisfactorily high value, given the hardness and subjectivity of the judging task (recall that images in a pair were explicitly chosen to be indistinguishable in terms of weight when wearing the same clothing type).
As a further safeguard against random guessing, we embedded a two-letter code in each image. To enter their vote, workers had to type the corresponding code. This way, workers had to actually look at the images---at the very least, to read the code---and could not simply click on one image randomly.

\section{Results}
\label{sec:Results}

Having described the design of the three studies, we now discuss the results for each study in turn.%
\footnote{
Code: \url{https://github.com/epfl-dlab/darks_and_stripes}}

\subsection{Observational study: weight estimation of real images}
\label{sec:Results:Observational study}

\begin{figure*}
    \begin{minipage}{0.55\linewidth}
  		\includegraphics[width=.5\linewidth, clip=true, trim=10 10 10 10]{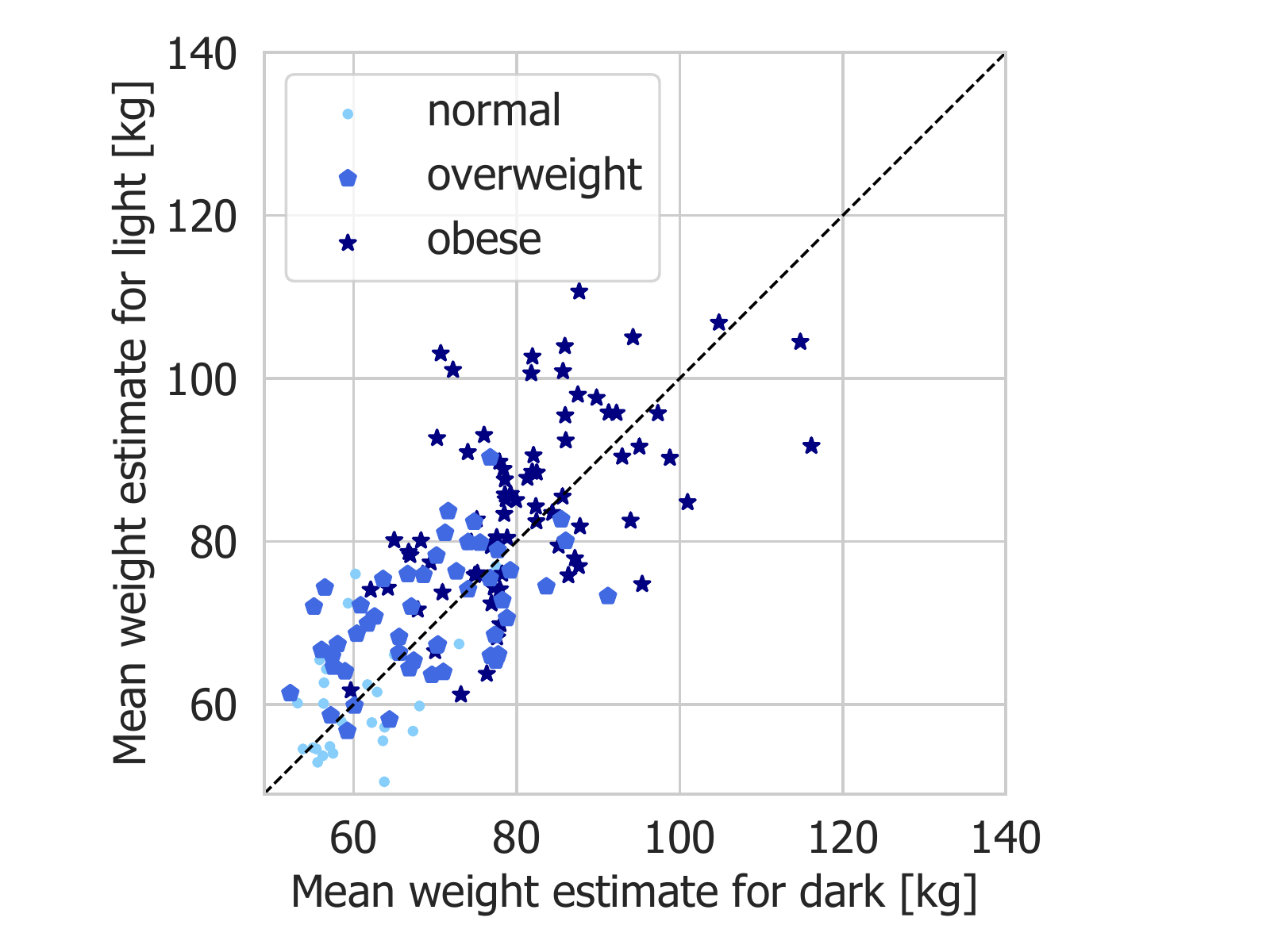}
  		\hspace{-6mm}
  		\includegraphics[width=.5\linewidth, clip=true, trim=10 10 10 10]{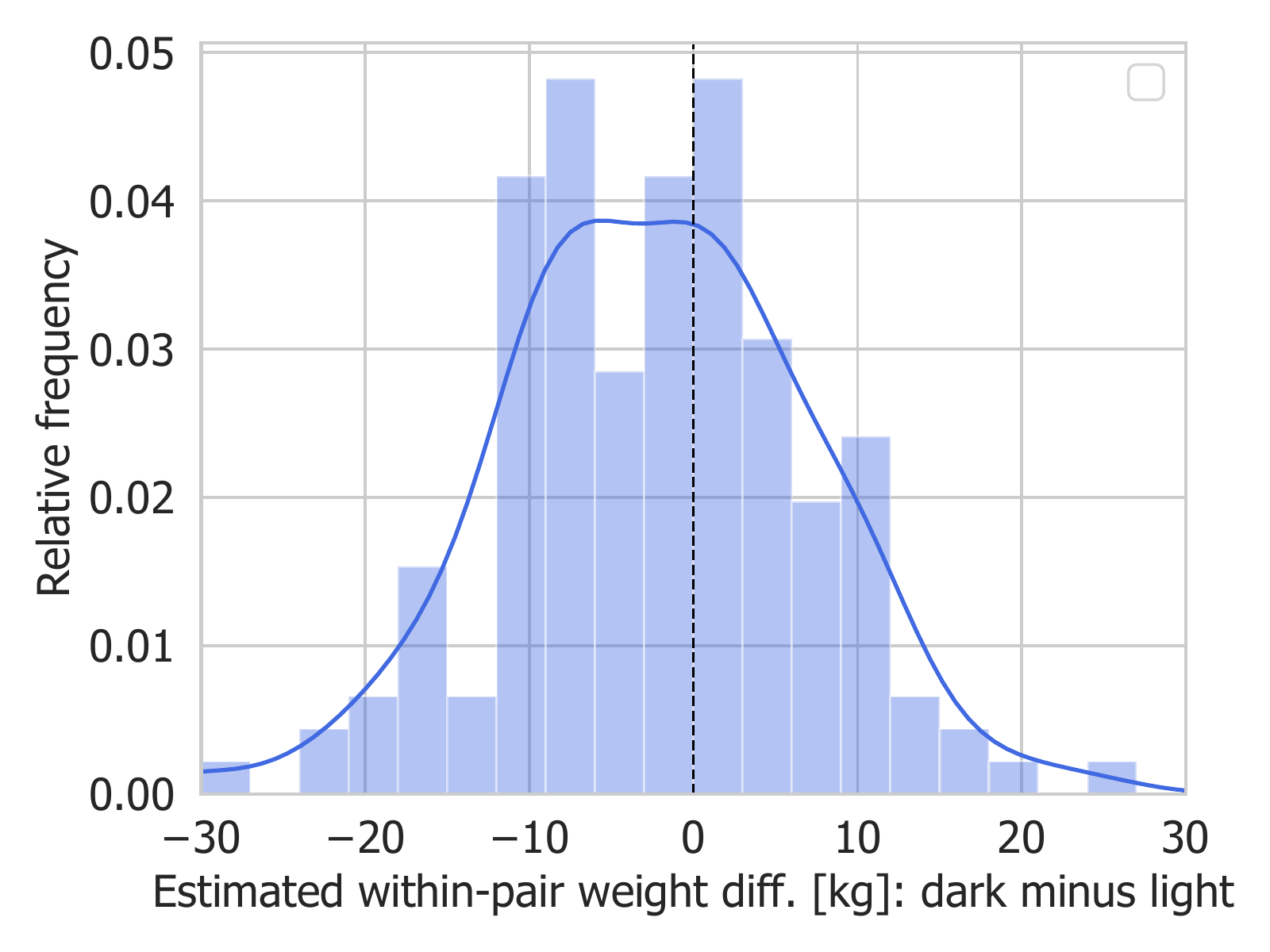}
  	\end{minipage}
  	{\small
    \begin{minipage}{0.45\linewidth}
    \begin{tabular}{lrrr}
        & \textbf{dark -- light} & \textbf{light -- striped} & \textbf{dark -- striped} \\
        \hline
        Num.\ pairs & 153 & 65 & 54 \\
        \hline
        Perceived & \textbf{$-$2.27 kg} & $-0.18$ kg & \textbf{$-$3.08 kg} \\
        weight diff. & $[-3.74,-0.83]$ & $[-2.54,2.09]$ & $[-5.38,-0.77]$ \\
        $\Delta w_\est$ & ($p=0.0039$) & ($p=0.68$) & ($p=0.0072$) \\
        \hline
        Perceived & $-0.11$ cm & $-0.21$ cm & $-0.01$ cm \\
        height diff. & $[-0.77,0.55]$ & $[1.05,0.64]$ & $[-0.97,0.98]$ \\
        $\Delta h_\est$ & ($p=0.88$) & ($p=0.56$) & ($p=0.79$) \\
        \hline
        Perceived & \textbf{$-$0.77} & $0.06$ & \textbf{$-$1.05} \\
        BMI diff. & $[-1.24,-0.31]$ & $[-0.69,0.77]$ & $[-1.74,-0.34]$ \\
        $\Delta b_\est$ & ($p=0.0023$) & ($p=0.70$) & ($p=0.0071$) \\
        \hline
    \end{tabular}
  	\end{minipage}
  	}
  
  \centerline{(a) Observational study: weight estimation of real images (\Secref{sec:Results:Observational study})}
  \bigskip

    \begin{minipage}{0.55\linewidth}
  		\includegraphics[width=.5\linewidth, clip=true, trim=10 10 10 10]{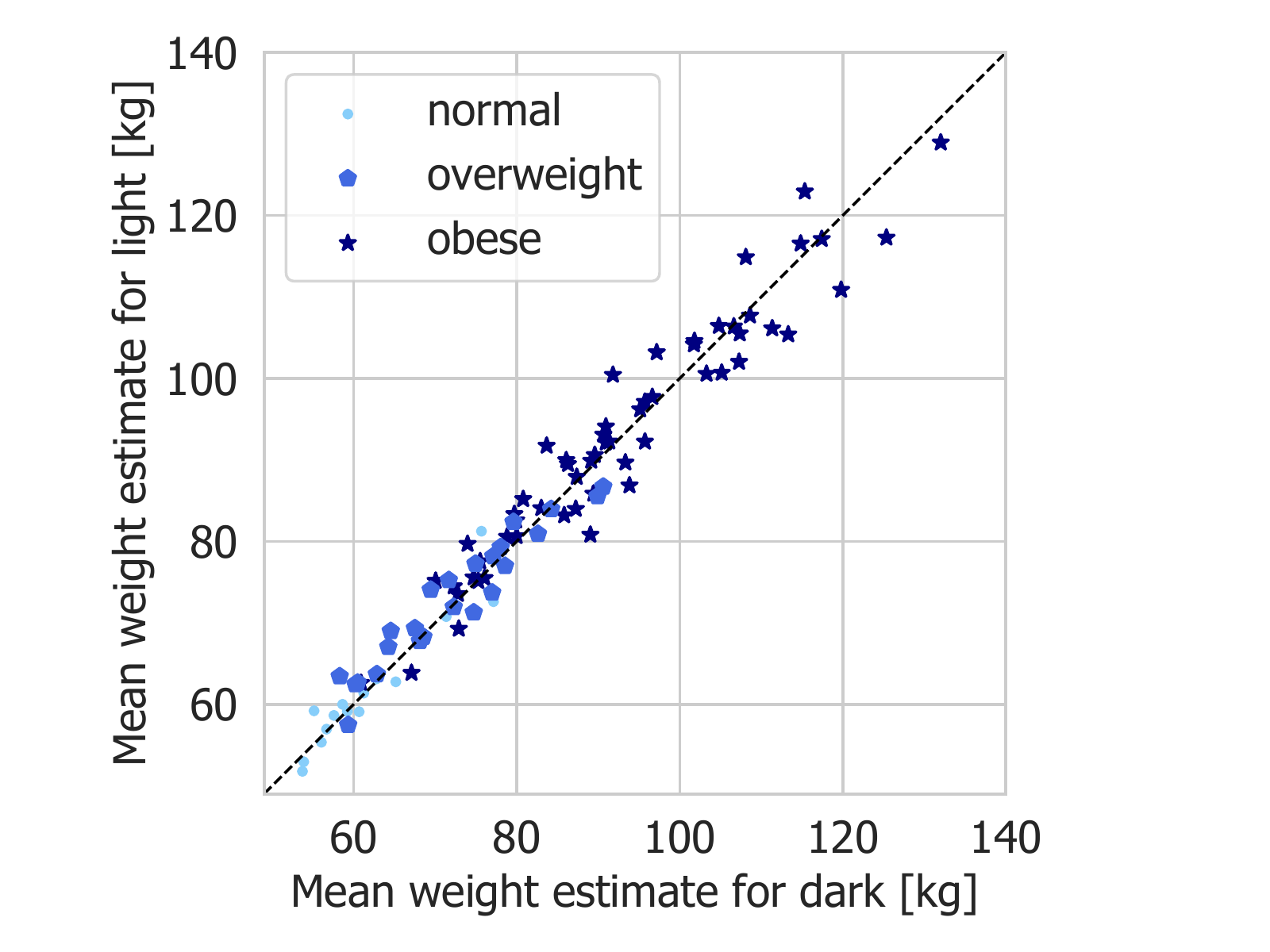}
  		\hspace{-6mm}
  		\includegraphics[width=.5\linewidth, clip=true, trim=10 10 10 10]{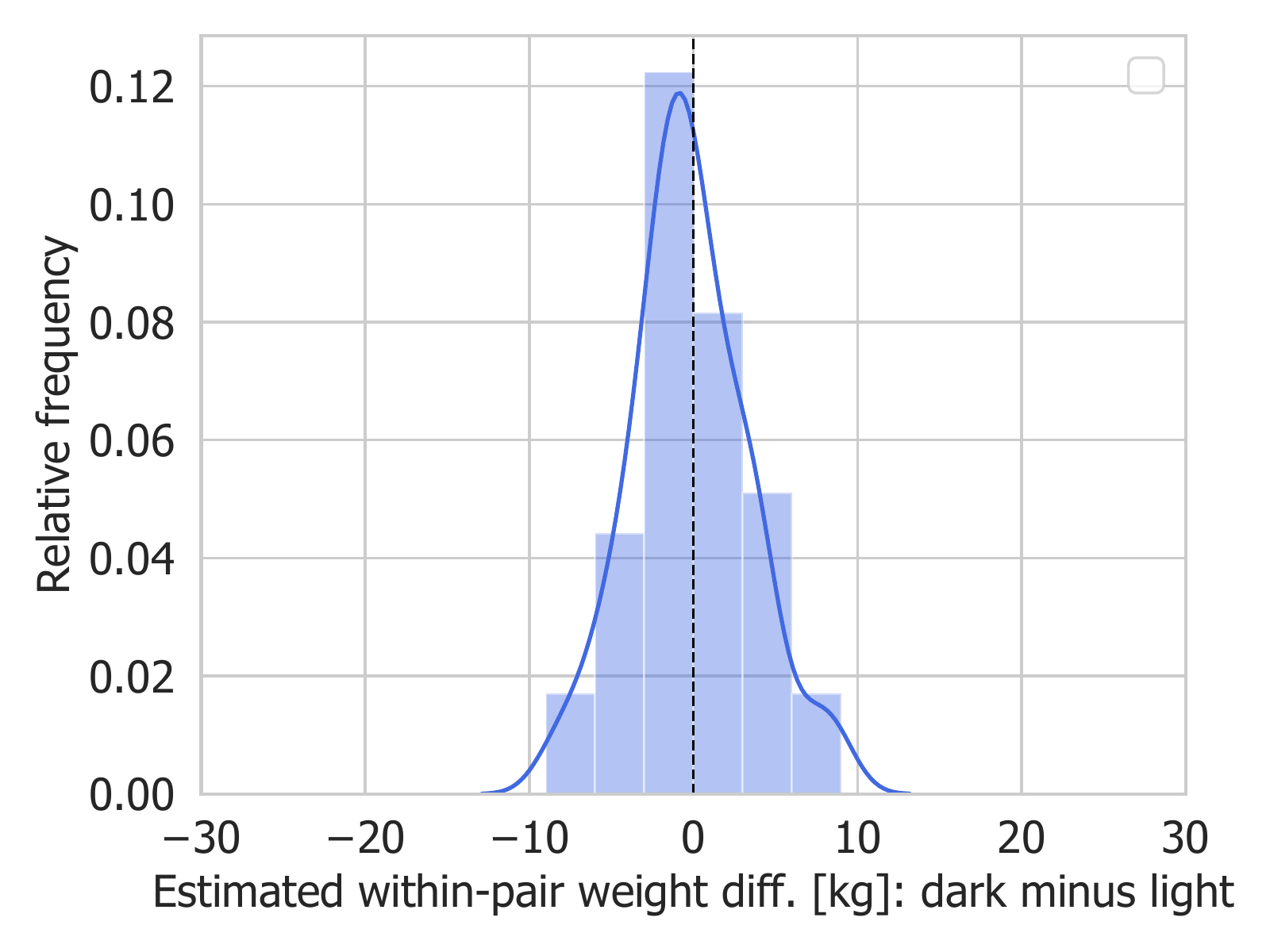}
  	\end{minipage}
  	{\small
    \begin{minipage}{0.45\linewidth}
    \begin{tabular}{lrrr}
        & \textbf{dark -- light} & \textbf{light -- striped} & \textbf{dark -- striped} \\
        \hline
        Num.\ pairs & 98 & 99 & 97 \\
        \hline
        Perceived & $-$0.23 kg & $-0.12$ kg & $-$0.13 kg \\
        weight diff. & $[-0.93,0.46]$ & $[-0.80,0.55]$ & $[-0.80,0.54]$ \\
        $\Delta w_\est$ & ($p=0.34$) & ($p=0.78$) & ($p=0.50$) \\
        \hline
        Perceived & $0.09$ cm & $0.22$ cm & $0.29$ cm \\
        height diff. & $[-0.20,0.37]$ & $[0.05,0.50]$ & $[0.01,0.56]$ \\
        $\Delta h_\est$ & ($p=0.58$) & ($p=0.087$) & ($p=0.052$) \\
        \hline
        Perceived & $-0.13$ & $-0.12$ & 0.17 \\
        BMI diff. & $[-0.37,0.11]$ & $[-0.35,0.11]$ & $[-0.40,0.06]$ \\
        $\Delta b_\est$ & ($p=0.15$) & ($p=0.50$) & ($p=0.061$) \\
        \hline
    \end{tabular}
  	\end{minipage}
    }

  \centerline{(b) Experimental \studyOne: weight estimation of manipulated images (\Secref{sec:Results:Experimental study 1})}

	\caption{
	Results of
	\textbf{(a)} observational study (\Secref{sec:Results:Observational study}) and
	\textbf{(b)} experimental \studyOne (\Secref{sec:Results:Experimental study 1}).
	\textbf{Left:} Scatter plot of weight estimates for all matched image pairs (dark \vs\ light).
	\textbf{Center:} Distribution of within-pair differences (\ie, distribution of $x-y$ from scatter plot).
	\textbf{Right:} Mean within-pair differences for weight, height, and BMI estimates, with bootstrapped 95\% CIs and $p$-values from Wilcoxon's signed rank test (values with $p<0.05$ in bold).
	For completeness, plots for all measurements (weight, height, BMI) and for comparisons of all clothing types (dark, light, striped) are available in
	\Figref{fig:results obs DL}--\ref{fig:results obs LS} (observational study) and
	\Figref{fig:results exp1 DL}--\ref{fig:results exp1 LS} (experimental \studyOne).
	}
  	\label{fig:results obs+exp1}
\end{figure*}

\xhdr{Effect of dark colors}
We begin by considering the impact of clothing color on weight perception as estimated in the observational study.
The weight estimates $(w_\est^{p_1}, w_\est^{p_2})$ (\cf\ \Eqnref{eq:w_est}) for all 153 matched pairs $p=(p_1,p_2)$ are visualized in the scatter plot of \Figref{fig:results obs+exp1}a (left).
The distribution of within-pair dark\hyp minus\hyp light differences $\Delta w_\est^p$ in perceived weight (\cf\ \Eqnref{eqn:pairwise difference}) is shown in \Figref{fig:results obs+exp1}a (center).
Visual inspection reveals a left skew of the distribution, indicating that the dark-clad person in a pair is perceived to weigh less than the light-clad person.

\begin{figure*}
    \centering
	\begin{subfigure}{.33\linewidth}
 		\centering
  		\includegraphics[width=\linewidth]{Plots/plot_scatter_w_pairwise_comp_dirty_dark_light.pdf}
	\end{subfigure}
	\begin{subfigure}{.33\linewidth}
  		\centering
  		\includegraphics[width=\linewidth]{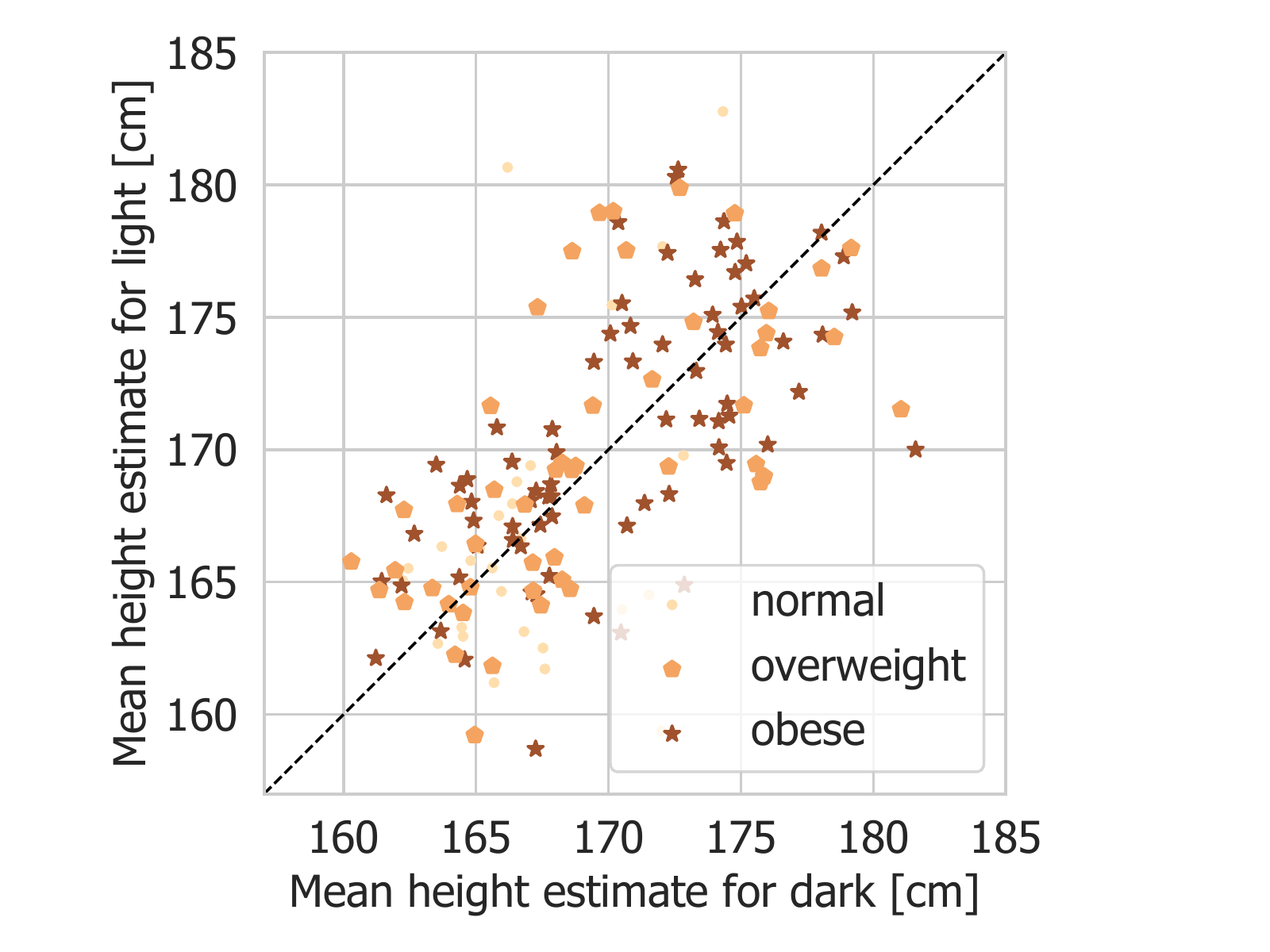}
  	\end{subfigure}
	\begin{subfigure}{.33\linewidth}
  		\centering
  		\includegraphics[width=\linewidth]{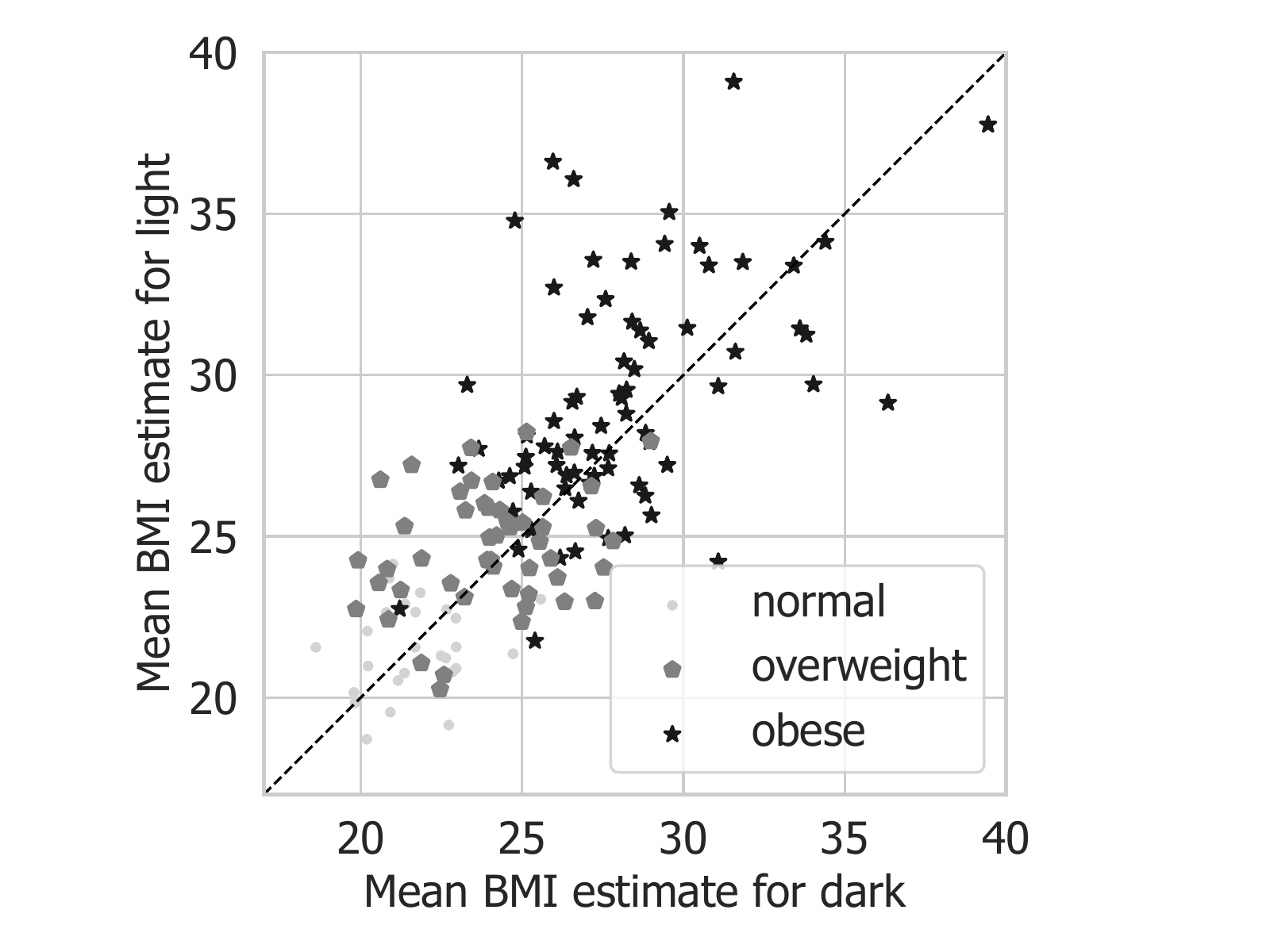}
 	\end{subfigure}

	\begin{subfigure}{.33\linewidth}
 		\centering
  		\includegraphics[width=\linewidth]{Plots/plot_dist_w_pairwise_est_dirty_dark_light.pdf}
  		\caption{Weight}
	\end{subfigure}
	\begin{subfigure}{.33\linewidth}
  		\centering
  		\includegraphics[width=\linewidth]{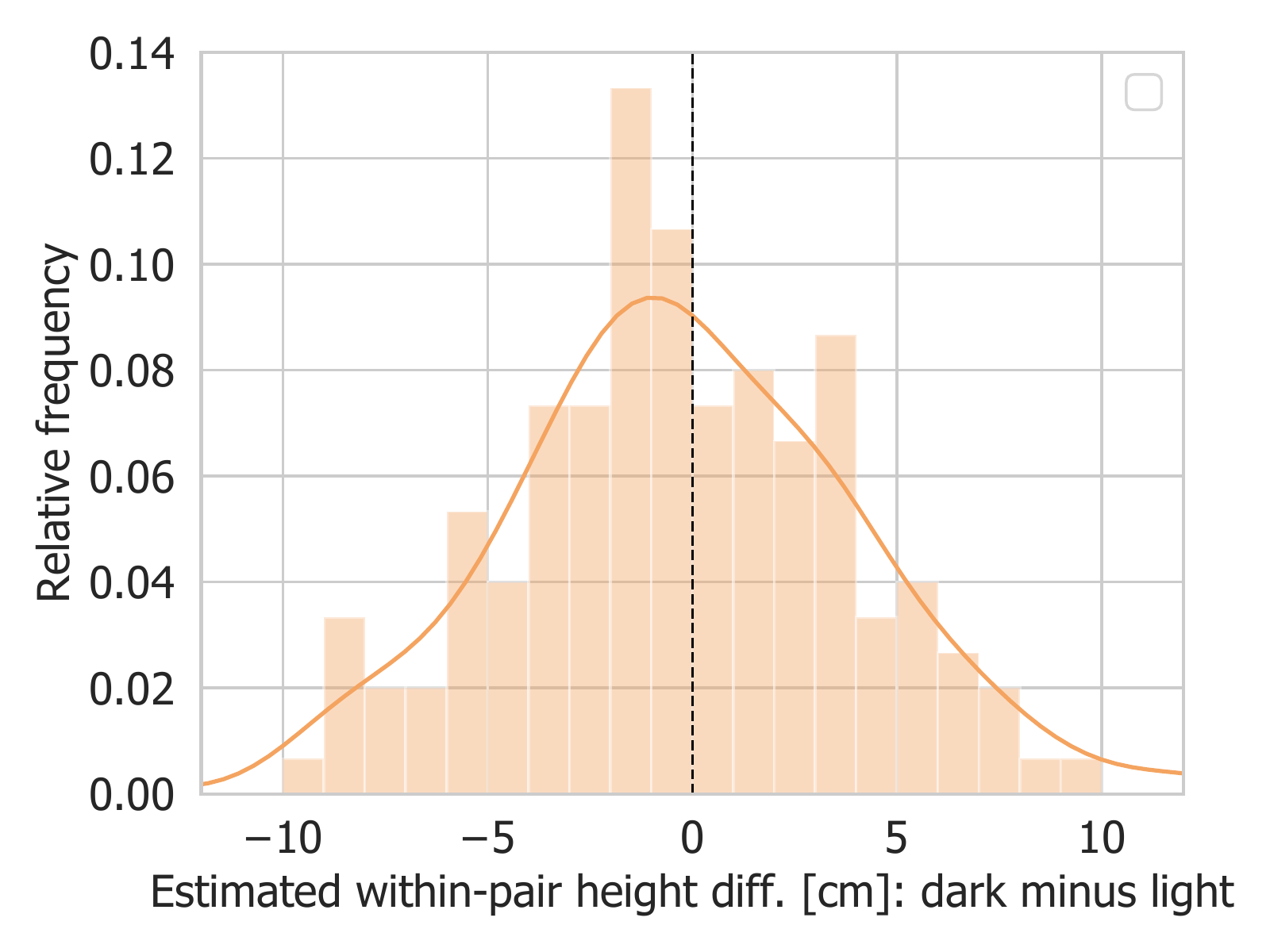}
  		\caption{Height}
  	\end{subfigure}
	\begin{subfigure}{.33\linewidth}
  		\centering
  		\includegraphics[width=\linewidth]{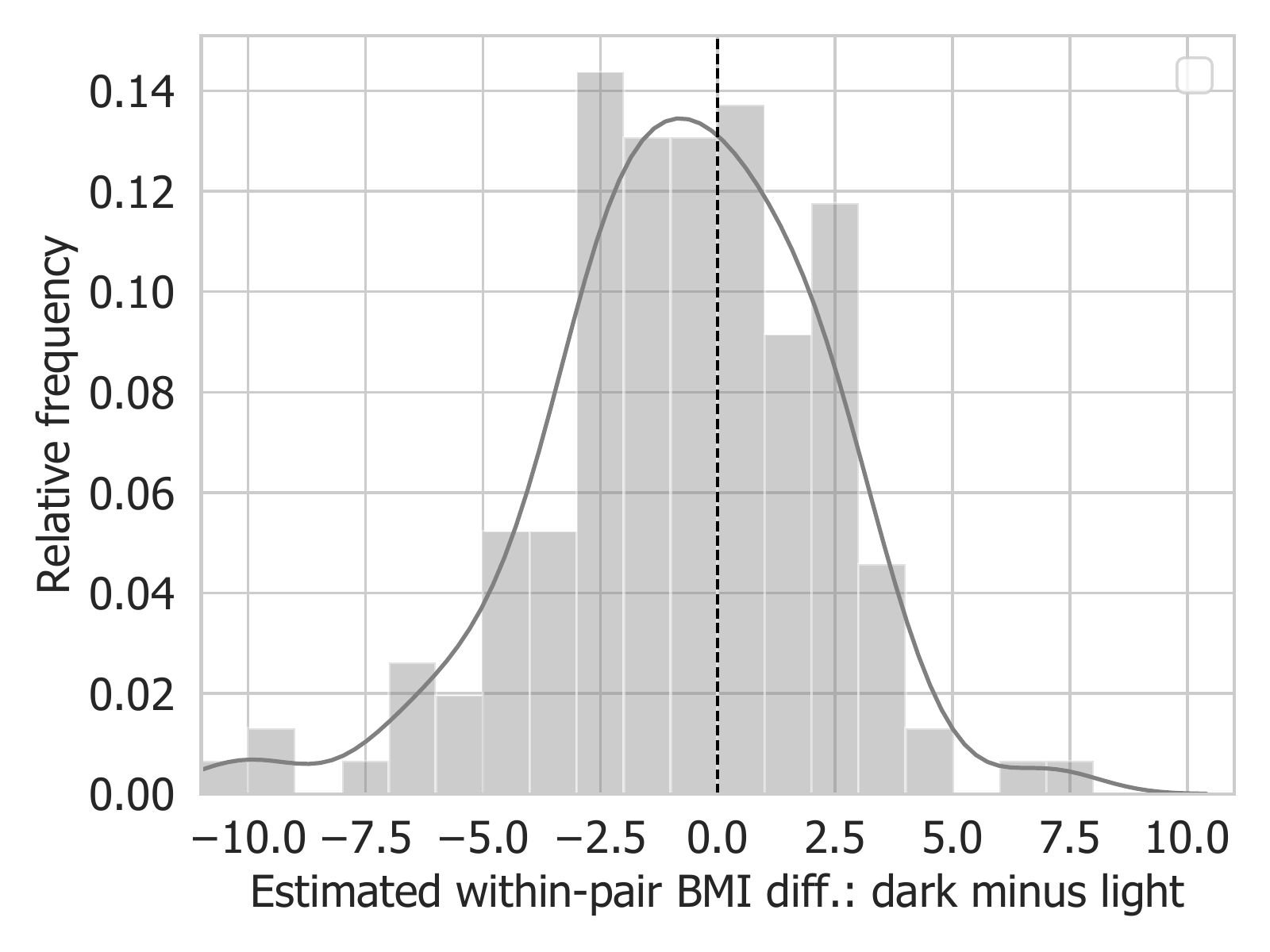}
  		\caption{BMI}
  	\end{subfigure}
    \vspace{2mm}
	\caption{
	Results of observational study, dark \vs\ light.
	}
  	\label{fig:results obs DL}
\end{figure*}

Whereas the plots of \Figref{fig:results obs+exp1}a pertain only to the perception of weight, the first column of the table in \Figref{fig:results obs+exp1}a summarizes the distribution of perceived within-pair differences for all of weight, height, and BMI ($\Delta w_\est$, $\Delta h_\est$, $\Delta b_\est$; \cf\ \Eqnref{eqn:avg pairwise difference}) in terms of averages.%
\footnote{
Moreover, plots for all of weight, height, and BMI are available in \Figref{fig:results obs DL}.
}
We observe that average weight and BMI estimates were significantly lower for dark than for light, by 2.27~kg ($p=0.004$ according to Wilcoxon's signed rank test) and 0.77 BMI points ($p=0.002$), respectively.
This implies that people who wear dark were, on average, perceived as weighing less.
In contrast, height estimates were not affected by clothing color ($p=0.88$).

\xhdr{Effect of horizontal stripes}
Next, we investigate the effect of horizontal stripes in two matched analyses:
one where we compare images with horizontal stripes to images with solid light colors (65 pairs),
and one where we compare images with horizontal stripes to images with solid dark colors (54 pairs).

\begin{figure*}
    \centering
	\begin{subfigure}{.33\linewidth}
 		\centering
  		\includegraphics[width=\linewidth]{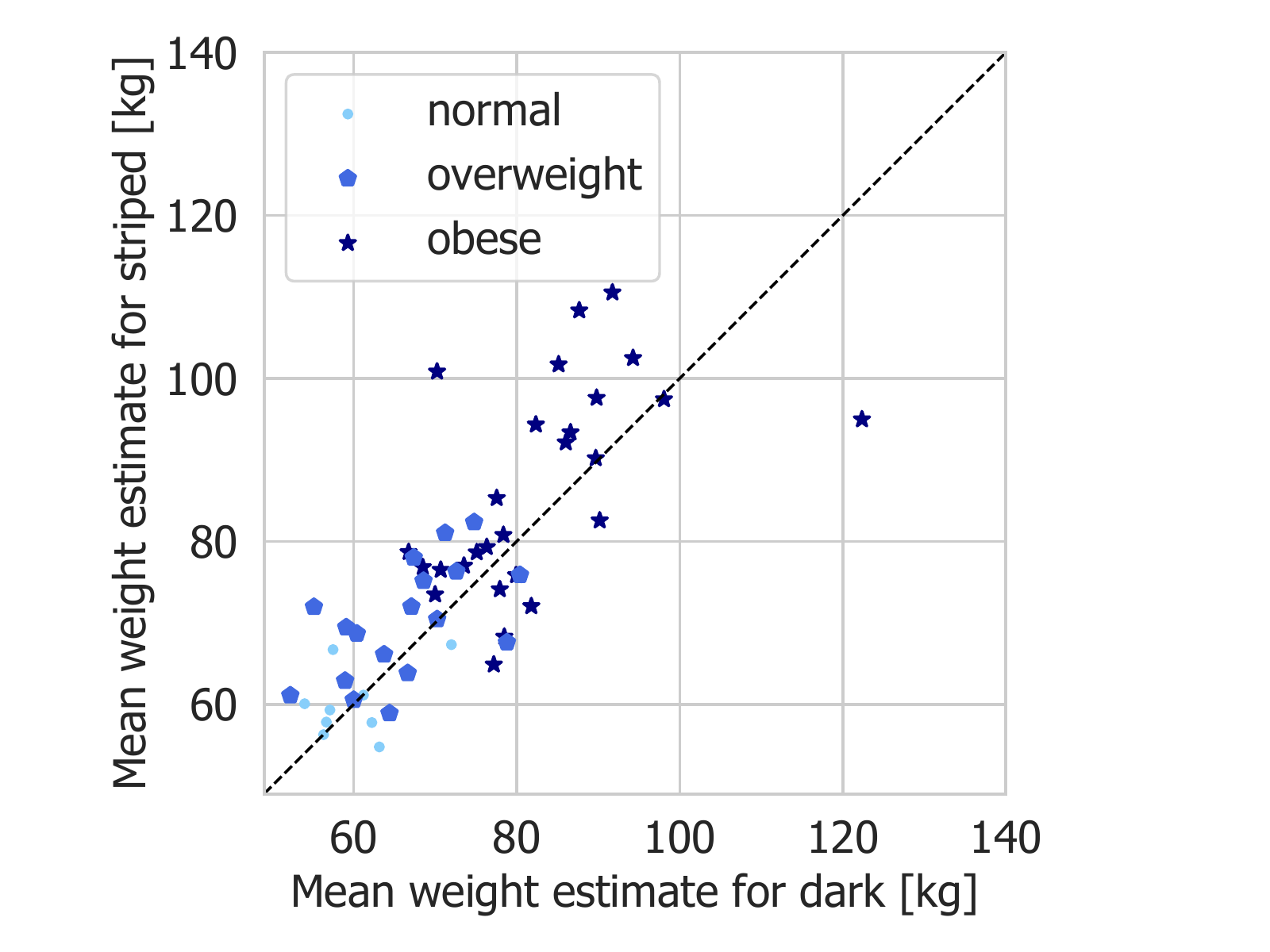}
	\end{subfigure}
	\begin{subfigure}{.33\linewidth}
  		\centering
  		\includegraphics[width=\linewidth]{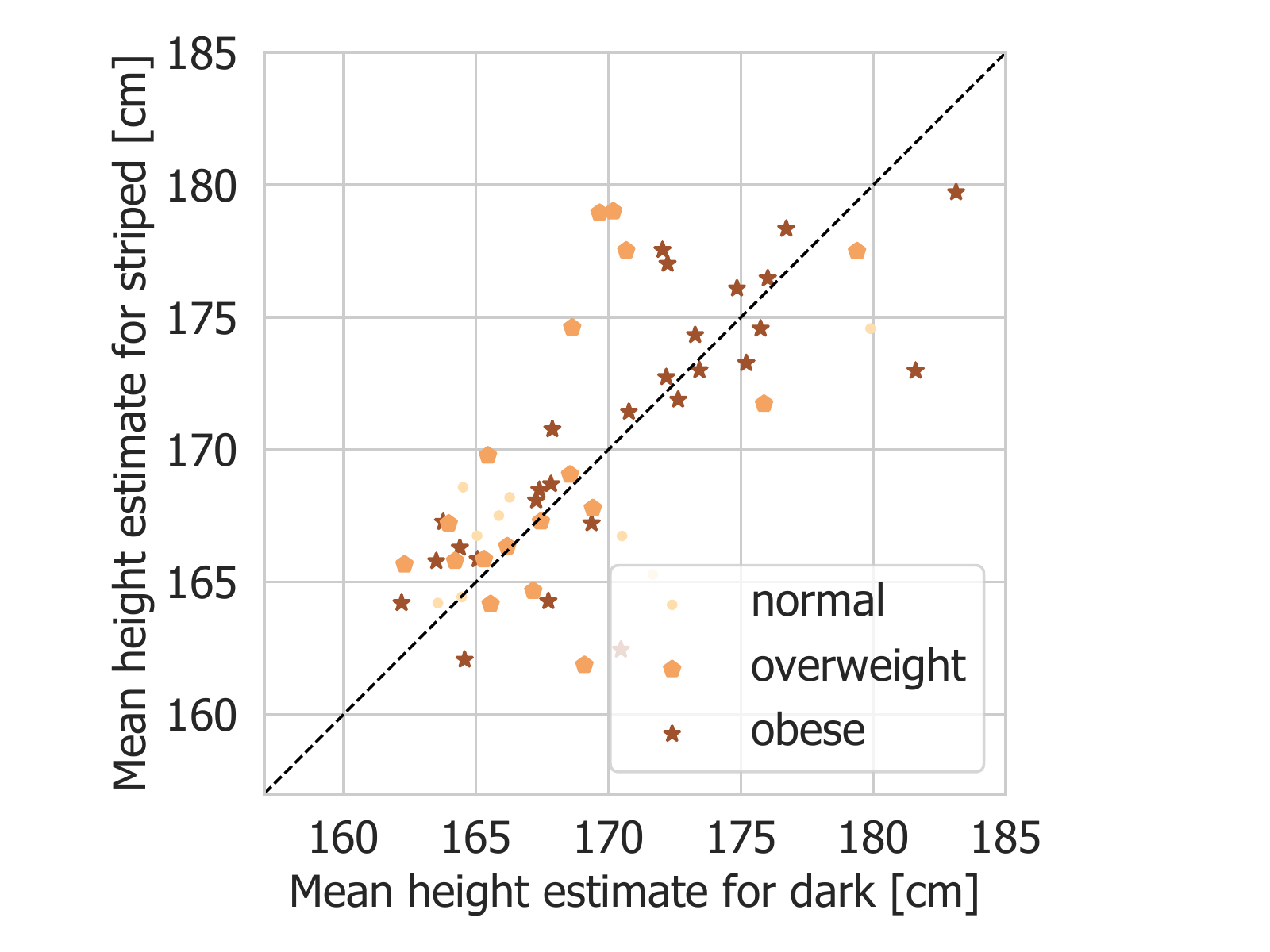}
  	\end{subfigure}
	\begin{subfigure}{.33\linewidth}
  		\centering
  		\includegraphics[width=\linewidth]{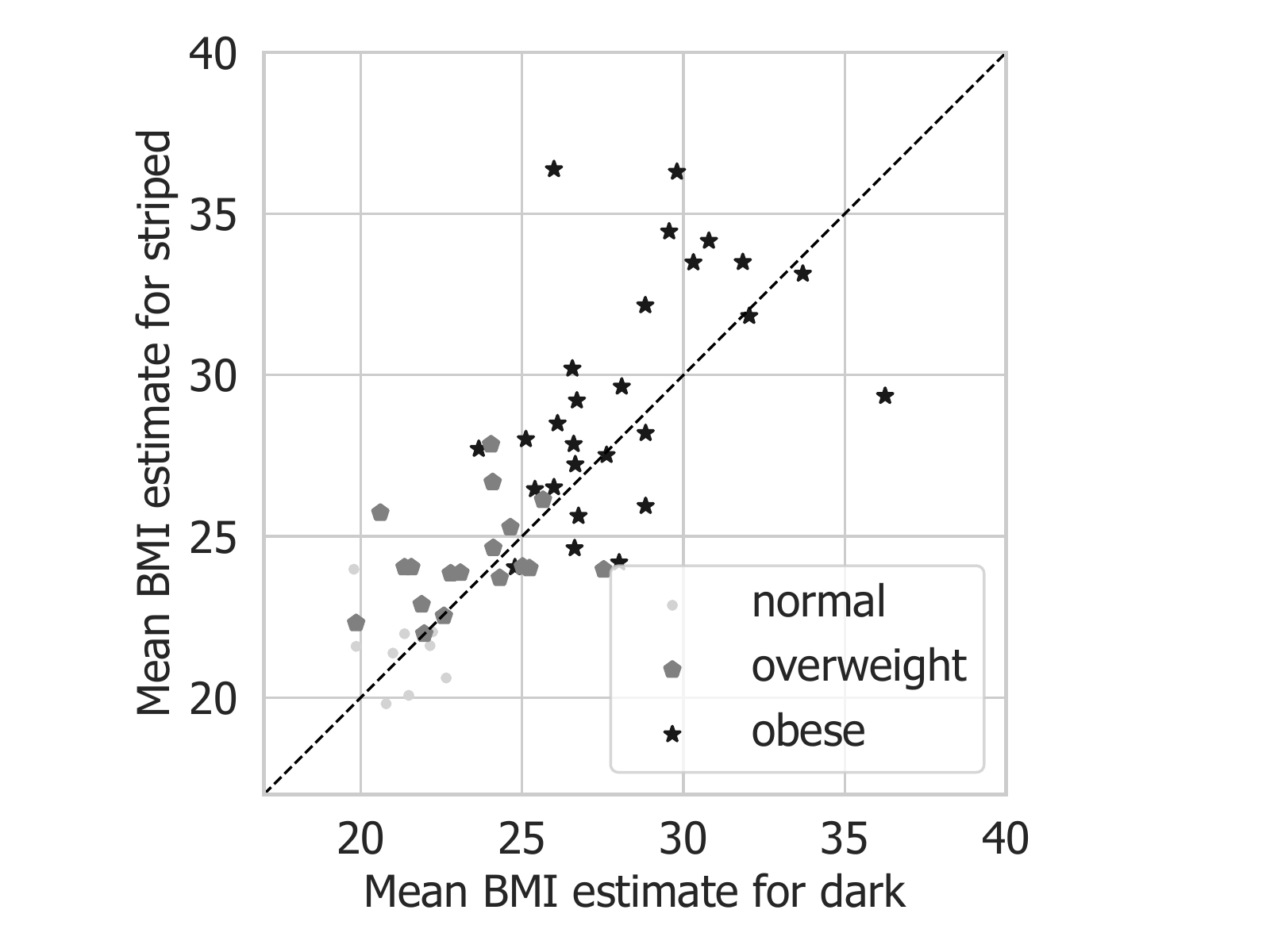}
 	\end{subfigure}

	\begin{subfigure}{.33\linewidth}
 		\centering
  		\includegraphics[width=\linewidth]{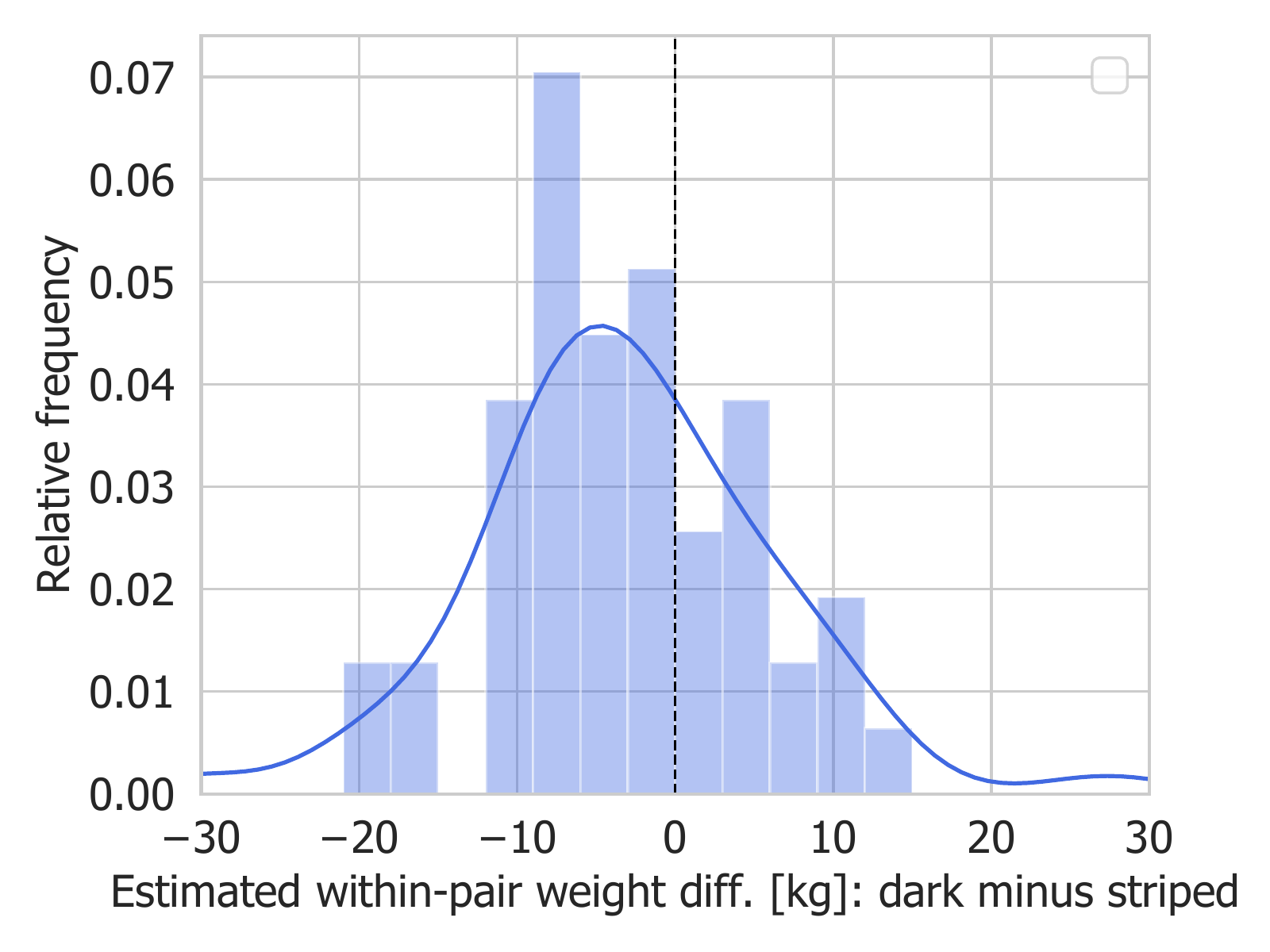}
  		\caption{Weight}
	\end{subfigure}
	\begin{subfigure}{.33\linewidth}
  		\centering
  		\includegraphics[width=\linewidth]{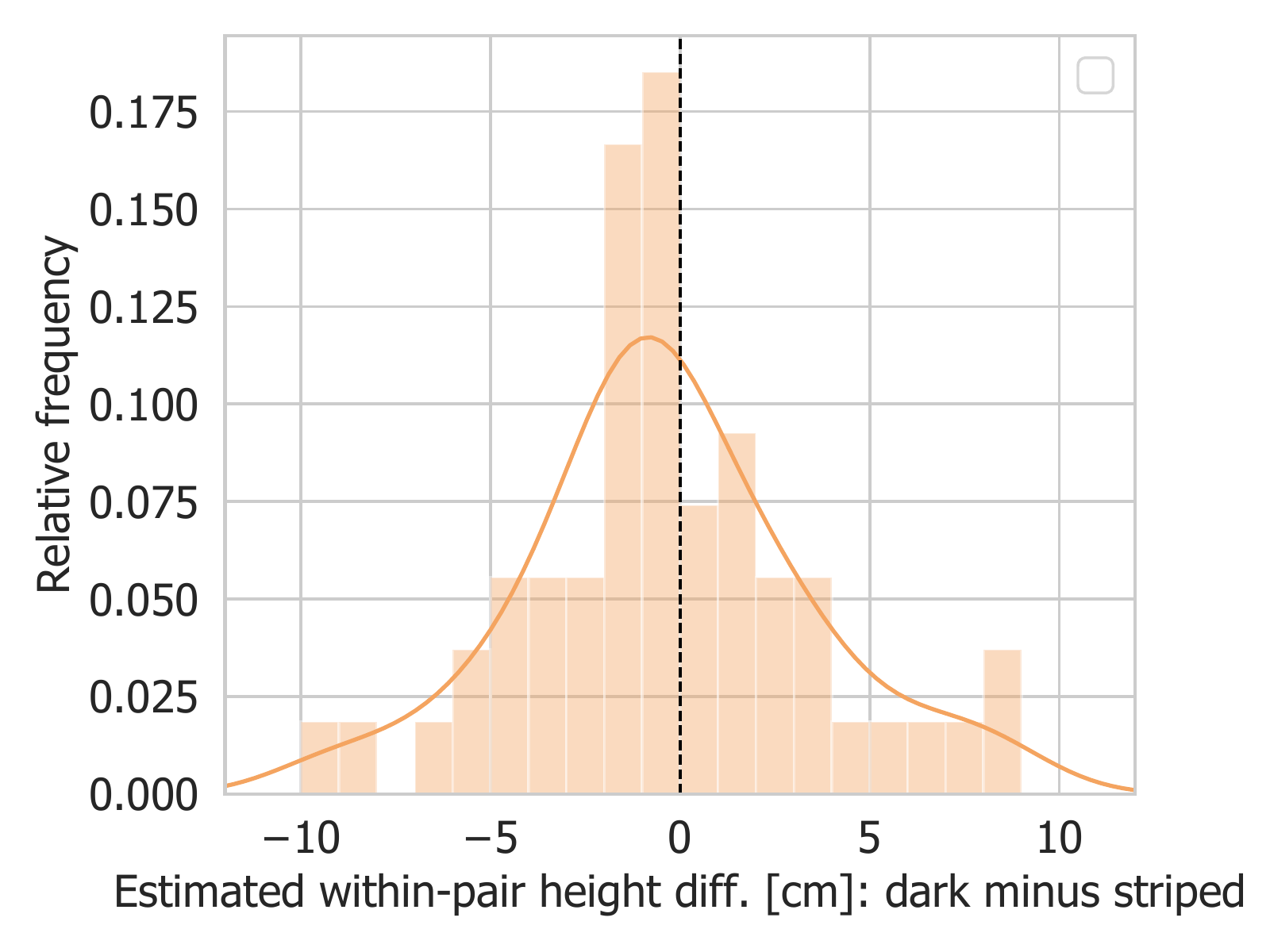}
  		\caption{Height}
  	\end{subfigure}
	\begin{subfigure}{.33\linewidth}
  		\centering
  		\includegraphics[width=\linewidth]{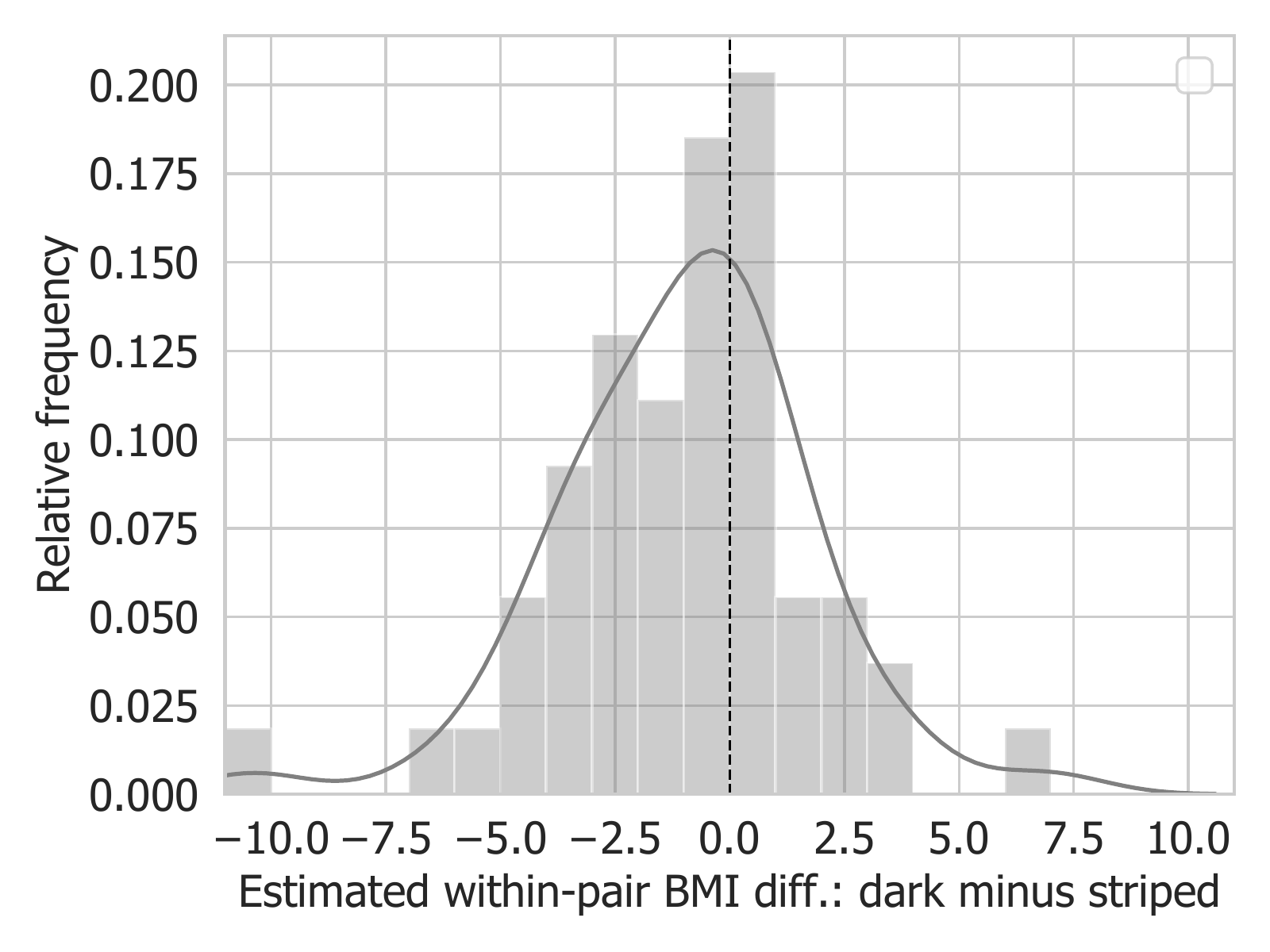}
  		\caption{BMI}
  	\end{subfigure}
    \vspace{2mm}
	\caption{
	Results of observational study, dark \vs\ striped.
	}
  	\label{fig:results obs DS}
\end{figure*}

\begin{figure*}
    \centering
	\begin{subfigure}{.33\linewidth}
 		\centering
  		\includegraphics[width=\linewidth]{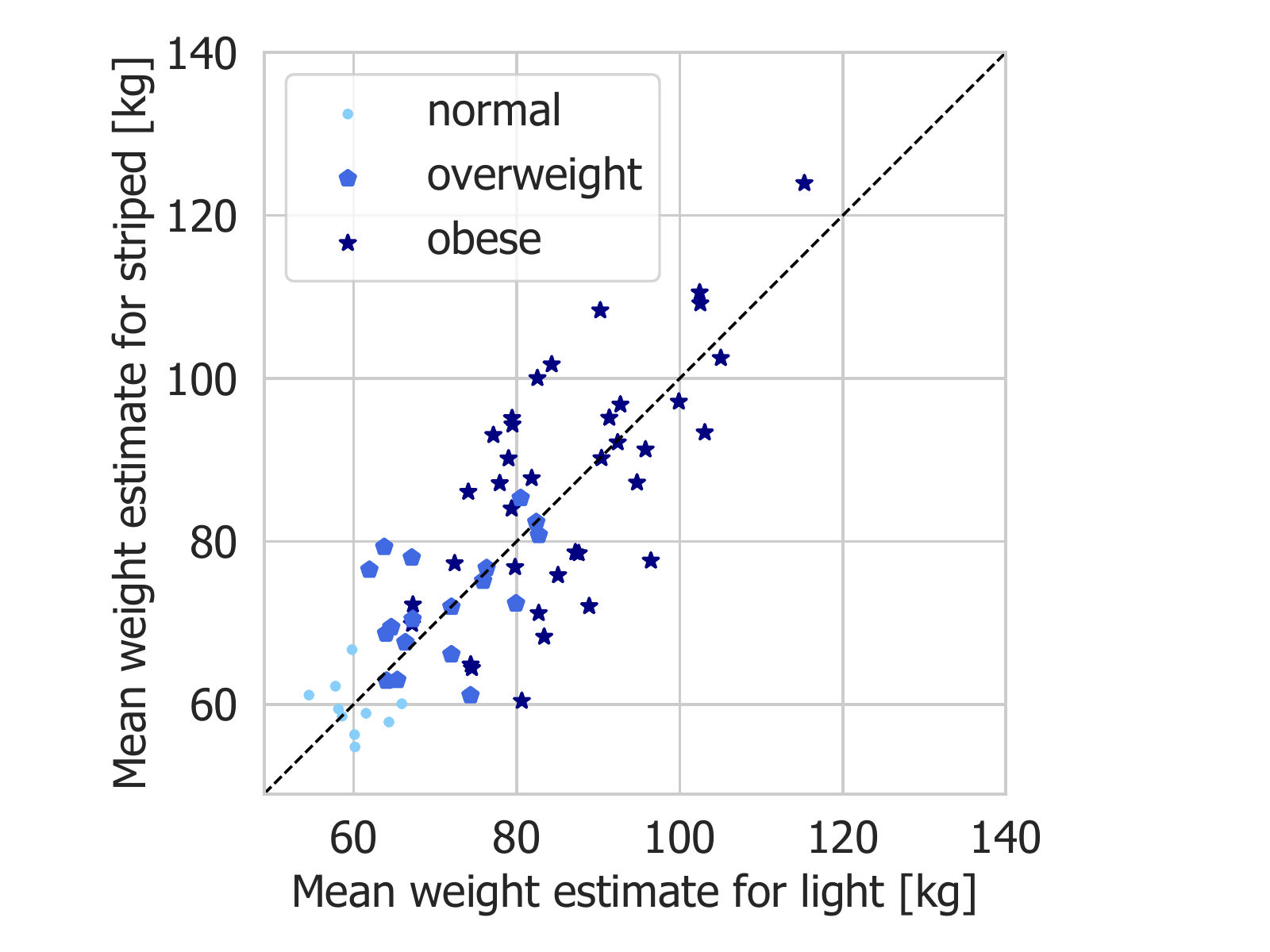}
	\end{subfigure}
	\begin{subfigure}{.33\linewidth}
  		\centering
  		\includegraphics[width=\linewidth]{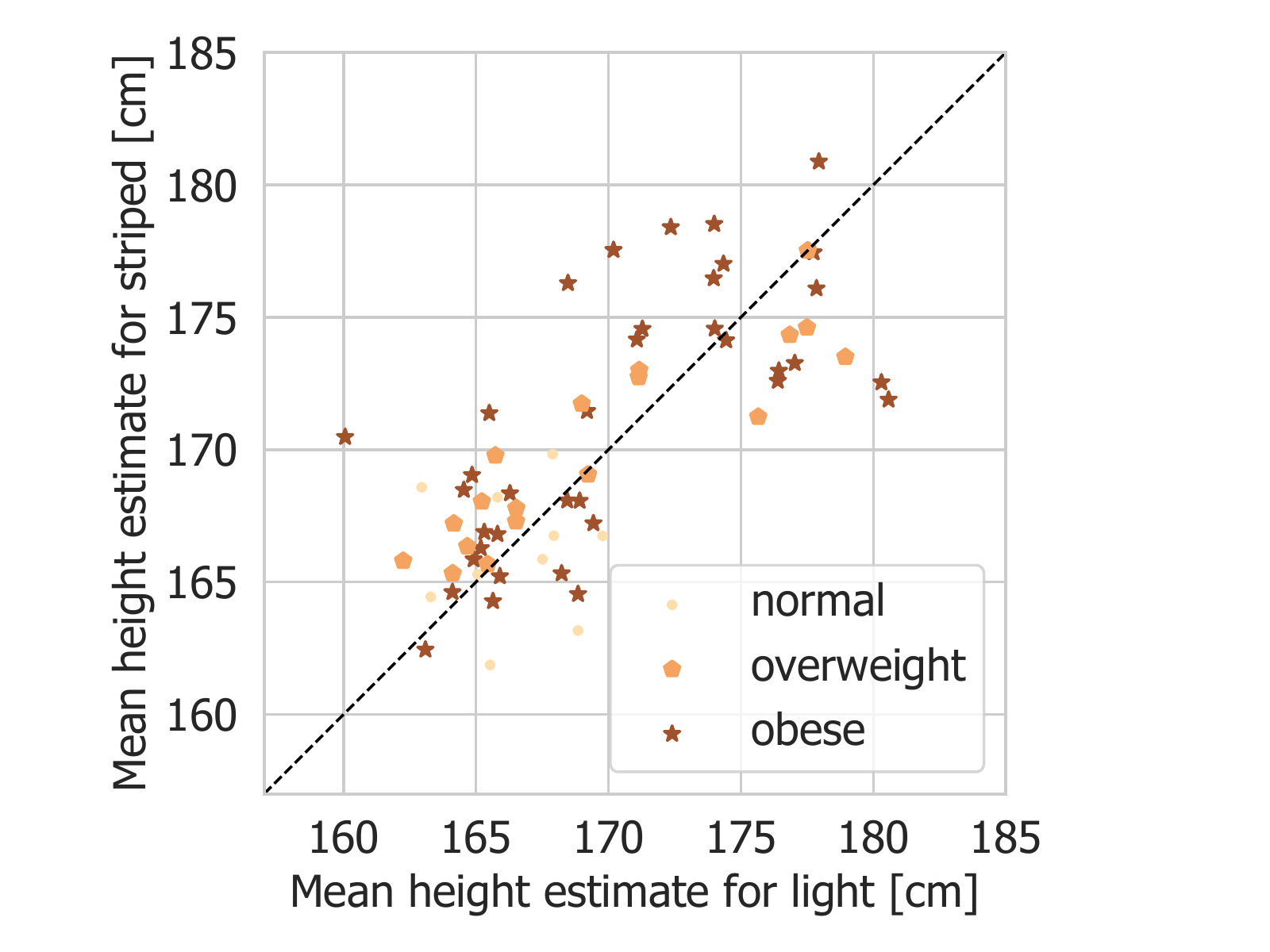}
  	\end{subfigure}
	\begin{subfigure}{.33\linewidth}
  		\centering
  		\includegraphics[width=\linewidth]{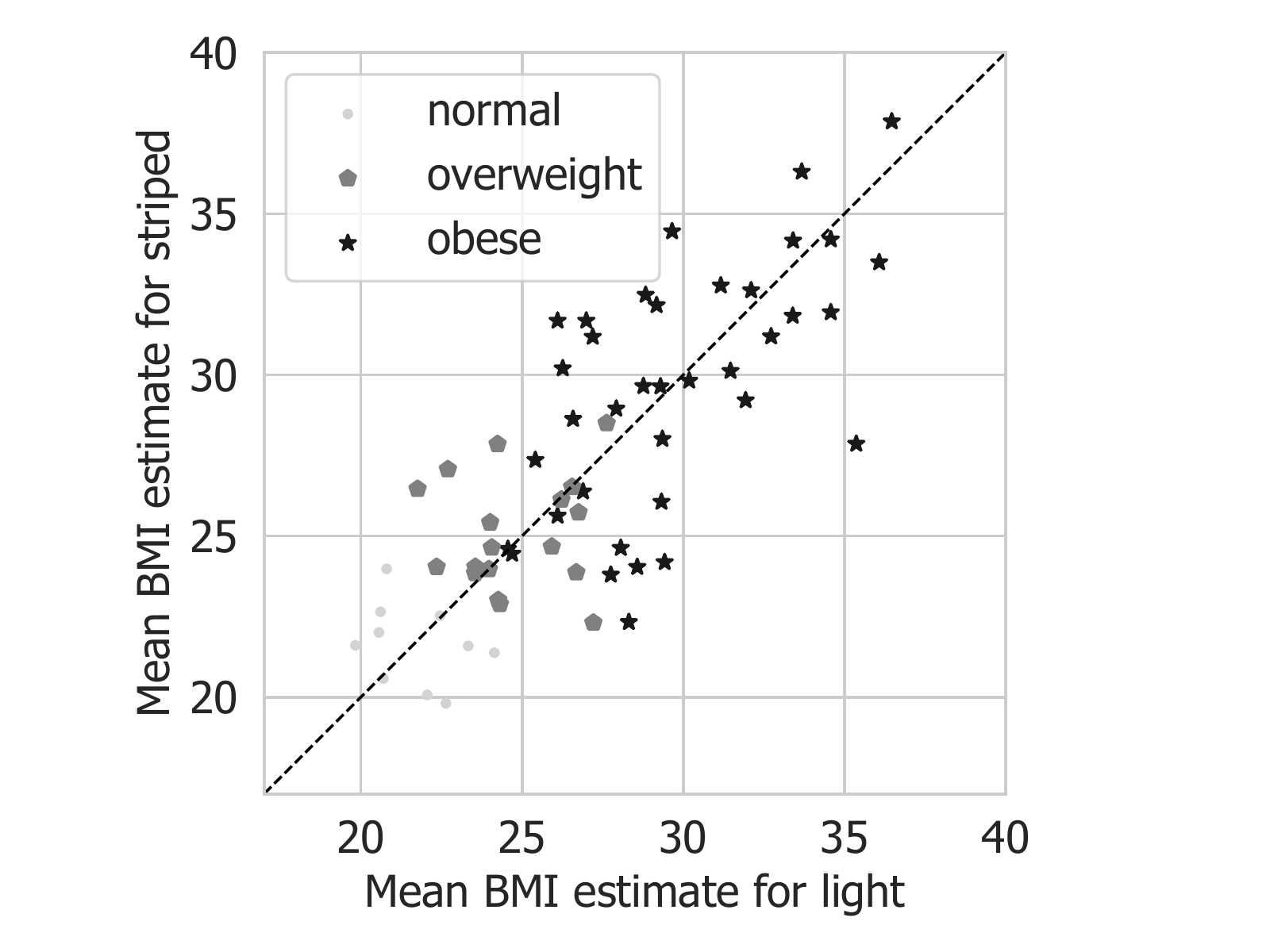}
 	\end{subfigure}

	\begin{subfigure}{.33\linewidth}
 		\centering
  		\includegraphics[width=\linewidth]{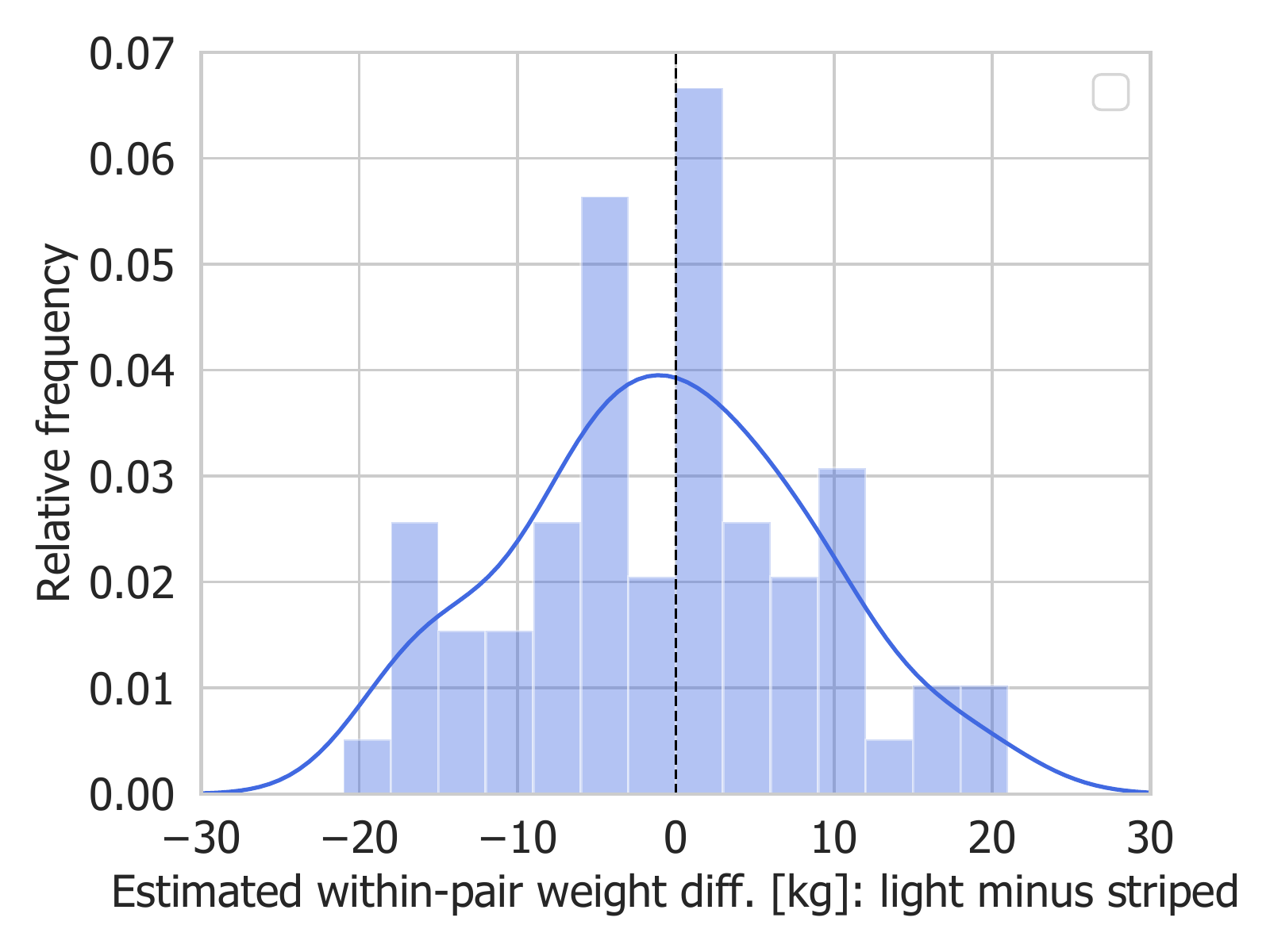}
  		\caption{Weight}
	\end{subfigure}
	\begin{subfigure}{.33\linewidth}
  		\centering
  		\includegraphics[width=\linewidth]{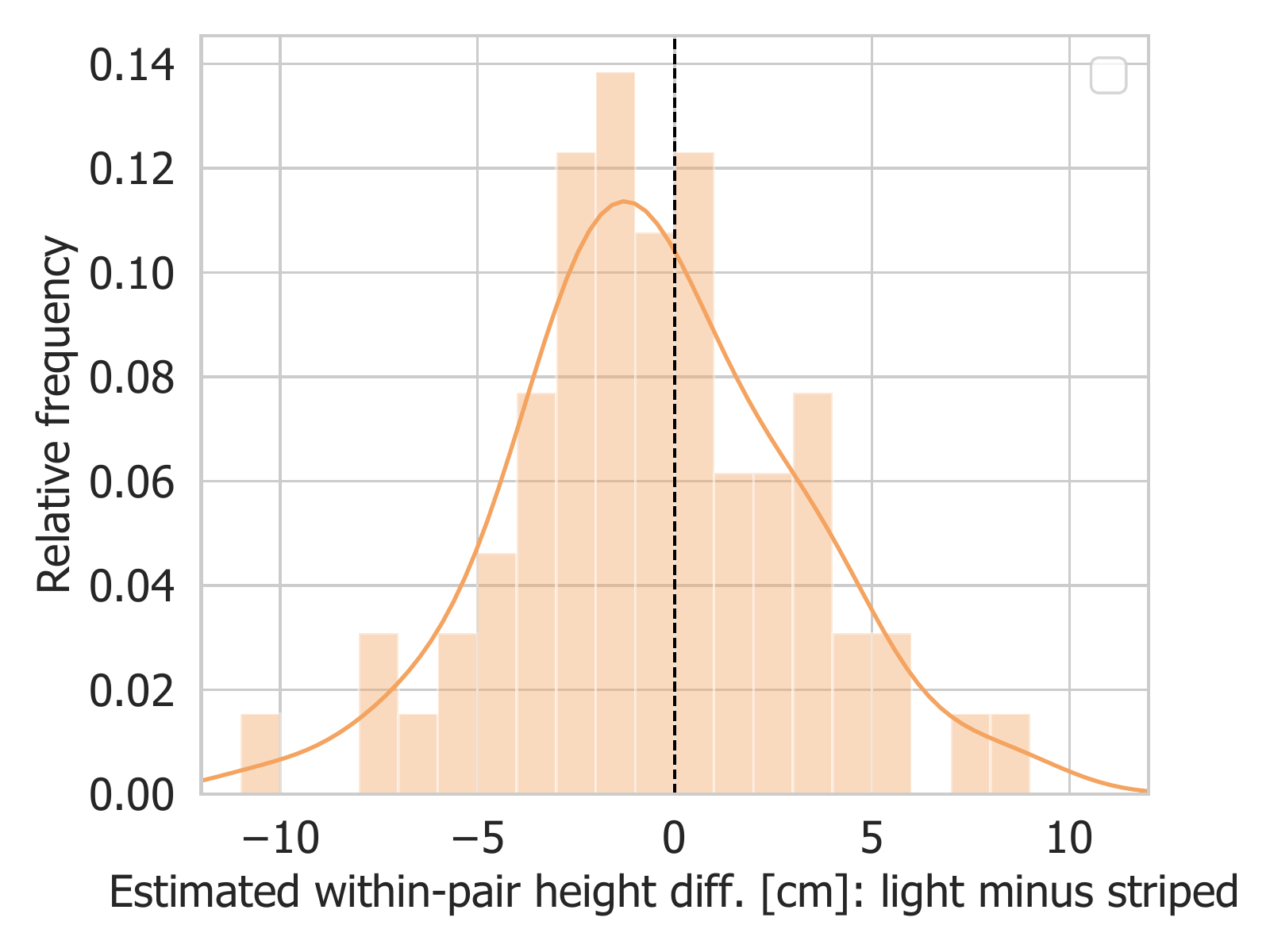}
  		\caption{Height}
  	\end{subfigure}
	\begin{subfigure}{.33\linewidth}
  		\centering
  		\includegraphics[width=\linewidth]{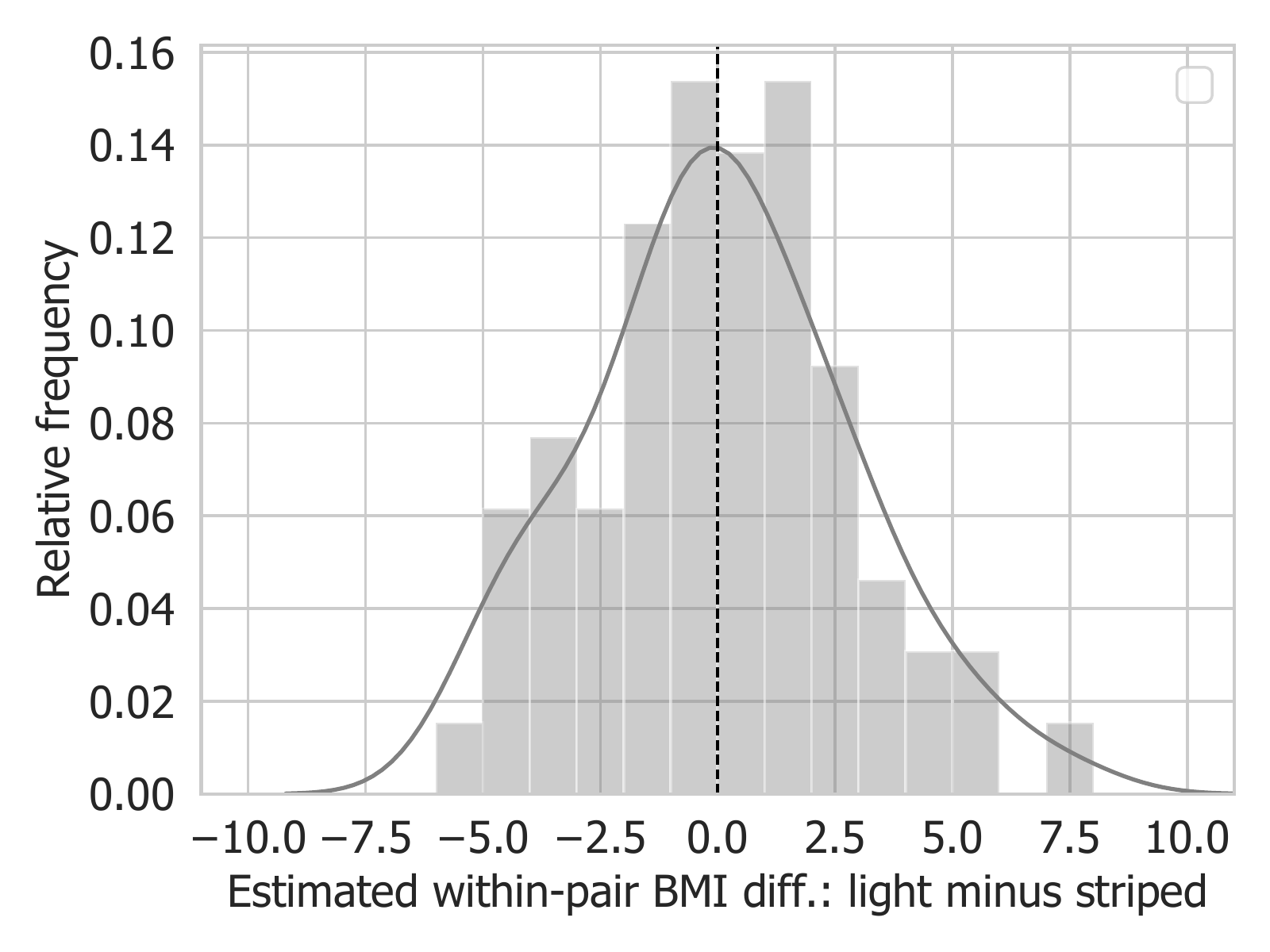}
  		\caption{BMI}
  	\end{subfigure}
    \vspace{2mm}
	\caption{
	Results of observational study, light \vs\ striped.
	}
  	\label{fig:results obs LS}
\end{figure*}

The results are summarized in the second and third columns of the table in \Figref{fig:results obs+exp1}a (also \cf\ \Figref{fig:results obs DS}--\ref{fig:results obs LS}).
We observe that crowd estimates do not differ significantly between horizontally striped and solid light clothes, neither for weight nor for height nor, as a consequence, for the derived BMI ($p=0.68, 0.56, 0.70$, respectively).
Individuals wearing dark clothes, on the contrary, were estimated as significantly less heavy ($p=0.007$), compared to people wearing horizontal stripes, with a mean difference of 3.08~kg in favor of dark.
The effect is also reflected in a difference of 1.05 BMI points ($p=0.007$) in favor of dark.

\xhdr{Summary}
Overall, the results from the observational study may be summarized as follows:
\begin{enumerate}
    \item Individuals wearing solid dark colors were judged to weigh less by a small but statistically significant amount of 2--3~kg, compared to individuals wearing solid light colors or horizontal stripes.
    \item Horizontal stripes and solid light colors were not significantly different in terms of weight perception.
    \item In terms of height perception, all three clothing types (solid light, solid dark, horizontal stripes) were indistinguishable.
\end{enumerate}


Taken together, the fact that weight, but not height, was perceived significantly lower for people wearing dark indicates that dark clothes reduce perceived body size---if the matching has balanced all confounding factors, an \textit{if} that we remove with the experimental studies, whose results we discuss next.

\subsection{Experimental \studyOne: weight estimation of manipulated images}
\label{sec:Results:Experimental study 1}

\begin{figure*}
    \centering
	\begin{subfigure}{.33\linewidth}
 		\centering
  		\includegraphics[width=\linewidth]{Plots/plot_scatter_w_pairwise_comp_clean_dark_light.pdf}
	\end{subfigure}
	\begin{subfigure}{.33\linewidth}
  		\centering
  		\includegraphics[width=\linewidth]{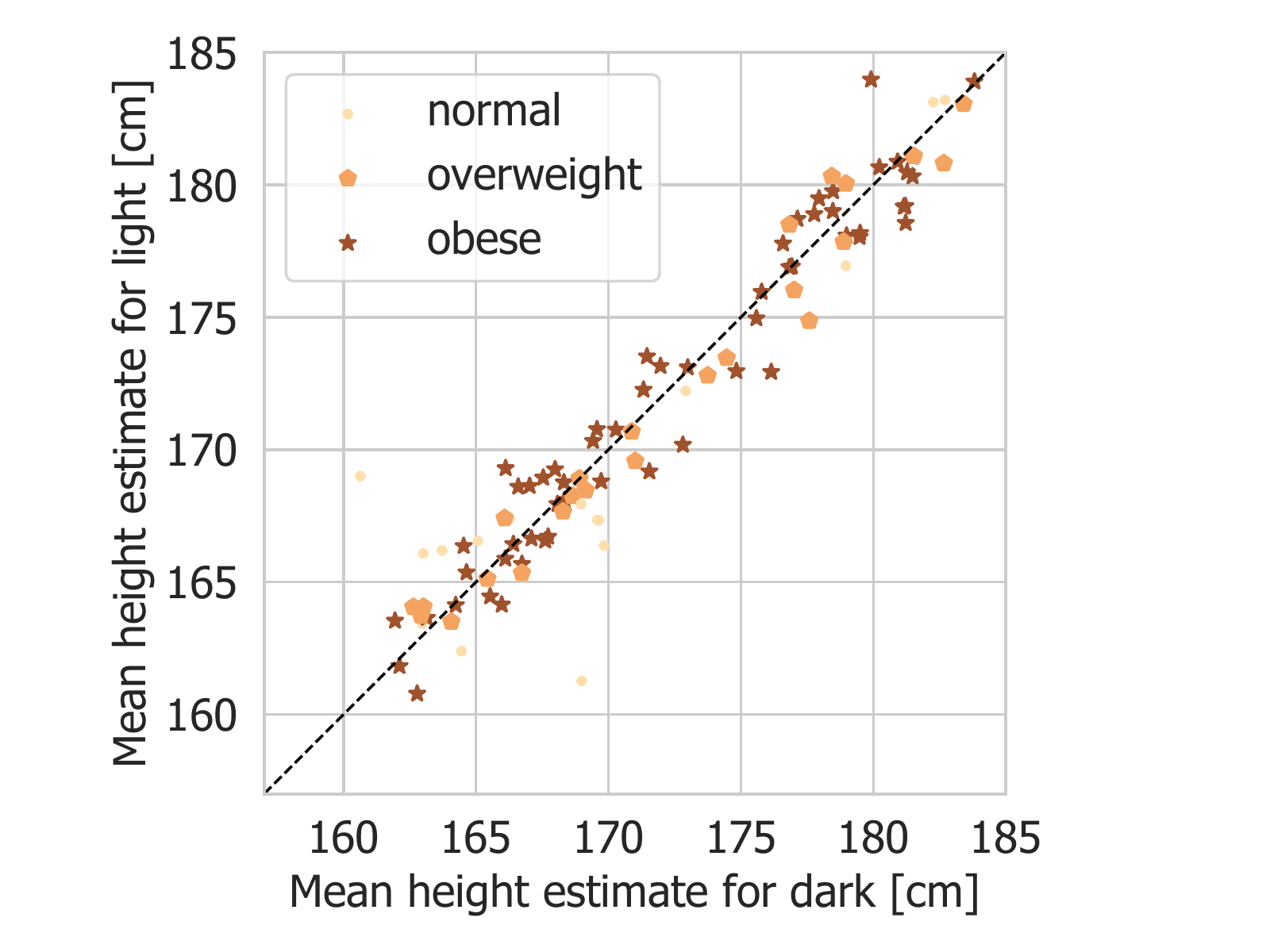}
  	\end{subfigure}
	\begin{subfigure}{.33\linewidth}
  		\centering
  		\includegraphics[width=\linewidth]{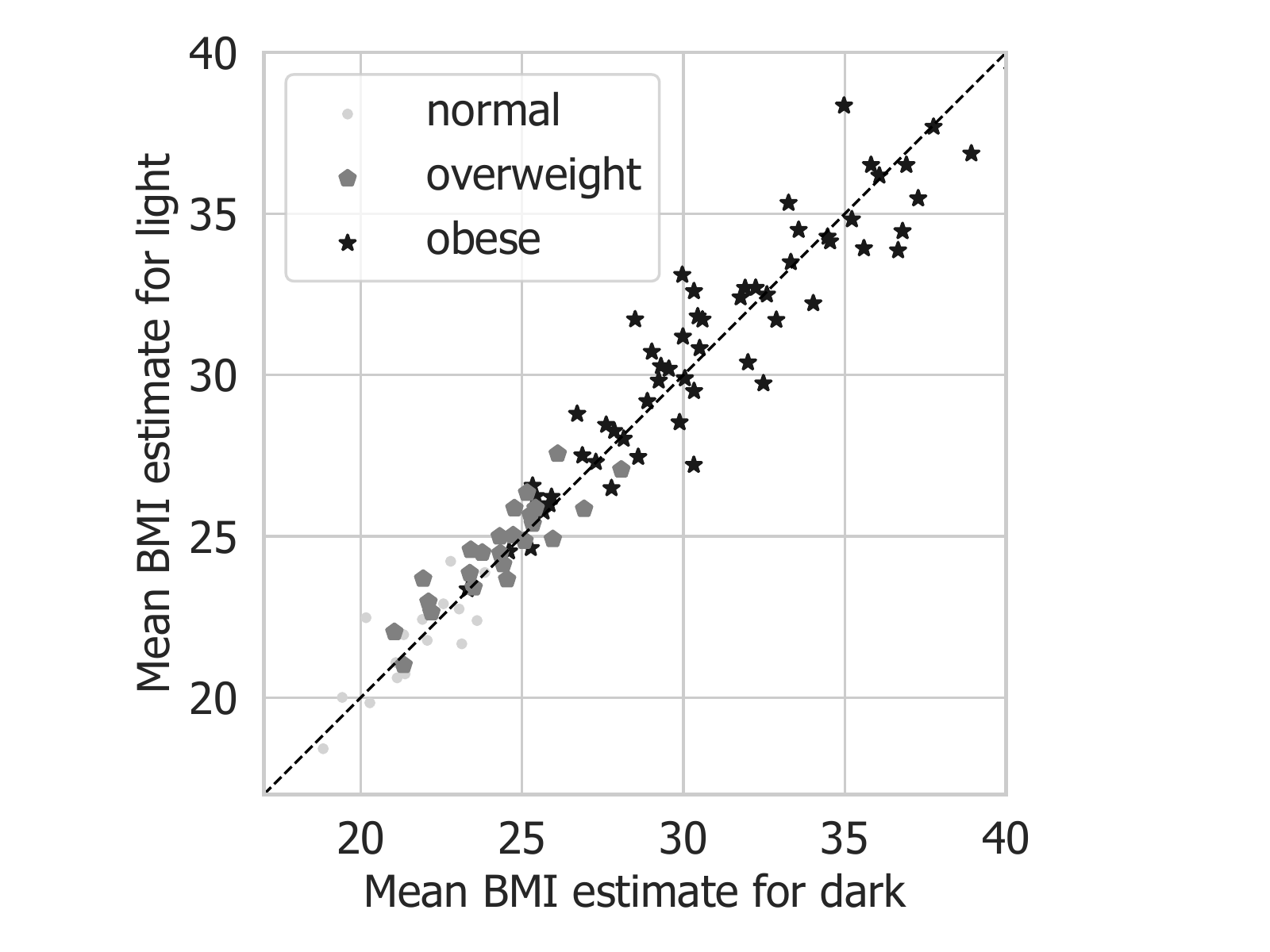}
 	\end{subfigure}

	\begin{subfigure}{.33\linewidth}
 		\centering
  		\includegraphics[width=\linewidth]{Plots/plot_dist_w_pairwise_est_clean_dark_light.pdf}
  		\caption{Weight}
	\end{subfigure}
	\begin{subfigure}{.33\linewidth}
  		\centering
  		\includegraphics[width=\linewidth]{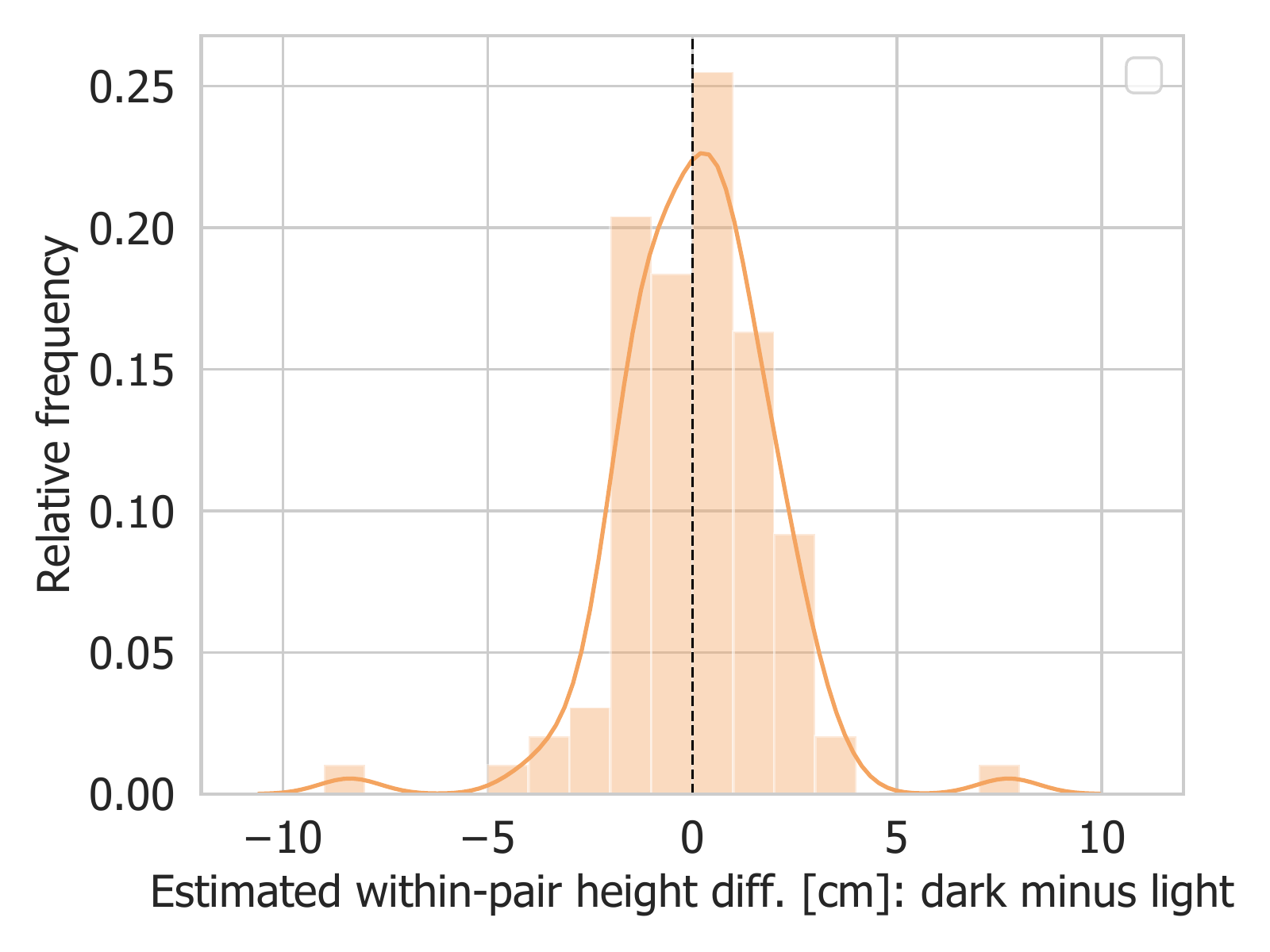}
  		\caption{Height}
  	\end{subfigure}
	\begin{subfigure}{.33\linewidth}
  		\centering
  		\includegraphics[width=\linewidth]{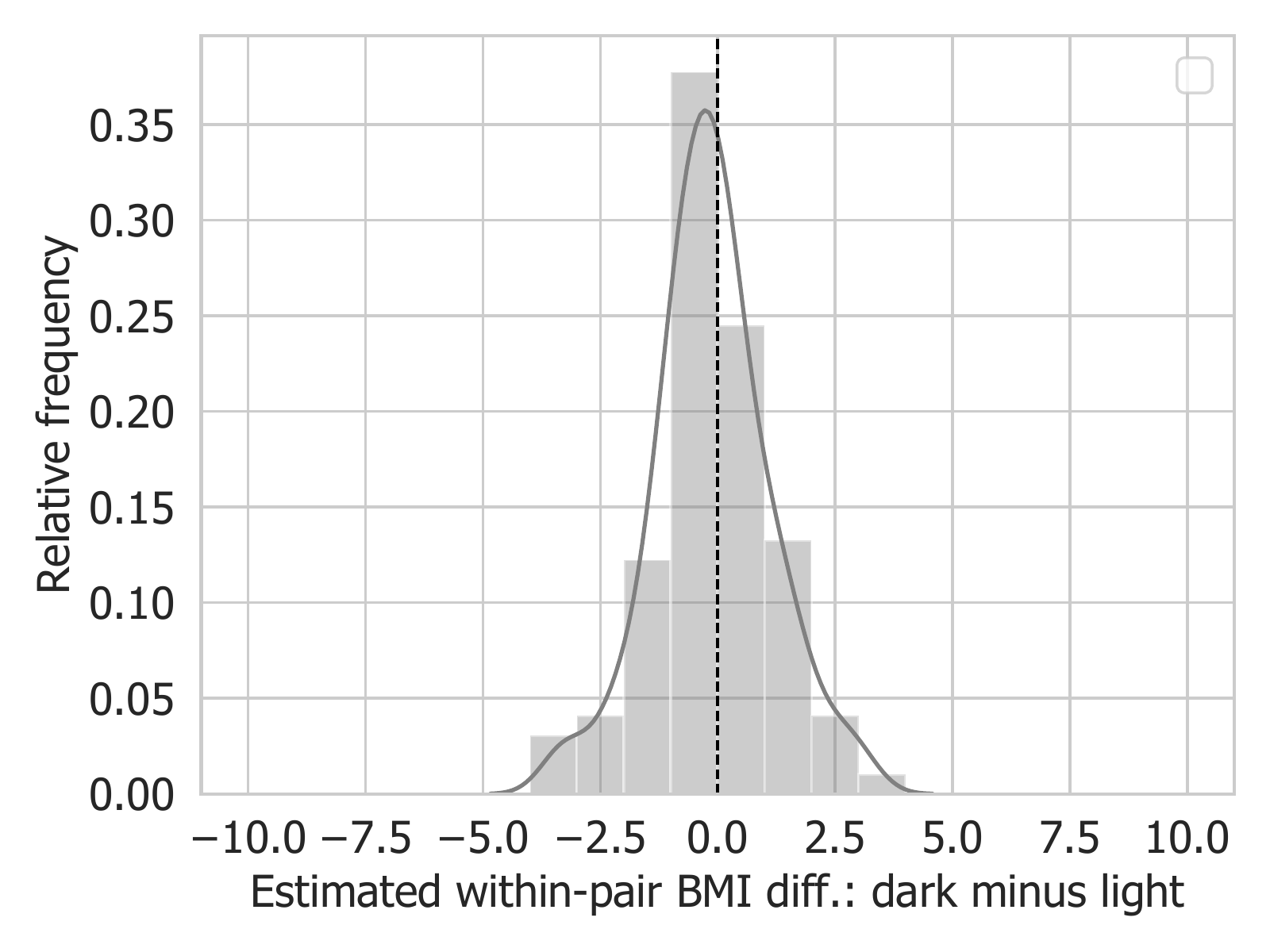}
  		\caption{BMI}
  	\end{subfigure}
    \vspace{2mm}
	\caption{
	Results of experimental \studyOne, dark \vs\ light.
	}
  	\label{fig:results exp1 DL}
\end{figure*}

\begin{figure*}
    \centering
	\begin{subfigure}{.33\linewidth}
 		\centering
  		\includegraphics[width=\linewidth]{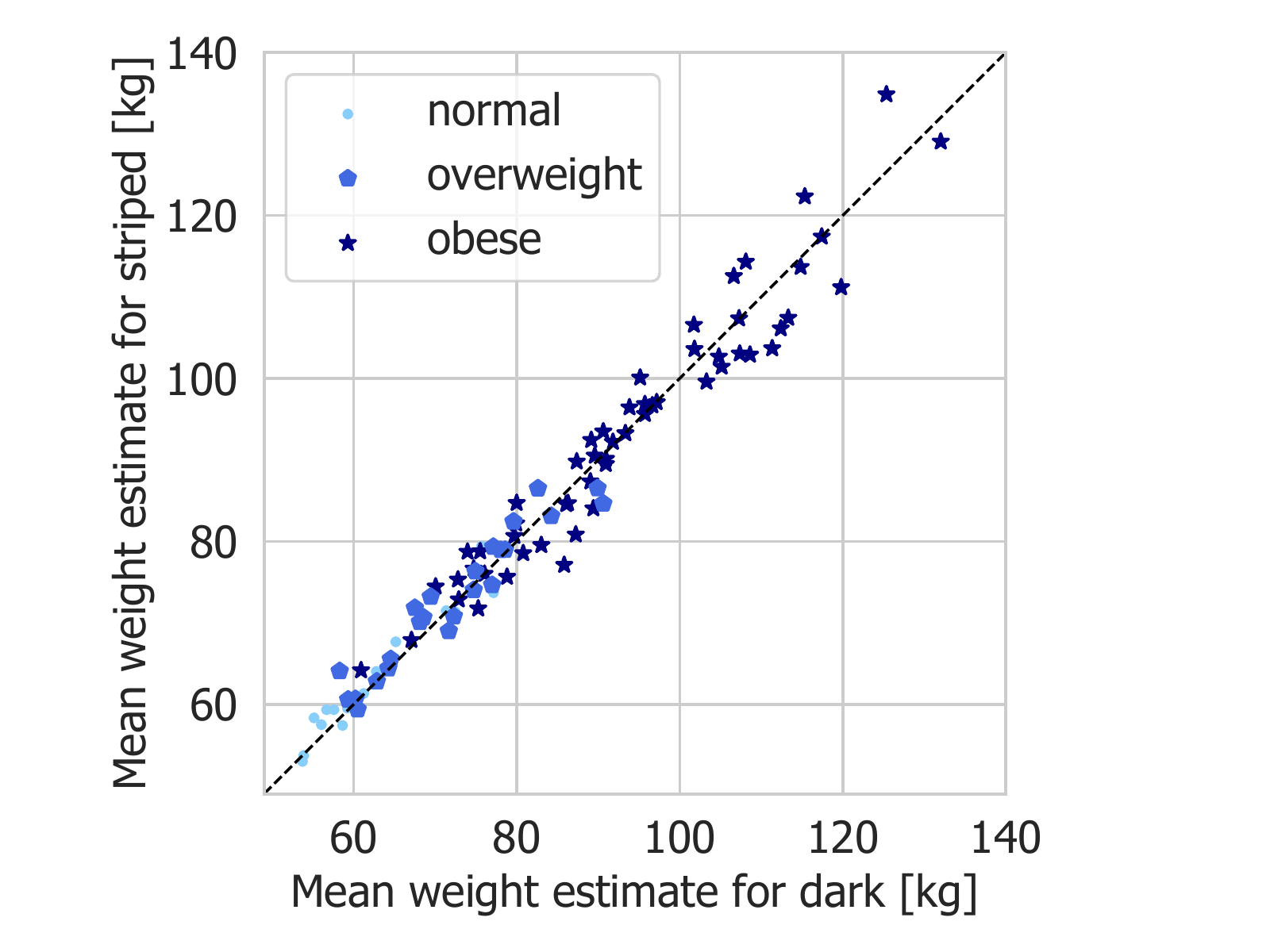}
	\end{subfigure}
	\begin{subfigure}{.33\linewidth}
  		\centering
  		\includegraphics[width=\linewidth]{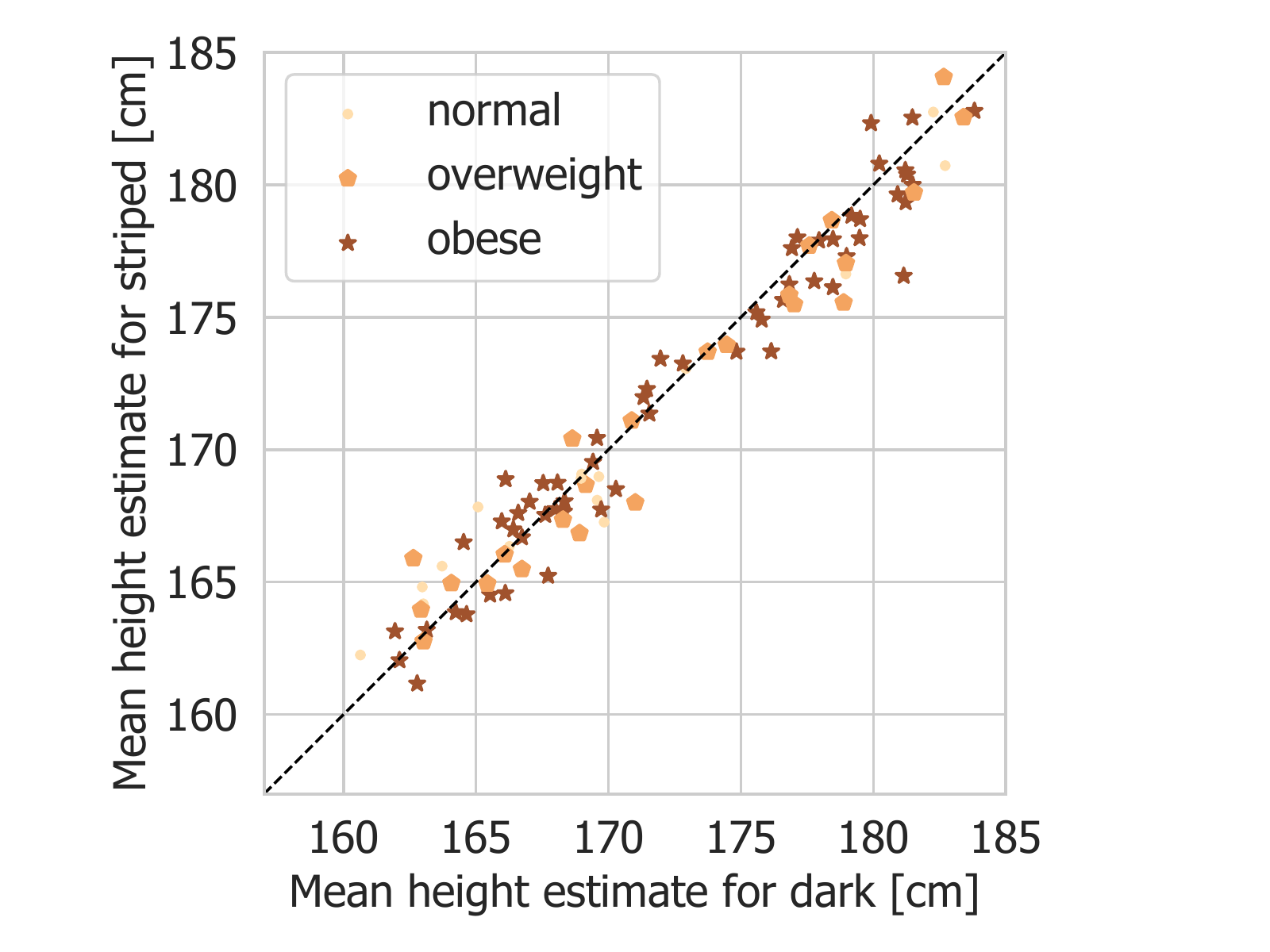}
  	\end{subfigure}
	\begin{subfigure}{.33\linewidth}
  		\centering
  		\includegraphics[width=\linewidth]{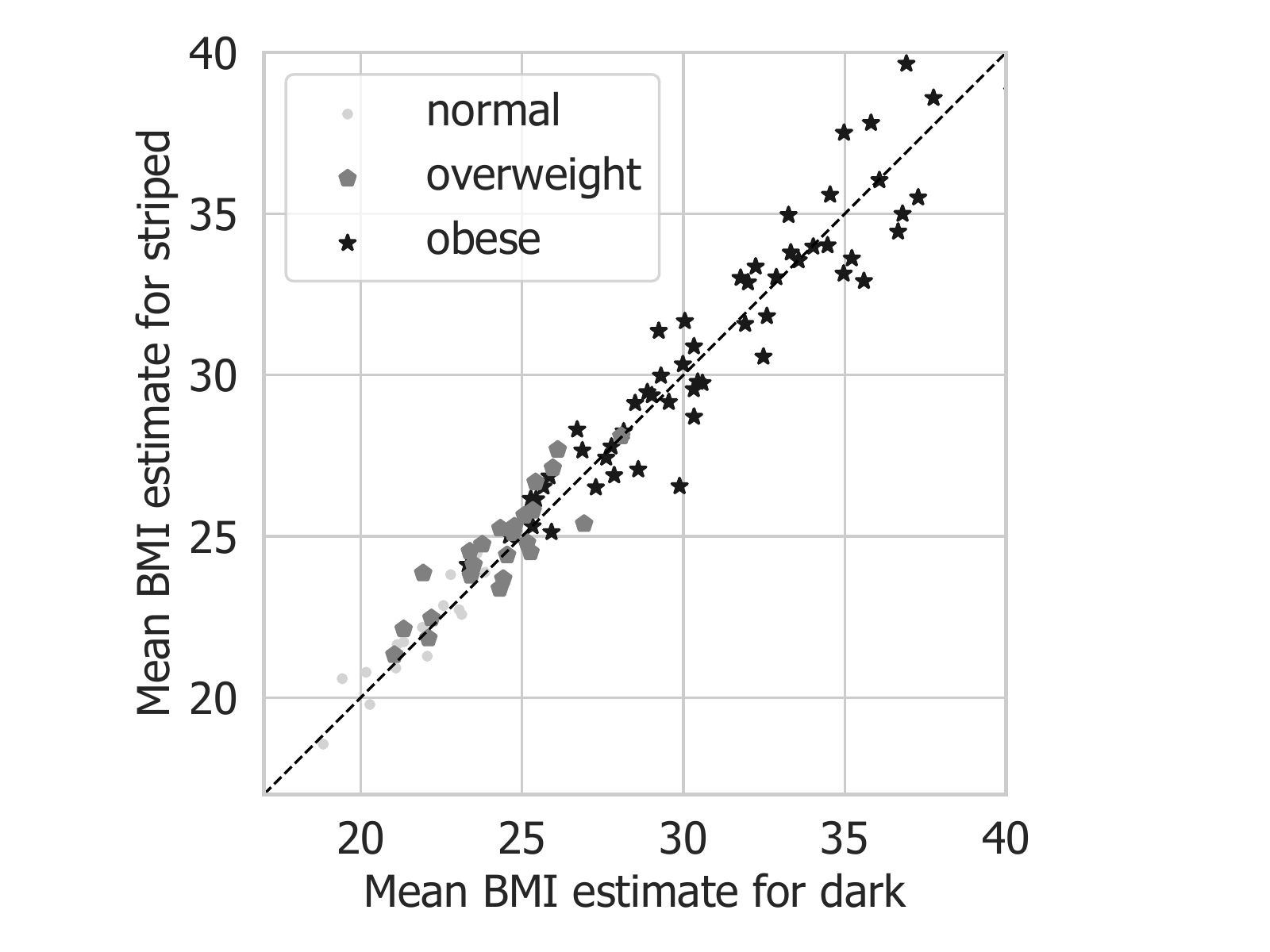}
 	\end{subfigure}

	\begin{subfigure}{.33\linewidth}
 		\centering
  		\includegraphics[width=\linewidth]{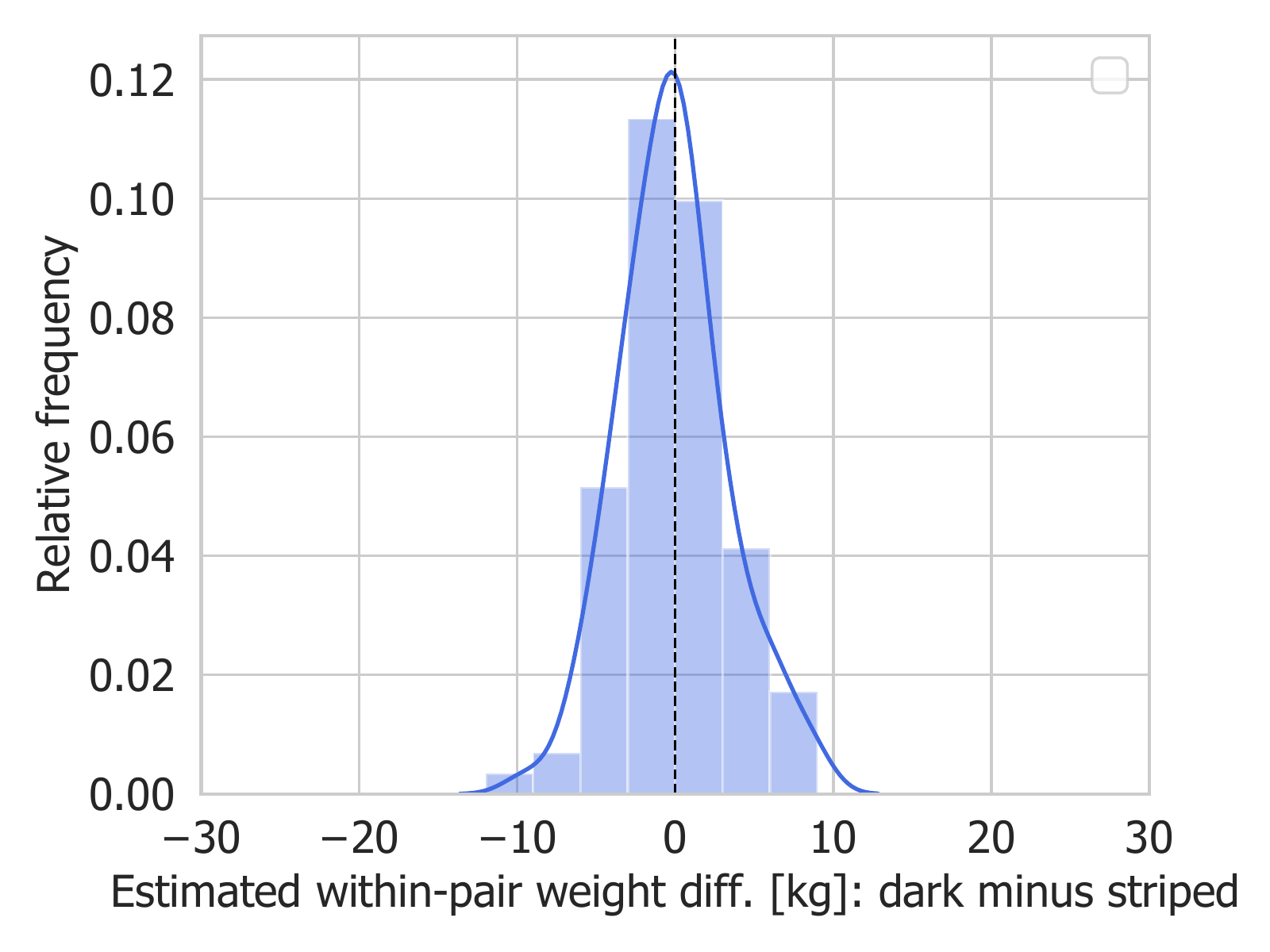}
  		\caption{Weight}
	\end{subfigure}
	\begin{subfigure}{.33\linewidth}
  		\centering
  		\includegraphics[width=\linewidth]{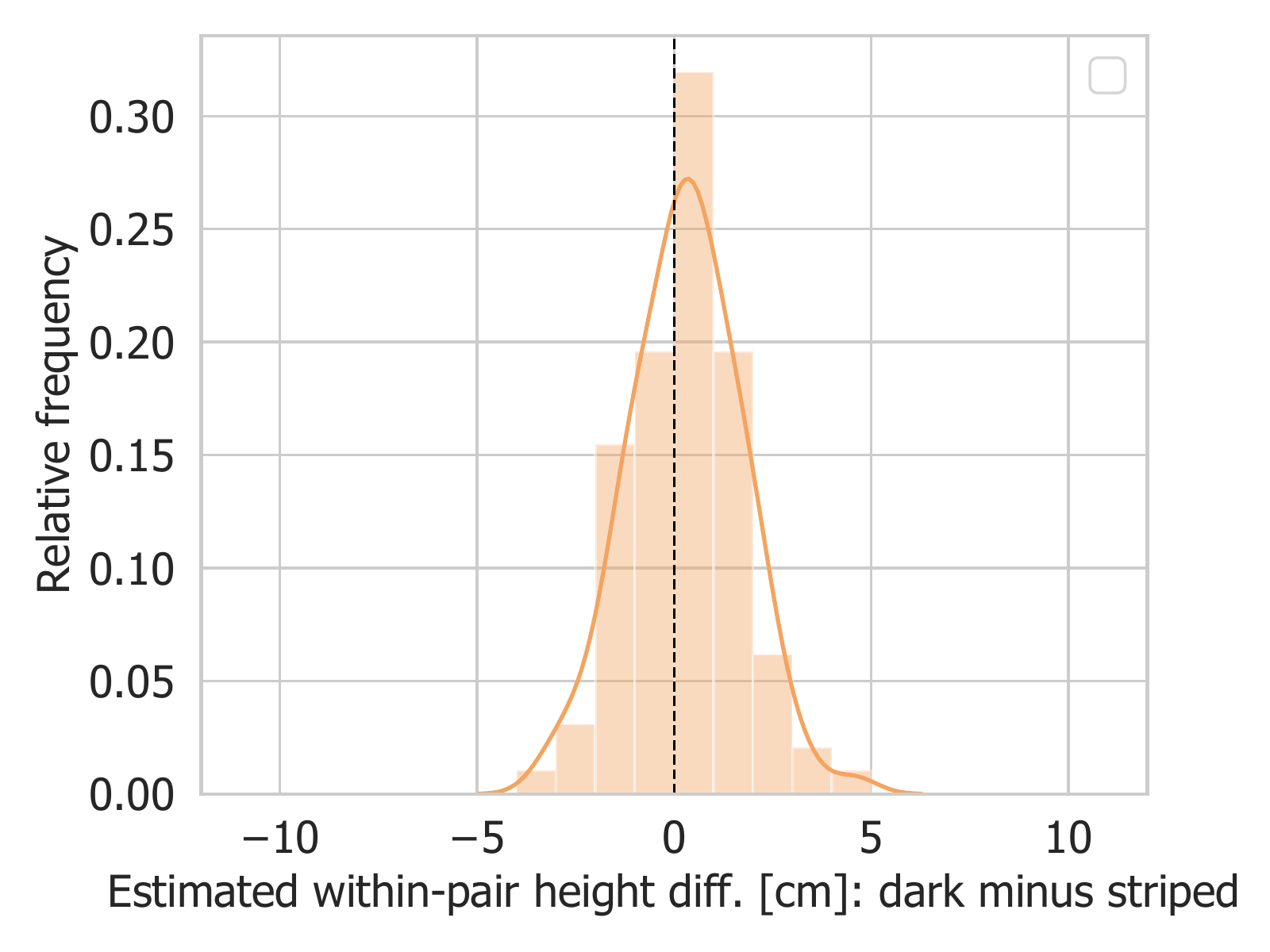}
  		\caption{Height}
  	\end{subfigure}
	\begin{subfigure}{.33\linewidth}
  		\centering
  		\includegraphics[width=\linewidth]{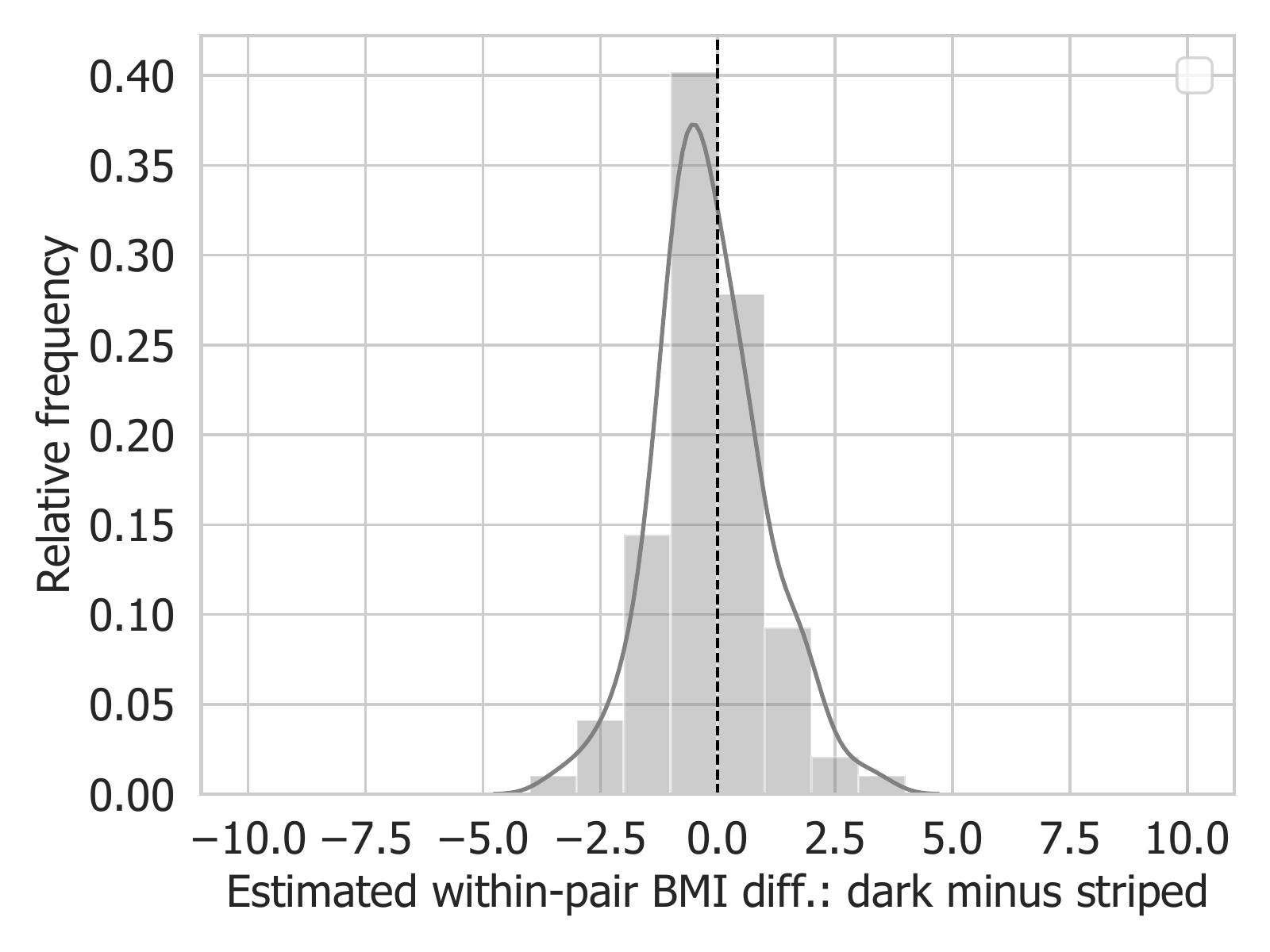}
  		\caption{BMI}
  	\end{subfigure}
    \vspace{2mm}
	\caption{
	Results of experimental \studyOne, dark \vs\ striped.
	}
  	\label{fig:results exp1 DS}
\end{figure*}

\begin{figure*}
    \centering
	\begin{subfigure}{.33\linewidth}
 		\centering
  		\includegraphics[width=\linewidth]{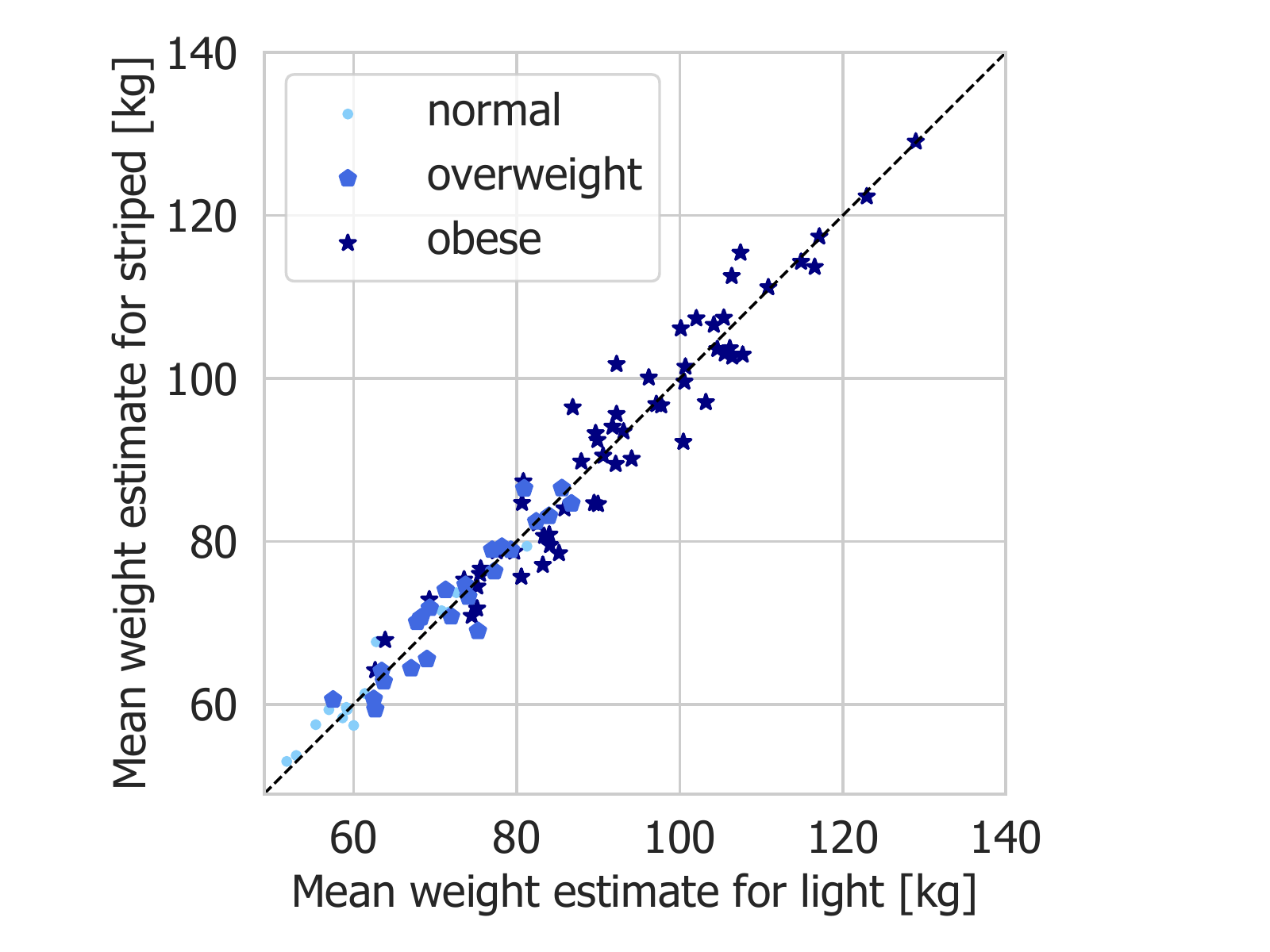}
	\end{subfigure}
	\begin{subfigure}{.33\linewidth}
  		\centering
  		\includegraphics[width=\linewidth]{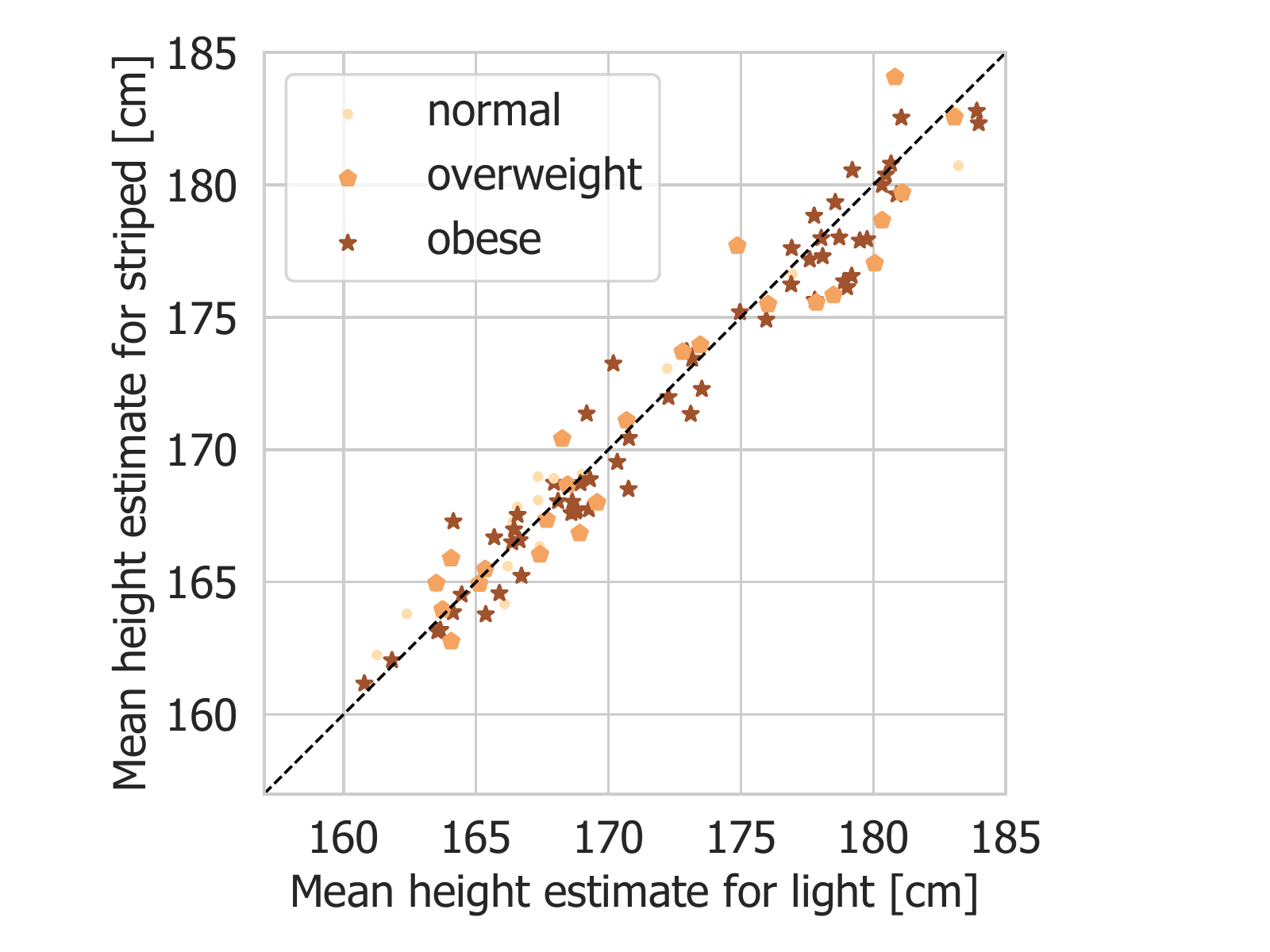}
  	\end{subfigure}
	\begin{subfigure}{.33\linewidth}
  		\centering
  		\includegraphics[width=\linewidth]{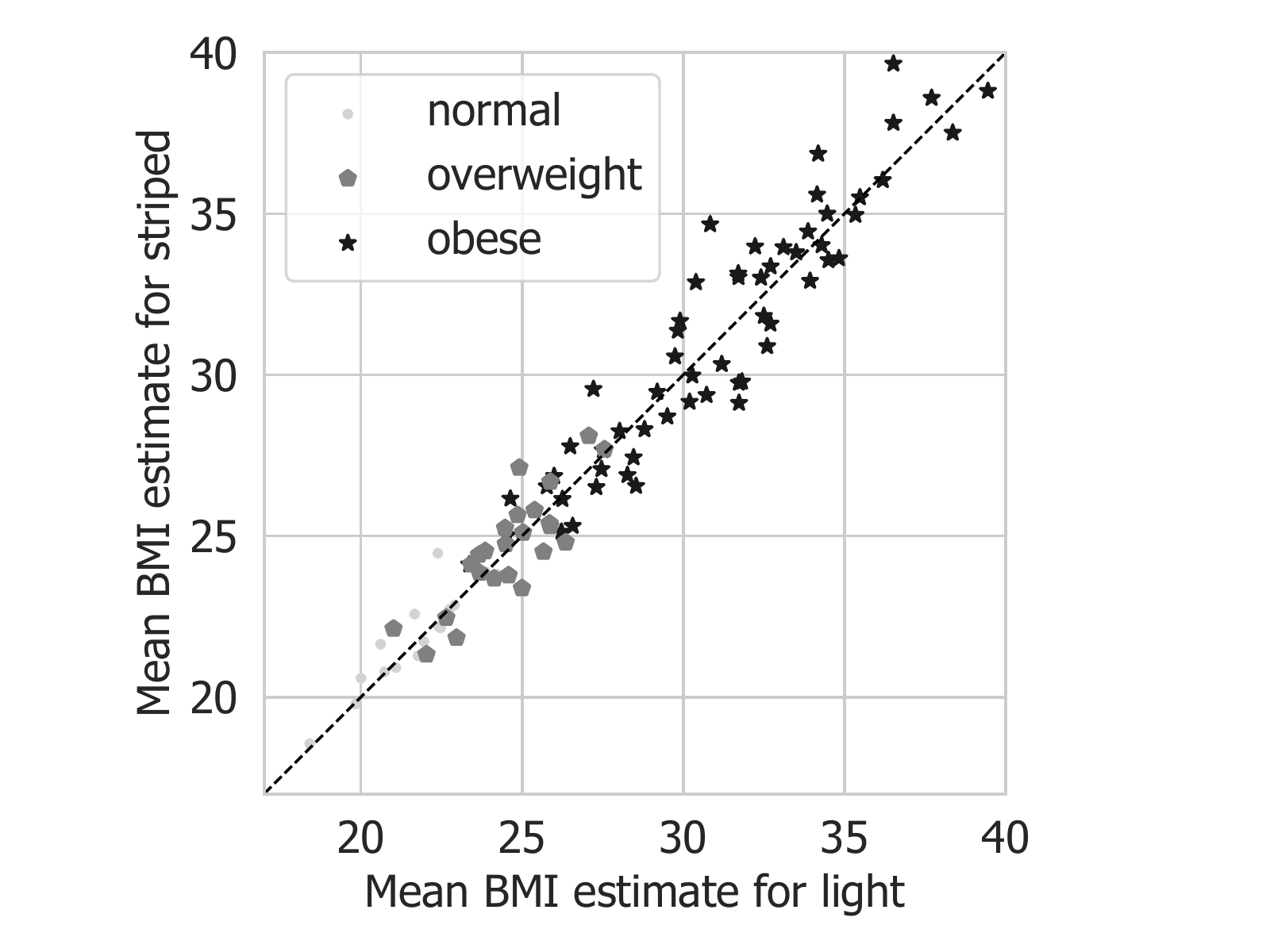}
 	\end{subfigure}

	\begin{subfigure}{.33\linewidth}
 		\centering
  		\includegraphics[width=\linewidth]{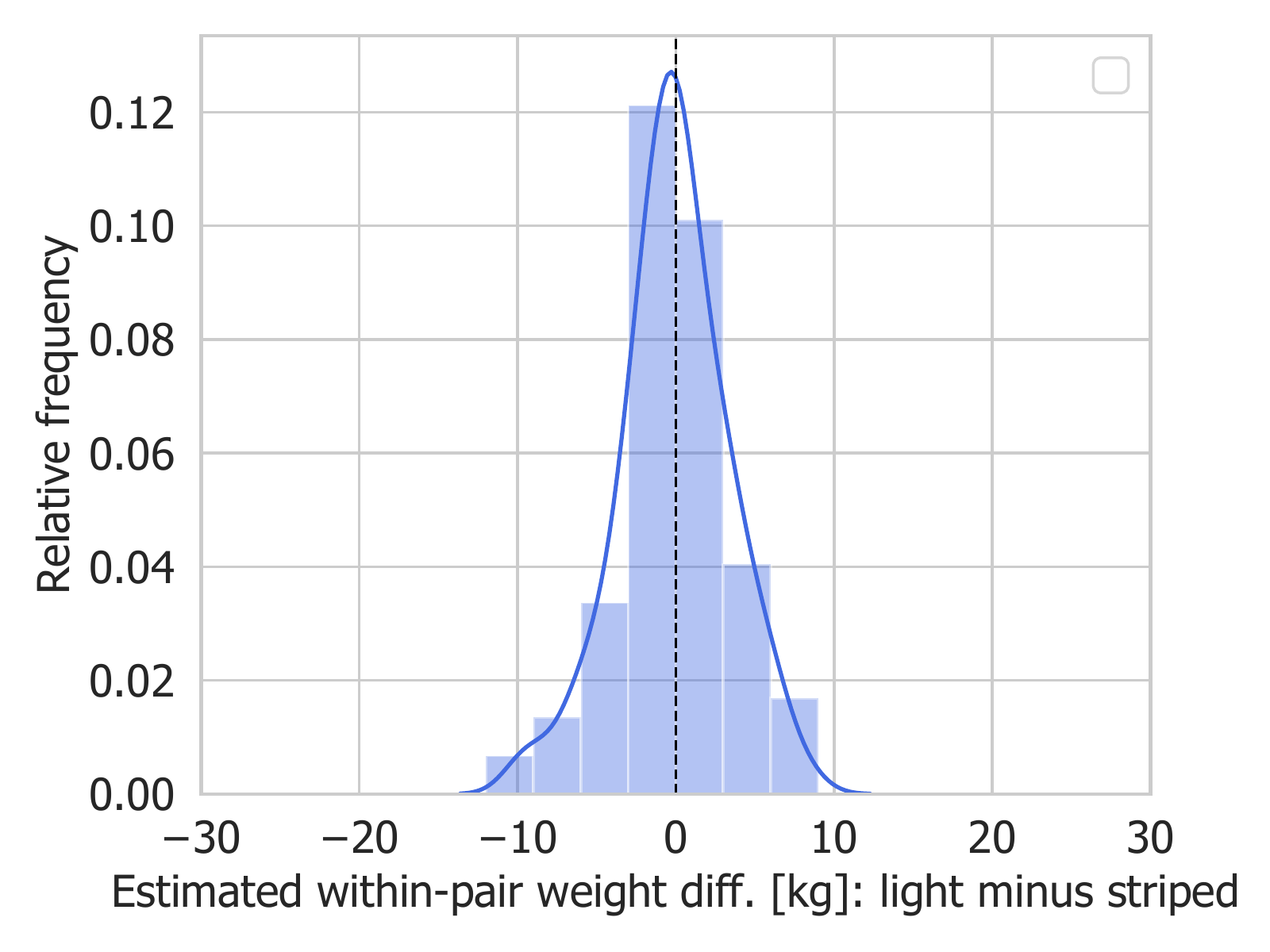}
  		\caption{Weight}
	\end{subfigure}
	\begin{subfigure}{.33\linewidth}
  		\centering
  		\includegraphics[width=\linewidth]{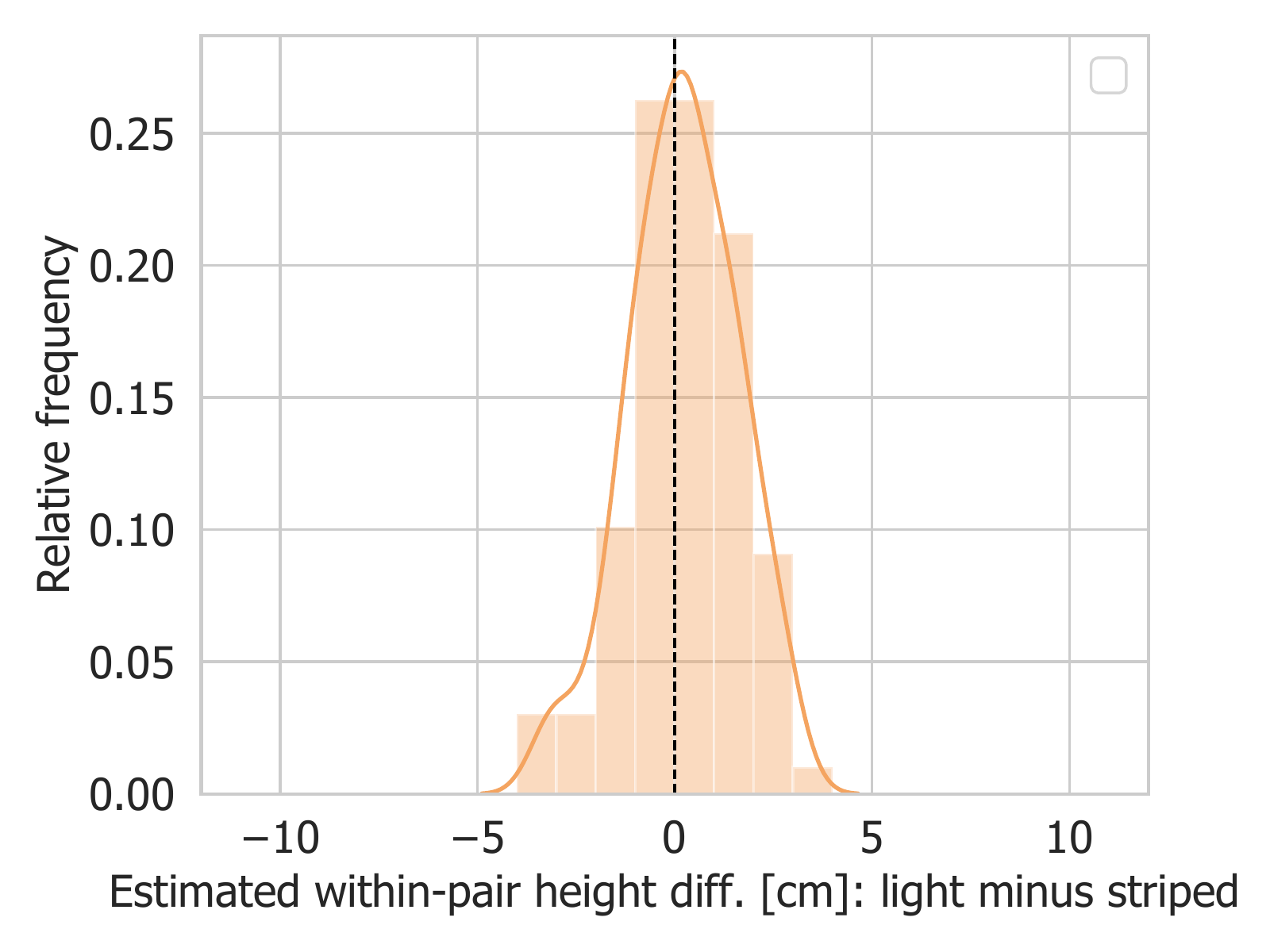}
  		\caption{Height}
  	\end{subfigure}
	\begin{subfigure}{.33\linewidth}
  		\centering
  		\includegraphics[width=\linewidth]{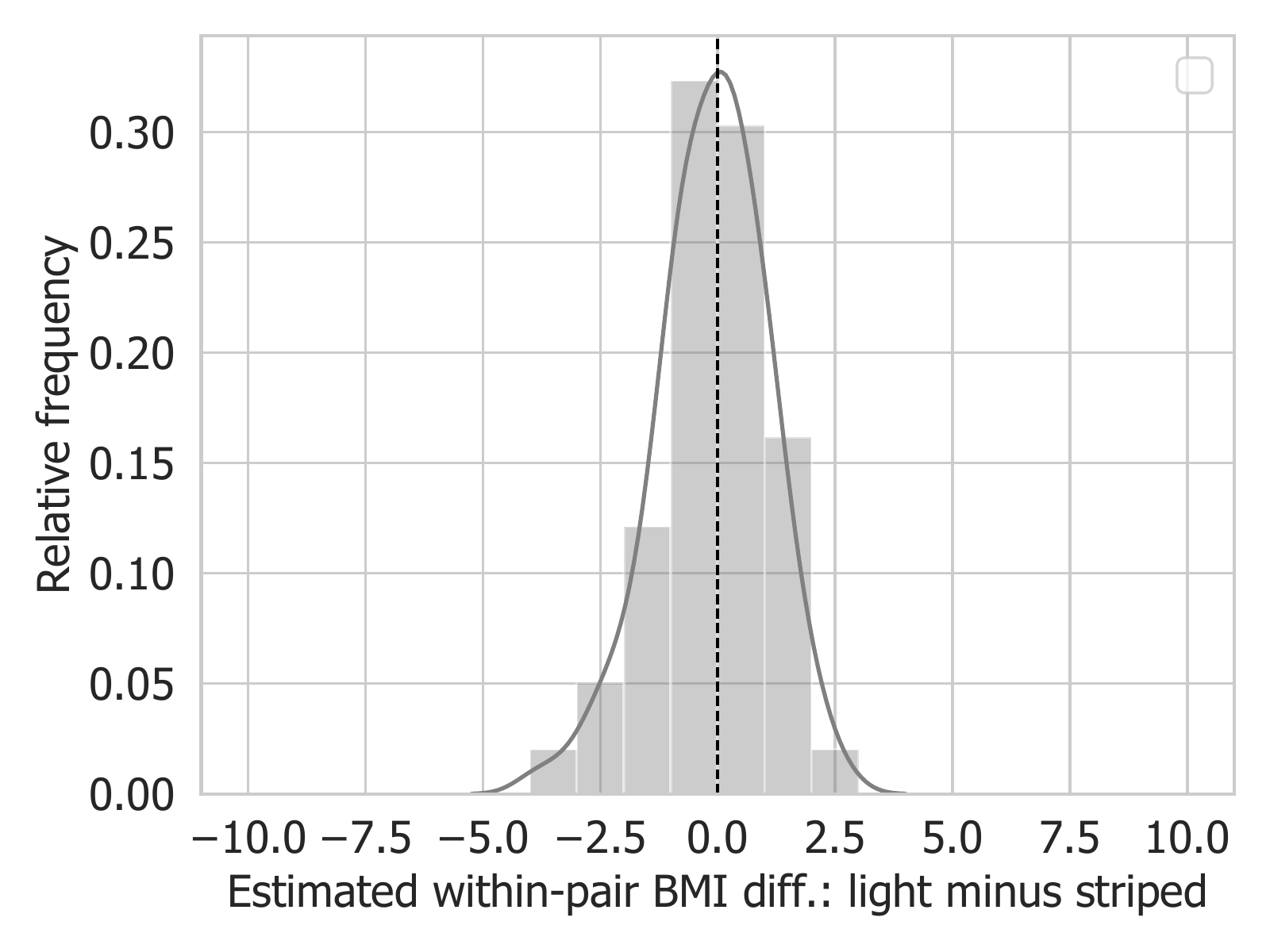}
  		\caption{BMI}
  	\end{subfigure}
    \vspace{2mm}
	\caption{
	Results of experimental \studyOne, light \vs\ striped.
	}
  	\label{fig:results exp1 LS}
\end{figure*}

The analysis of experimental \studyOne is conceptually identical to that of the observational study.
The only difference between the two studies consists in the datasets used:
whereas the observational study compared images that were matched in pairs after the photographs had been taken, experimental \studyOne started from images that had been created to form nearly identical pairs to begin with.%
\footnote{
The complete set of plots for experimental \studyOne is available in \Figref{fig:results exp1 DL}--\ref{fig:results exp1 LS}.
}

\xhdr{Effect of dark colors}
As before, we visualize the weight estimates for all pairs as a scatter plot (\Figref{fig:results obs+exp1}b, left),
display the distribution of within-pair differences of weight estimates (\Figref{fig:results obs+exp1}b, center),
and summarize the results for all of weight, height, and BMI in a table (first column of table in \Figref{fig:results obs+exp1}b).
We observe that the differences were much smaller in this experimental setup, compared to the observational setup (\Secref{sec:Results:Observational study}), presumably both because the matched images were more similar to each other and because we had retained only pairs of persons whose weight and height were judged similarly under identical clothing conditions (difference under 10~kg or 10~cm, respectively; \cf\ \Secref{sec:Research design:Experimental study 1}).
Although the average within-pair difference in perceived weight between solid light and solid dark clothing had the same sign as in the observational study, the effect was much smaller ($-0.23$~kg \vs\ $-2.27$~kg) and not statistically significant ($p=0.34$ according to Wilcoxon's signed rank test).

\xhdr{Effect of horizontal stripes}
The results for the comparisons of striped with light and dark clothes are summarized in the second and third columns of the table of \Figref{fig:results obs+exp1}b.
As in the observational setup, light and striped were statistically indistinguishable ($p=0.78$), and the average weight estimate for dark was smaller than for striped, but the difference was again much smaller than in the observational setup ($-0.13$~kg \vs\ $-3.08$~kg) and not statistically significant ($p=0.50$).

\xhdr{Summary}
The above results regarding weight estimation are inconclusive:
on the one hand, the effects point in the same direction as in the observational study, indicating that dark clothing may decrease perceived weight,
but, possibly due to much more closely matched image pairs, the measured effects are considerably smaller here and not statistically significant.
This could either mean that there is no effect or that there is a small effect that could not be detected by the present methodology due to a small sample size.
It is this disambiguity that led us to design experimental \studyTwo, which had increased power by moving from absolute to relative weight estimation.

\subsection{Experimental \studyTwo: pairwise weight comparison of manipulated images}
\label{sec:Results:Experimental study 2}

In experimental \studyTwo, we analyzed the same set of manipulated images as in experimental \studyOne, but using a fundamentally different methodology, based on relative pairwise weight comparison as opposed to absolute weight estimation.

Based on the formulaic result of the observational study---``$D < L \approx S$''---, which was qualitatively but insignificantly supported by experimental \studyOne,
we focused on comparing solid dark to solid light clothes, and solid light to horizontally striped clothes.
In other words, the question was:
Can we confirm that indeed ``$D < L$'' and ``$L \approx S$''?

\xhdr{Effect of dark colors}
We start with the results of the comparison of dark \vs\ light.
Recall from \Secref{sec:Research design:Experimental study 2} that, for a given person $i$, the effect of wearing solid dark rather than solid light is captured by the vote-share difference~$\delta_i$.
Averaging $\delta_i$ over all persons $i$, we obtain the overall causal effect $\delta$ of wearing dark, rather than light, on weight perception.
It amounted to $\delta=2.7\%$, a small but statistically significant effect ($p=0.0069$; 95\% CI $[0.88\%, 4.6\%]$).
In words, when a fixed person~$i$ is compared to another, \textit{a-priori} similar\hyp looking person~$j$ 100~times, $i$ can increase the number of times they are perceived to weigh less than $j$ by 2.7 if they wear dark, rather than light.

\begin{figure}[tb]
    \begin{minipage}{0.5\linewidth}
        \includegraphics[width=\linewidth]{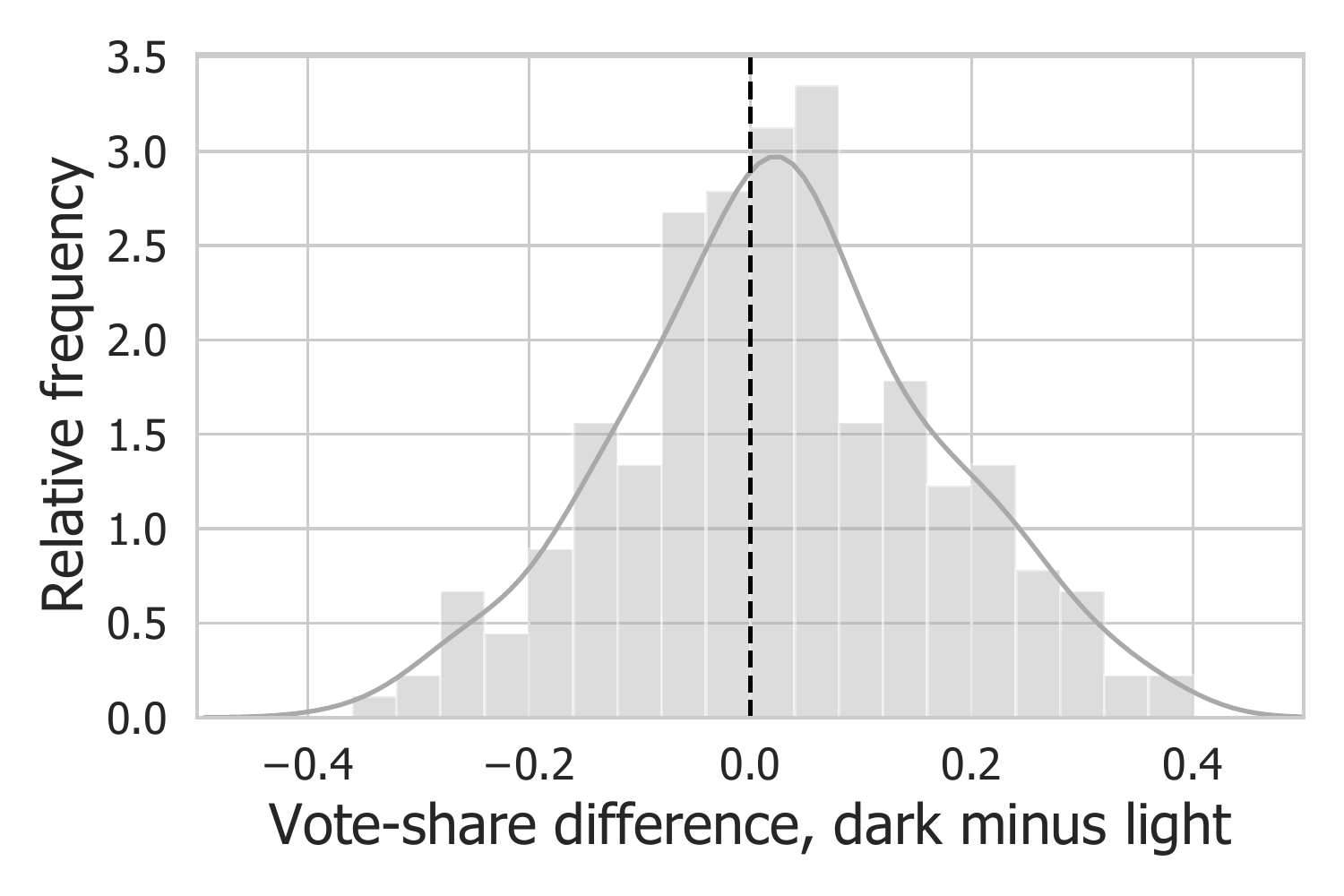}
  	\end{minipage}
    \begin{minipage}{0.48\linewidth}
    \resizebox{\linewidth}{!}{%
	\begin{tabular}{l | c | c}
		    & $L_2$ & $D_2$ \\
		 \hline
		$L_1$ & .525 & .511 \\
		& $[.499, .540]$ & $[.480, .527]$\\
		\hline
		$D_1$ & .566 & .548 \\	 
		& $[.532, .581]$ & $[.520, .563]$ \\	 
	\end{tabular}
	}
  	\end{minipage}

  \centerline{(a) Dark ($D$) \vs\ light ($L$)}
  \bigskip

    \begin{minipage}{0.5\linewidth}
        \includegraphics[width=\linewidth]{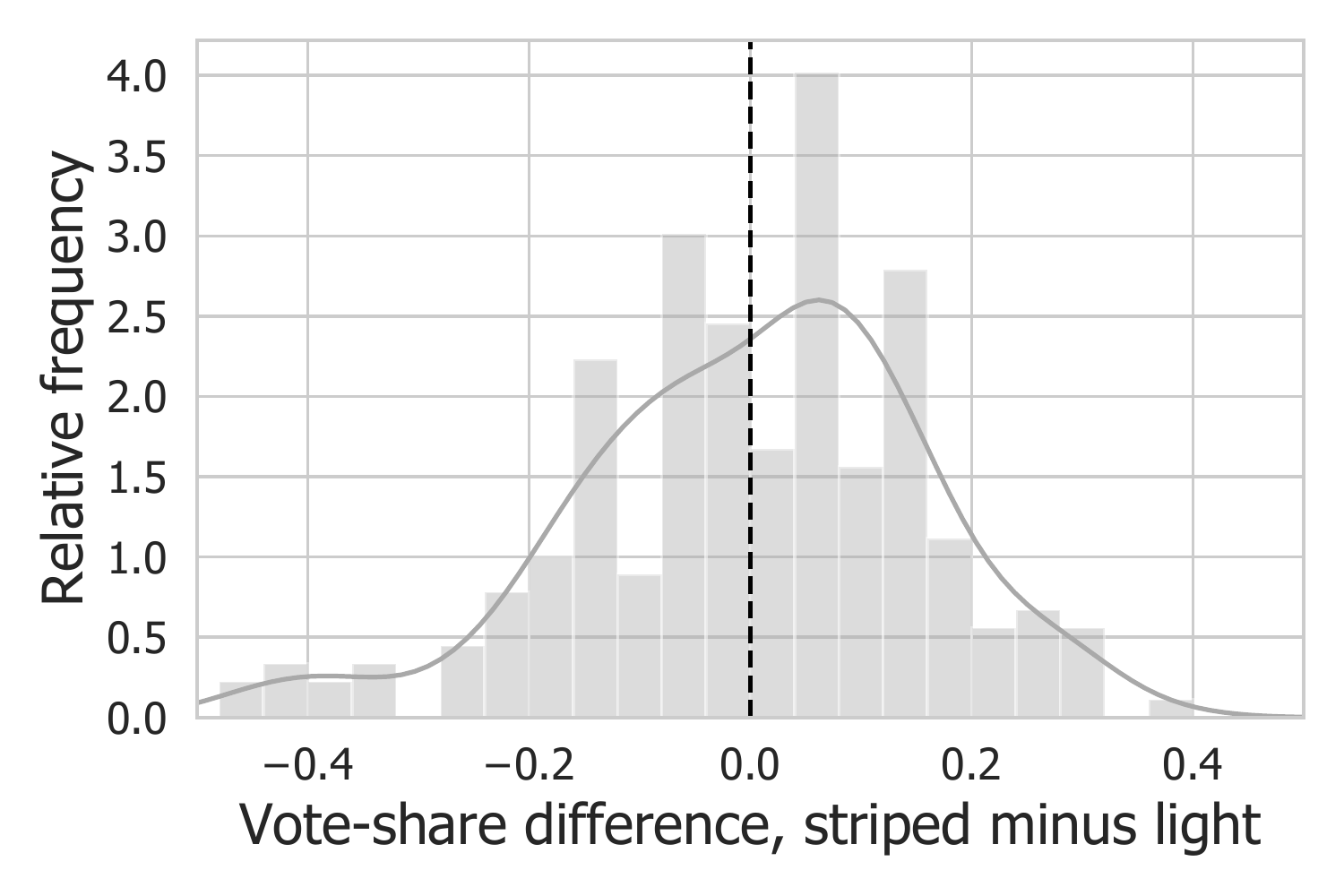}
  	\end{minipage}
  	{\small
    \begin{minipage}{0.48\linewidth}
    \resizebox{\linewidth}{!}{%
	\begin{tabular}{l | c | c}
		    & $L_2$ & $S_2$ \\
		 \hline
		$L_1$ & .513 & .536 \\
		& $[.487, .528]$ & $[.495, .552]$\\
		\hline
		$S_1$ & .537 & .549 \\
		& $[.510, .551]$ & $[.520, .564]$ \\	  
	\end{tabular}
	}
  	\end{minipage}
    }

  \centerline{(b) Horizontally striped ($S$) \vs\ light ($L$)}

	\caption{
	Results of experimental \studyTwo (\Secref{sec:Results:Experimental study 2}), comparing
	\textbf{(a)} dark ($D$) \vs\ light ($L$) and
	\textbf{(b)} horizontally striped ($S$) \vs\ light~($L$).
	\textbf{Left:} Distribution of vote-share differences ($D-L$ and $S-L$) in pairwise comparisons, where votes indicate who in the pair appears to weigh less.
	\textbf{Right:} Person~1's vote shares for all pair configurations (with 95\% CIs), where rows (columns) indicate clothing type of person~1 (person~2) in pairs; values above 0.5 indicate that person~1 was estimated to weigh less than person~2 more frequently than \textit{vice versa}.
	}
  	\label{fig:results exp2}
\end{figure}

The distribution of $\delta_i$ for the set of all people $i$ is visualized as a histogram in \Figref{fig:results exp2}a.
The advantage of dark over light is discernible as a slight right\hyp shift of the histogram.

Recall that, in experimental \studyTwo, each pair of people was rated multiple times in each of 4 configurations:
$LL$, $LD$, $DL$, $DD$.
Given this structure, we may simply count, for each configuration, how often person~1 in a pair was rated as weighing less.
The results, given in the table of \Figref{fig:results exp2}a, show that both person~1 and person~2 were always perceived as weighing less when they wore dark than when they wore light:
moving vertically down, which corresponds to person~1 switching from light to dark, increases person~1's win rate;
and similarly, moving horizontally right, which corresponds to person~2 switching from light to dark, increases person~2's win rate (manifested in the table of \Figref{fig:results exp2}a as a decrease in person~1's win rate).

Note that the overall causal effect $\delta$ can also be induced from the table of \Figref{fig:results exp2}a as the average of the 4 bottom-minus-top ($D_1L_2-L_1L_2$ and $D_1D_2-L_1D_2$) and left-minus-right ($L_1L_2-L_1D_2$ and $D_1L_2-D_1D_2$) differences, which also amounts to $\delta=2.7\%$.


Also note that the fact that all numbers in the table of \Figref{fig:results exp2}a are slightly greater than 50\% implies that there was position bias: although the order of the two people in each pair was randomized, crowd workers on average rated the first person as weighing less.
We emphasize that this does not alter our conclusions, as we work with differences of vote shares, rather than with raw vote shares directly.

\xhdr{Effect of horizontal stripes}
Repeating the above analysis for the comparison of horizontally striped \vs\ solid light clothes, we cannot determine any significant effect, with an estimated overall ``stripes-minus-light'' effect of $\delta=0.044\%$ ($p=0.58$; 95\% CI $[-2.0\%, 2.1\%]$).

The distribution of the individual $\delta_i$ and the number of vote shares for each of the 4 configurations are displayed in \Figref{fig:results exp2}b.

\xhdr{Summary}
We conclude that experimental \studyTwo confirmed the qualitative results from the observational study and from experimental \studyOne, namely that solid dark clothes make a person seem to weigh less than solid light clothes do (``$D<L$''), and that solid light clothes and horizontally striped clothes are indistinguishable in terms of weight perception (``$L \approx S$'').


\section{Discussion}

This work is concerned with the causal effects of clothing color and patterns on perceived weight.
Based on photographs taken under natural conditions and weight estimates obtained via crowdsourcing, we conducted a series of observational as well as experimental studies, all arriving at the same qualitative conclusions:
solid dark colors slightly but significantly decrease weight perception, compared to solid light colors or horizontal stripes,
whereas horizontal stripes neither increase nor decrease weight perception significantly, compared to solid light colors.

We reached these conclusions using an inherently computational approach:
through a combination of Web-based image collection, crowdsourced label collection from thousands of study participants, expert image manipulation, and automated image processing, we managed to scale our studies up by an order of magnitude, compared to previous studies in this area.
Only this way did the small effect of solid dark clothes become measurable: by switching from solid light to solid dark clothes, a person can increase their chance of appearing to weigh less than a similar\hyp looking reference person by only 2.7 percentage points on average.

On the contrary, we found no evidence of the anecdotal disadvantage of horizontal stripes:
by switching from solid light to horizontally striped clothes, a person's perceived weight did not change in a statistically significant way.

At first, it may seem that horizontal stripes had a height\hyp decreasing, rather than a weight\hyp increasing effect:
from the results of experimental \studyOne (\cf\ table in \Figref{fig:results obs+exp1}b) it appears that the strongest effect present (if any) is the slightly smaller \textit{height} estimates for horizontal stripes \vs\ solid colors (light and dark).
Although the estimated effects are small (0.22 and 0.29~cm for light and dark, respectively) and fall slightly short of the conventional 5\% significance level ($p=0.087$ and 0.052, respectively), we investigated further by running the more powerful experimental \studyTwo also for height in addition to weight.
The results were negative; no further support for a hypothetical shortening effect of horizontal stripes was found.

In addition to increasing the number of images and votes, we also went beyond previous studies in terms of the nature of the images.
Whereas prior work had mostly focused on staged photographs created specifically for the respective studies \cite{swami2012}, we started from real photographs collected from the Web, taken under natural conditions independent of the goals of our research.
This data collection process made our findings more robust against idiosyncrasies that might arise with pictures taken in narrow research contexts.
By further augmenting our image dataset through targeted manipulations performed by a professional graphic designer, our methodology marries the advantages of both realistic data and experimental control.

Nonetheless it should be noted that, since the images were gleaned from an online weight loss forum, they mostly depict overweight and obese individuals (\Tabref{tab:BMI}), a fact that should be taken into account when interpreting our findings.
Future work should specifically investigate whether the effects are identical for individuals with an underweight or normal BMI.

Although the results of all three studies presented here were consistent, it is important to point out that the studies were not pre\hyp registered and were designed and evaluated sequentially.
(The data used in the observational study was originally collected for another research project \cite{martynov2020human}.)
In order to further decrease the likelihood of false\hyp positive findings, we thus encourage future work to replicate our results in pre\hyp registered studies.

In terms of further limitations, we emphasize a point made in \Secref{sec:Weight- and height-labeled images}, namely that this paper focused on upper\hyp body clothing.
The reasons were twofold: on the one hand, a large fraction of images in our collection contain upper bodies only; on the other hand, the upper body, including the abdominal area, are anecdotally particularly important for weight perception.
Future endeavors should attempt to lend data-driven support for this assumption and should quantify the relative importance of upper- \vs\ lower-body clothing for weight perception.

Moreover, the present research considered only horizontal stripes, foregoing the study of vertical stripes, despite the fact that folk wisdom commonly claims that vertical stripes make a person appear to weigh less.
The reason for ignoring images with vertical stripes was simply that they do not occur in our dataset in sufficient numbers.
Future work should make an explicit effort to collect a larger number of photographs with vertical stripes and apply our methodology to them.
For instance, several large-scale fashion datasets are available~\cite{liu2016deepfashion,xiao2017fashion}, and automatically sifting them through for vertical stripes seems feasible given the computational techniques for color and stripe detection introduced in this work (\Secref{sec:Annotated image data}).

Despite any limitations, we would like to highlight the robustness of our results, as apparent in their consistency across the three studies.
Our observations are rendered considerably more reliable by the fact that each of the three studies is different in its own way:
The original Reddit images are unedited and thus maximally realistic in nature, but contain numerous unmeasured covariates (\eg, background, face shape, body pose, camera angle, size and fit of clothes), which might in the worst case explain away the effects measured in the observational study (\Secref{sec:Research design:Observational study}).
On the contrary, the manipulated images used in the experimental studies (\Secref{sec:Research design:Experimental study 1}--\ref{sec:Research design:Experimental study 2}) are free of confounds, but were all produced by the same graphic designer and could thus conceivably be biased by consistent minor flaws not caught by our \textit{post-hoc} filtering.
And finally, the two experimental studies were based on fundamentally different crowd labels, absolute weight guesses in the case of \studyOne, and relative weight comparisons in the case of \studyTwo.
The similarity of the results obtained from all three studies, each with its own strengths and weaknesses, puts our conclusions on more solid ground.

This study opens multiple interesting directions for future research.
For instance, we did not attempt to answer the question of causal mechanisms. In other words, \textit{why} do dark clothes make individuals appear to weigh slightly less?
Although mathematical models have been proposed~\cite{helmholtz1867,thompson2011}, empirical evidence sometimes contradicts the models \cite{ashida2013helmholtz}.
Asking crowd workers why they think a person in an image appears to weigh less could shed new light on the question of causal pathways.

Finally, we foresee numerous practical applications of the results and techniques introduced in this work.
On the one hand, our results provide a solid empirical base for everyday fashion advice.
On the other hand, machine\hyp learning\hyp driven fashion technologies~\cite{liu2016deepfashion,zhu2017your} are on the rise;
combining them with scalable crowdsourcing methods as leveraged in this paper could lead to tools for rapidly sensing how large populations would perceive certain clothing items.

\section*{Ackowledgments}
\noindent
We thank Klaus Sch\"onenberger, Jean-Philippe Thiran, Fernando P\'erez Cruz, and their teams for valuable discussions and input. We also thank Ingmar Weber and Ferda Olfi for giving us access to their image data.
This project was partly funded by a grant from the Swiss Data Science Center.
West's lab is further supported by Microsoft, Google, Facebook.

\bibliography{bibref}{}

\begin{thebibliography}{10}

\bibitem{adaval2019seeing}
Rashmi Adaval, Geetanjali Saluja, and Yuwei Jiang.
\newblock Seeing and thinking in pictures: A review of visual information
  processing.
\newblock {\em Consumer Psychology Review}, 2(1):50--69, 2019.

\bibitem{ashida2013helmholtz}
Hiroshi Ashida, Kana Kuraguchi, and Kiyofumi Miyoshi.
\newblock Helmholtz illusion makes you look fit only when you are already fit,
  but not for everyone.
\newblock {\em i-Perception}, 4(5):347--351, 2013.

\bibitem{cao2016}
Zhe Cao, Tomas Simon, Shih-En Wei, and Yaser Sheikh.
\newblock Realtime multi-person {2D} pose estimation using part affinity
  fields.
\newblock In {\em Proc.\ IEEE Conference on Computer Vision and Pattern
  Recognition}, 2017.

\bibitem{chittilappilly2016survey}
Anand~Inasu Chittilappilly, Lei Chen, and Sihem Amer-Yahia.
\newblock A survey of general-purpose crowdsourcing techniques.
\newblock {\em IEEE Transactions on Knowledge and Data Engineering},
  28(9):2246--2266, 2016.

\bibitem{cornelissen2016visual}
Katri~K. Cornelissen, Lucinda~J. Gledhill, Piers~L. Cornelissen, and Martin~J.
  Tov{\'e}e.
\newblock Visual biases in judging body weight.
\newblock {\em British Journal of Health Psychology}, 21(3):555--569, 2016.

\bibitem{feldon2000}
Leah Feldon.
\newblock {\em Does This Make Me Look Fat? The Definitive Rules for Dressing
  Thin for Every Height, Size, and Shape}.
\newblock Villard Books, New York, 2000.

\bibitem{frith2004clothing}
Hannah Frith and Kate Gleeson.
\newblock Clothing and embodiment: Men managing body image and appearance.
\newblock {\em Psychology of Men \& Masculinity}, 5(1):40, 2004.

\bibitem{kittur2008crowdsourcing}
Aniket Kittur, Ed~H. Chi, and Bongwon Suh.
\newblock Crowdsourcing user studies with {Mechanical Turk}.
\newblock In {\em Proc.\ SIGCHI Conference on Human Factors in Computing
  Systems}, 2008.

\bibitem{klepp2017}
Ingun~Grimstad Klepp and Mari Rysst.
\newblock Deviant bodies and suitable clothes.
\newblock {\em Fashion Theory}, 21(1):79--99, 2017.

\bibitem{kocabey2017face}
Enes Kocabey, Mustafa Camurcu, Ferda Ofli, Yusuf Aytar, Javier Marin, Antonio
  Torralba, and Ingmar Weber.
\newblock Face-to-{BMI}: Using computer vision to infer body mass index on
  social media.
\newblock In {\em Proc.\ International AAAI Conference on Web and Social
  Media}, 2017.

\bibitem{kremkow2014}
Jens Kremkow, Jianzhong Jin, Stanley~J. Komban, Yushi Wang, Reza Lashgari,
  Xiaobing Li, Michael Jansen, Qasim Zaidi, and Jose-Manuel Alonso.
\newblock Neuronal nonlinearity explains greater visual spatial resolution for
  darks than lights.
\newblock {\em Proceedings of the National Academy of Sciences},
  111(8):3170--3175, 2014.

\bibitem{kundt}
August Kundt.
\newblock Untersuchungen \"uber {A}ugenma{\ss} und optische {T}\"auschungen.
\newblock {\em Annalen der Physik}, 196(9):118--158, 1863.

\bibitem{lean1995waist}
M.~E.~J. Lean, T.~S. Han, and C.~E. Morrison.
\newblock Waist circumference as a measure for indicating need for weight
  management.
\newblock {\em BMJ}, 311(6998):158--161, 1995.

\bibitem{liu2016deepfashion}
Ziwei Liu, Ping Luo, Shi Qiu, Xiaogang Wang, and Xiaoou Tang.
\newblock Deepfashion: Powering robust clothes recognition and retrieval with
  rich annotations.
\newblock In {\em Proc.\ IEEE Conference on Computer Vision and Pattern
  Recognition}, 2016.

\bibitem{luz2015survey}
Nuno Luz, Nuno Silva, and Paulo Novais.
\newblock A survey of task-oriented crowdsourcing.
\newblock {\em Artificial Intelligence Review}, 44(2):187--213, 2015.

\bibitem{maisey1999}
Douglas~S. Maisey, Ellen L.~E. Vale, Piers~L. Cornelissen, and Martin~J.
  Tov\'ee.
\newblock Characteristics of male attractiveness for women.
\newblock {\em The Lancet}, 353(9163):1500, 1999.

\bibitem{martynov2020human}
Kirill Martynov, Kiran Garimella, and Robert West.
\newblock Human biases in body measurement estimation.
\newblock {\em arXiv preprint arXiv:2009.07828}, 2020.

\bibitem{savitzky1964smoothing}
Abraham Savitzky and Marcel J.~E. Golay.
\newblock Smoothing and differentiation of data by simplified least squares
  procedures.
\newblock {\em Analytical Chemistry}, 36(8):1627--1639, 1964.

\bibitem{swami2012}
Viren Swami and Amy~Sunshine Harris.
\newblock The effects of striped clothing on perceptions of body size.
\newblock {\em Social Behavior \& Personality}, 40(8):1239--1244, 2012.

\bibitem{thompson2011}
Peter Thompson and Kyriaki Mikellidou.
\newblock Applying the {H}elmholtz illusion to fashion: Horizontal stripes
  won't make you look fatter.
\newblock {\em i-Perception}, 2:69--76, 2011.

\bibitem{tovee1998}
Martin~J. Tov\'ee, Susa Reinhardt, Joanne~L. Emery, and Piers~L. Cornelissen.
\newblock Optimum body-mass index and maximum sexual attractiveness.
\newblock {\em The Lancet}, 352(9127):548, 1998.

\bibitem{helmholtz1867}
Heinrich von Helmholtz.
\newblock {\em Treatise on Physiological Optics Vol. III}.
\newblock Dover Publications, 1867.

\bibitem{voracek2003}
Martin Voracek and Maryanne Fisher.
\newblock Shapely centrefolds? {T}emporal change in body measures: Trend
  analysis.
\newblock {\em BMJ}, 325(7378):1447--1448, 2003.

\bibitem{oppel}
Nicholas~J. Wade, Dejan Todorovi\'c, David Phillips, and Bernd Lingelbach.
\newblock Johann {J}oseph oppel (1855) on geometrical optical illusions: A
  translation and commentary.
\newblock {\em i-Perception}, 8(3), 2017.

\bibitem{winakor1987effect}
Geitel Winakor and Rebecca Navarro.
\newblock Effect of achromatic value of stimulus on responses to women's
  clothing styles.
\newblock {\em Clothing and Textiles Research Journal}, 5(2):40--48, 1987.

\bibitem{xiao2017fashion}
Han Xiao, Kashif Rasul, and Roland Vollgraf.
\newblock Fashion-{MNIST}: A novel image dataset for benchmarking machine
  learning algorithms.
\newblock {\em arXiv preprint arXiv:1708.07747}, 2017.

\bibitem{yuen2011survey}
Man-Ching Yuen, Irwin King, and Kwong-Sak Leung.
\newblock A survey of crowdsourcing systems.
\newblock In {\em Proc. IEEE International Conference on Privacy, Security,
  Risk and Trust, and IEEE International Conference on Social Computing}, 2011.

\bibitem{zhu2017your}
Shizhan Zhu, Raquel Urtasun, Sanja Fidler, Dahua Lin, and Chen Change~Loy.
\newblock Be your own {P}rada: Fashion synthesis with structural coherence.
\newblock In {\em Pro.\ IEEE International Conference on Computer Vision},
  2017.

\end{thebibliography}
\bibliographystyle{plain}

\begin{strip}
\end{strip}

\begin{biography}[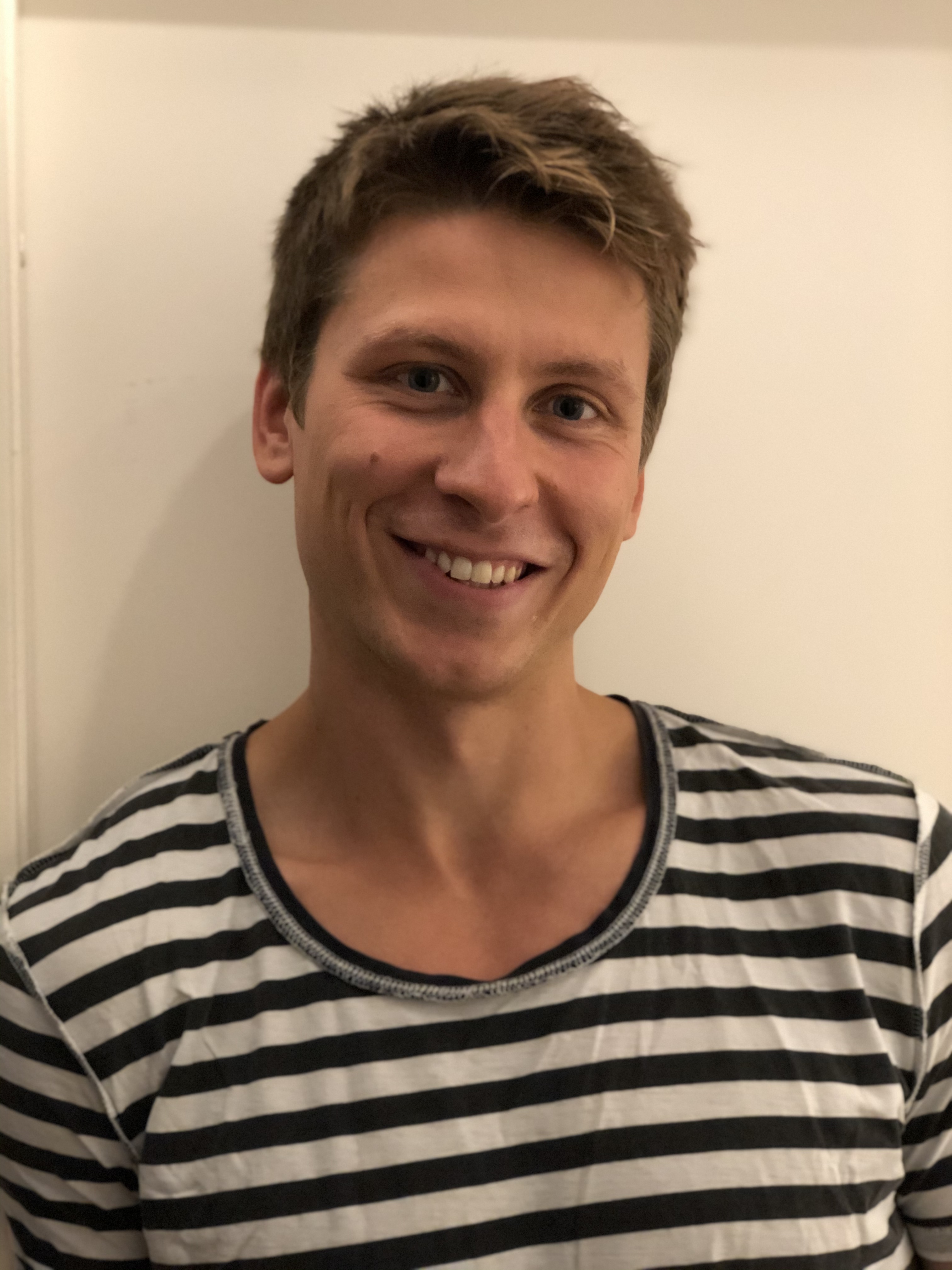]
\noindent
\textbf{Kirill Martynov} Kirill Martynov is a software engineer at Google. Before joining Google, he worked as a research assistant at the Data Science Lab of EPFL, at the Group of Mathematical Modeling of TU M\"unchen, and at the Corporate Technology Department of Siemens. Kirill received his Master's and Bachelor's degrees in Computational Science from TU M\"unchen.
\end{biography}

\begin{biography}[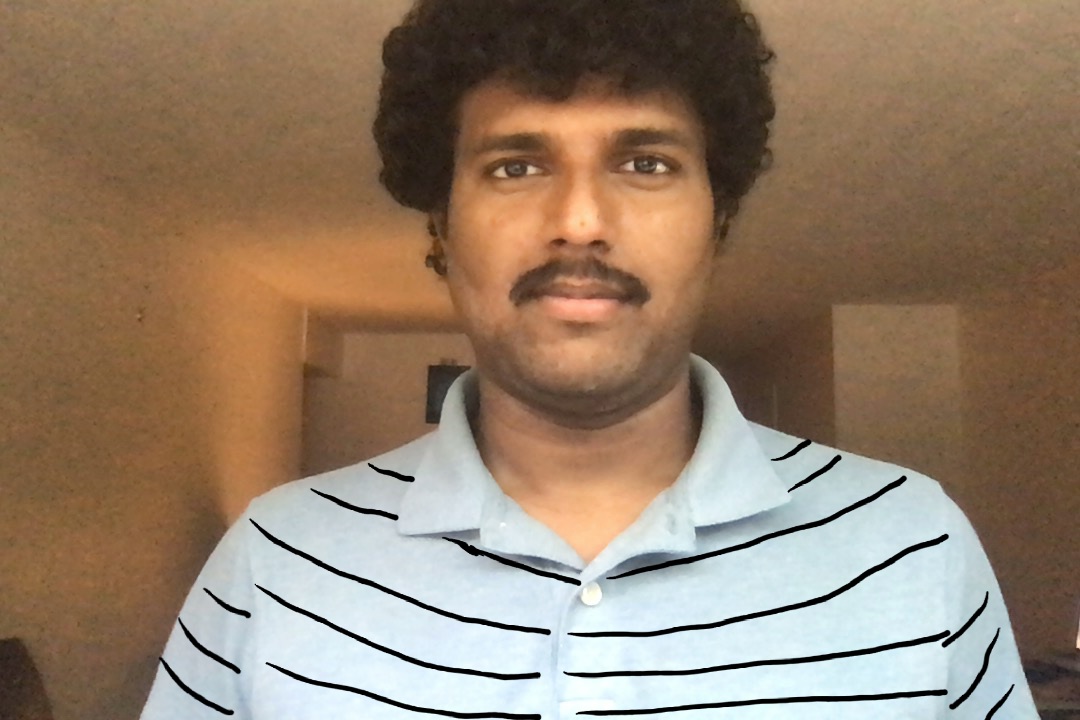]
\noindent
\textbf{Kiran Garimella} Kiran Garimella is the Michael Hammer postdoc at MIT. Before joining MIT, he was a postdoc at EPFL. His research focuses on using digital data for social good, including areas like polarization, misinformation, and human migration. His work on studying and mitigating polarization on social media won the best student paper awards at WSDM 2017 and WebScience 2017. Kiran received his PhD at Aalto University, Finland, and Masters \& Bachelors from IIIT Hyderabad, India. Prior to his PhD, he worked as a Research Engineer at Yahoo!\ Research, Barcelona, and QCRI, Doha.

\end{biography}
\vskip 22mm
\begin{biography}[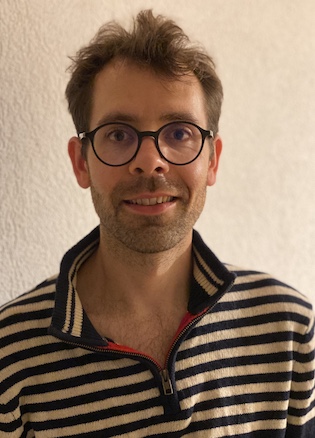]
\noindent
\textbf{Robert West} Robert West is an assistant professor of computer science at EPFL, where he heads the Data Science Lab. His research aims to understand, predict, and enhance human behavior in social and information networks by developing techniques in computational social science, data mining, network analysis, machine learning, and natural language processing. He holds a PhD in computer science from Stanford University, an MSc from McGill University, and a Dipl.-Inf.\ from TU M\"unchen.
\end{biography}

\begin{strip}
\end{strip}

\end{document}